\documentclass[onecolumn,preprintnumbers,amsmath,amssymb,superscriptaddress,showkeys]{revtex4}

\usepackage{graphicx}
\usepackage[latin1]{inputenc}
\usepackage{dcolumn}
\usepackage{bm}
\usepackage{epsfig}
\usepackage{color}

\begin{document}


\title{Bipartisanship breakdown, functional networks and forensic analysis in Spanish 2015 and 2016 national elections}

%

\author{Juan Fern\'andez-Gracia} 
\email{jfernand@hsph.harvard.edu}
\affiliation{Harvard T.H. Chan School of Public Health, Harvard University, 677 Huntington Ave, Boston, MA 02115 (USA)}
\affiliation{Instituto de Fisica Interdisciplinar y Sistemas Complejos (IFISC), CSIC-UIB, Mallorca (Spain)}
\author{Lucas Lacasa} 
\email{l.lacasa@qmul.ac.uk}
\affiliation{School of Mathematical Sciences, Queen Mary University of London, Mile End Road E14NS London (UK)\\}%

\date{\today}

\begin{abstract}
In this paper we present a social network and forensic analysis of the vote counts of Spanish national elections that took place in December 2015 and their sequel in June 2016. Vote counts are extracted at the level of municipalities, yielding an unusually high resolution dataset with over 8000 samples. We initially consider the phenomenon of Bipartisanship breakdown by analysing spatial distributions of several Bipartisanship indices. We find that such breakdown is more prominent close to cosmopolite and largely populated areas and less important in rural areas where Bipartisanship still prevails, and its evolution mildly consolidates in the 2016 round, with some evidence of Bipartisanship reinforcement which we hypothesize to be due to psychological mechanisms of risk aversion. We subsequently consider a functional network analysis by which we are able to cluster together municipalities with similar voting statistics, finding an effective partition of municipalities which remarkably coincides with the first-level political and administrative division of autonomous communities. On a third step we explore to which extent vote data are faithful by applying forensic techniques to vote statistics. To tackle this question we first address the frequencies of the first and second significant digits in vote counts and explore the conformance of these distributions at three different levels of aggregation to Benford's law for each of the main political parties. The results and interpretations are mixed and vary across different levels of aggregation, finding a general good quantitative agreement at the national scale for both municipalities and precincts but finding systematic nonconformance at the level of individual precincts. As a complementary metric, we further explore
the co-occurring statistics of voteshare and turnout, finding a mild tendency in the clusters of the conservative party to smear out towards the area of high turnout and voteshare, what has been previously interpreted as a possible sign of incremental fraud. In every case results are qualitatively similar between 2015 and 2016 elections.
\end{abstract}

\maketitle

\section{Introduction and datasets}
\begin{figure*}
\centering
\includegraphics[width=0.8\columnwidth]{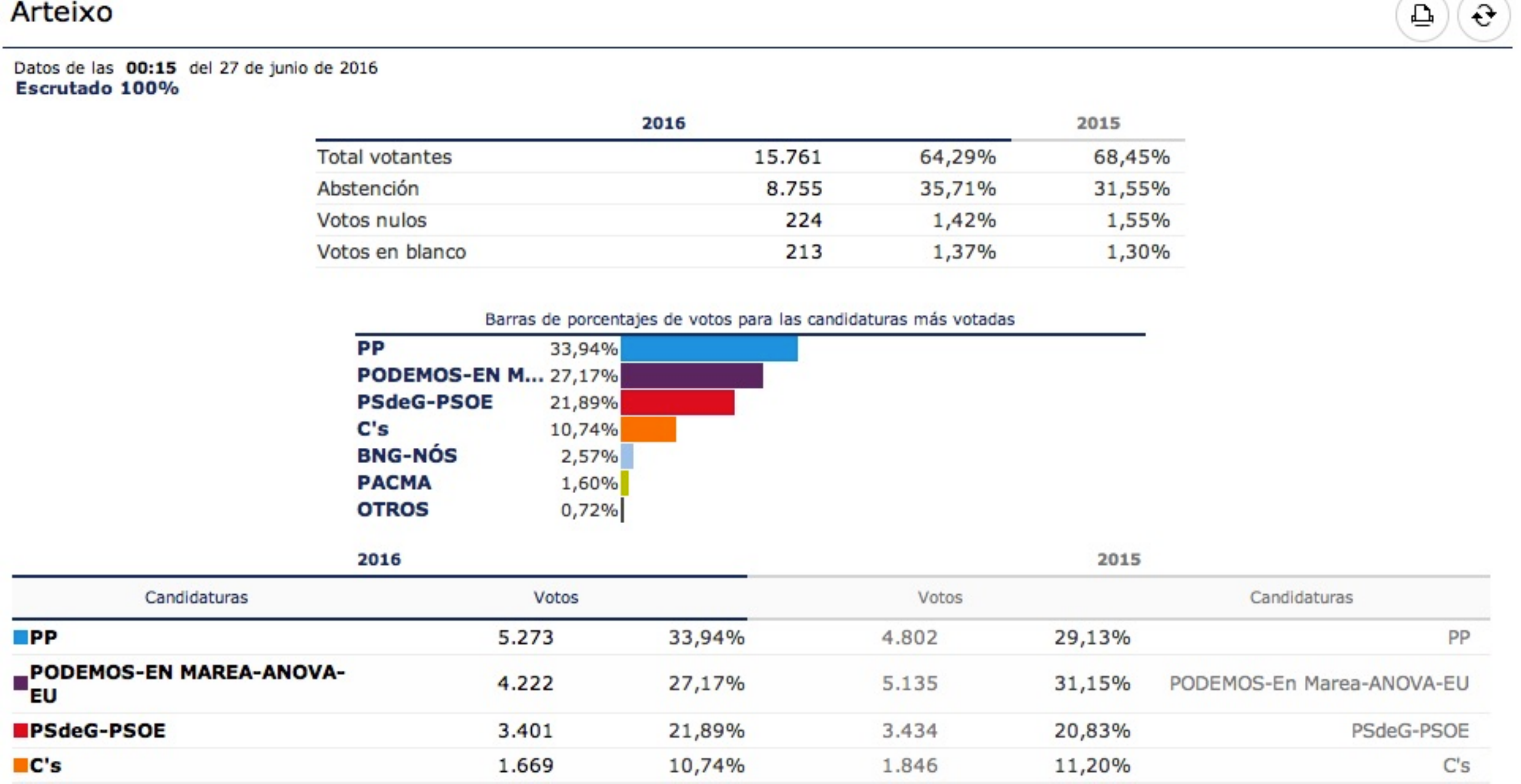}
\caption{Sample municipality (Arteixo) along with vote count statistics, as reported in the Ministerio del Interior official webpage \cite{ministerio}.}
\label{fotico}
\end{figure*}

In the last decade and in parallel with the improvement of computational resources and the possibility of accessing, storing and manipulating massive digital records easily, the political science community has engaged with the task of producing quantitative and systematic methods to detect irregularities in electoral results \cite{book}. In this work we analyze the vote count statistics obtained in the Spanish national elections that took place in December 2015 as well as in their sequel of June 2016. Since the end of 2014, the emergence of new parties such as the anti-austerity Podemos and the rise of other ones such as Ciudadanos (C's) challenged an already decadent bipartisanship system, as was evidenced by the highly fragmented total voteshare in 2015. These results further defined a new type of political equilibrium in Spain, where the quest for alliances across parties was required to form a workable majority. Unfortunately this situation was not achieved and the parliament was unable to build the necessary coalitions to make such a workable majority, what triggered the onset of new elections only six months after the previous ones, in June 2016. These special and unique conditions, together with the fact that the polls and electoral surveys preceding and on the day of the elections showed an unusually high discrepancy with the actual results motivates the use of some of the recently developed techniques for elections forensic analysis to scrutinize any source of irregularity in these elections.\\ 
High resolution vote count data (at several levels of aggregation down to the level of municipalities) have been extracted from the official webpage of the Ministerio Del Interior \cite{ministerio} (Spanish ministry of home affairs) for both 2015 and 2016 elections. For concreteness we have focused on vote counts on congress and discarded senate (see Fig.~\ref{fotico} for a guide of the type of data available from the ministry of home affairs website).

\noindent The high quality of the dataset under study allows us to address important social and political questions in a scientific way. In this work we have considered three specific questions, namely (i) can we confirm that the bipartisanship system is challenged? (ii) Can a social network analysis reveal quantitative information on the voting profiles and similarities across municipalities and regions? And (iii) can we scrutinize these data against forensic techniques? And if so, is there any evidence of fraud?\\

\noindent To address the first question we will define a set of Bipartisanship indices and will explore the spatial distribution of these over the Iberian peninsula of Spain (at the fine-grained level of municipalities), while for the second question we will build on state of the art community detection methods (Infomap) applied on a functional-like network of municipalities extracted via cosine similarity.\\
The third question will be addressed via two different studies. The first one addresses the deviation or conformance of vote counts statistics to the so-called Benford's law \cite{nigrini, mebane} that predicts that the first significant digits in some datasets (including vote counts) should follow an inverse-logarithmic distribution. The rationale for this analysis is that statistically significant deviations between the empirical distribution and the theoretical one point us toward electoral irregularities. These irregularities might in turn be due either to unintentional mismanagement of the voting process and/or to fraud. This type of analysis only flags the existence of such irregularities and gives no judgment on what was the cause for such irregularity. To complement this study, we then explore the presence and detection of sources of incremental and extreme fraud from the co-occurring statistics of vote and turnout numbers, following a recent study \cite{klimek}.\\

\noindent The rest of the paper goes as follows: in section II we introduce similarity indices and explore the spatial distribution of these. Having access to the 2015 and 2016 election statistics, we will be able to explore the social effect of distrust and the possible longitudinal progression of Bipartisanship breakdown. In section III we perform the social network analysis of the data, creating functional networks via cosine similarity computed on the vote profile at the level of municipalities. Then, in section IV we focus on forensic methods. We introduce Benford's law along with the precise types of statistical tests that have been proposed in the realm of election forensics, and we present the results obtained from these tests for both the December 2015 and June 2016 elections at three different levels of aggregation. The main results and interpretations on this first study are reported in this section and additional material and analysis are shown in an appendix. In this section we also present the second forensic analysis that addresses the co-occurring statistics of vote and turnout numbers. Finally, in section V we provide some discussion and conclude.

\section{Bipartisanship indices}
Votecounts can be aggregated at several spatial levels (municipalities, precincts, etc). In general, for a specific region (e.g. a given municipality) vote data consist of a vector of vote percentage ${\bf v}=(v_1,v_2,\dots)$ where $v_i$ is the percentage of votes to party $i$, and $\sum_{i=1}^N v_i=1$, $N$ being the total number of parties with representation in that region. Whereas in a majority of municipalities the main national-wide parties PP, PSOE, Cs and PODEMOS have representation, other smaller, regional (or otherwise) parties also appear with different frequency in the different municipalities. In other words, $N$ (total number of parties) might fluctuate from municipality to municipality.\\

\begin{figure*}
\centering
\includegraphics[width=0.32\columnwidth]{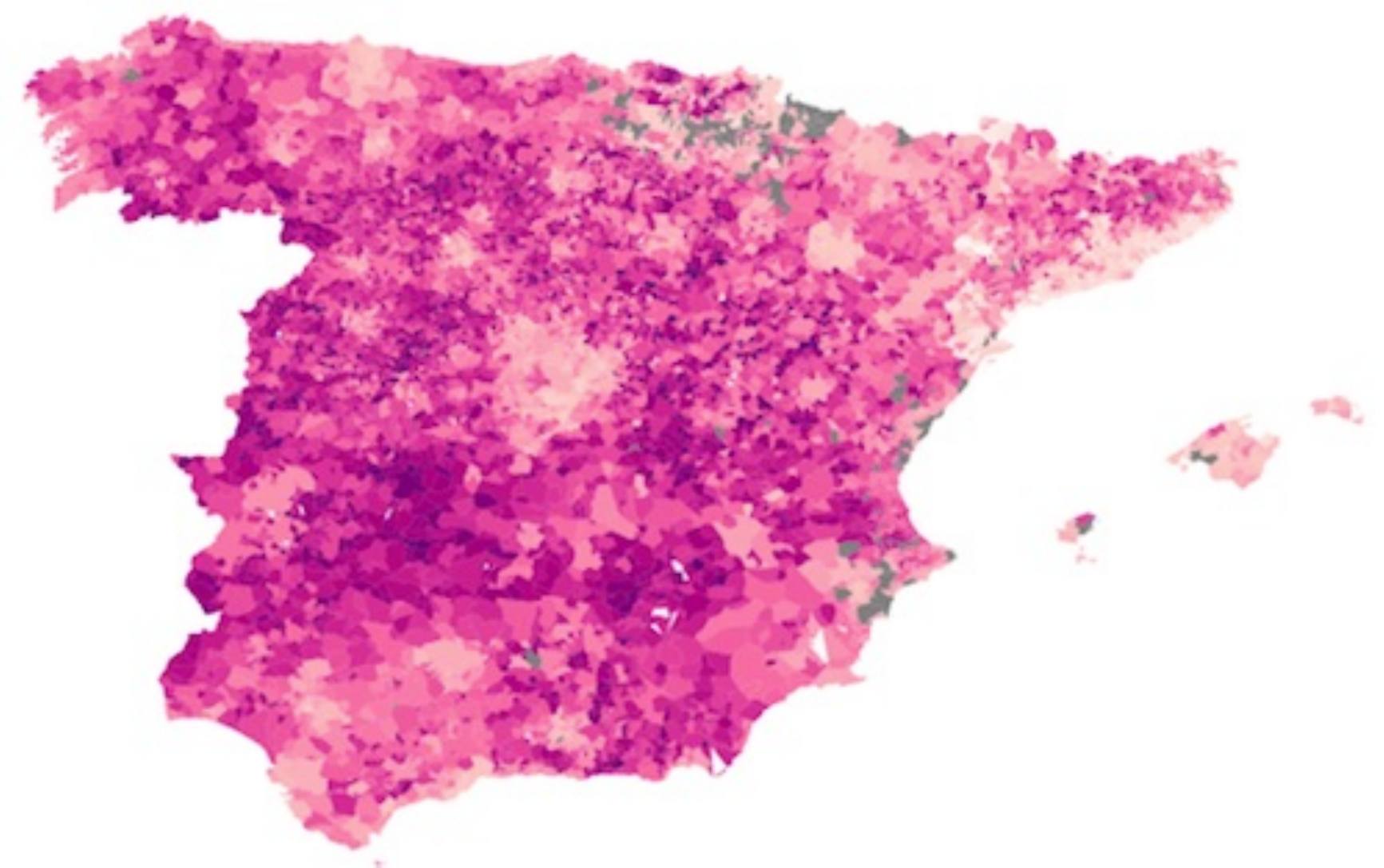}
\includegraphics[width=0.32\columnwidth]{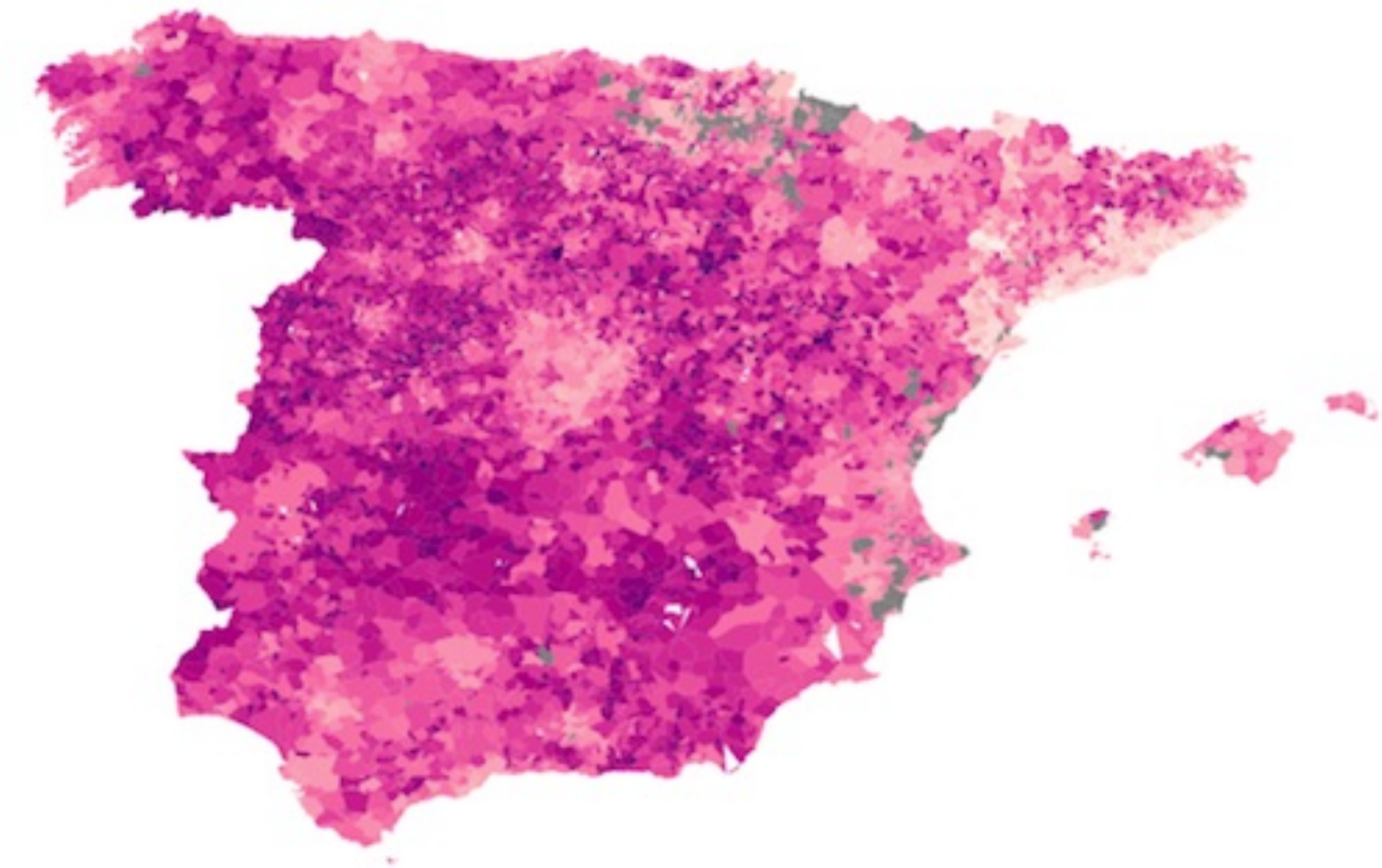}
\includegraphics[width=0.32\columnwidth]{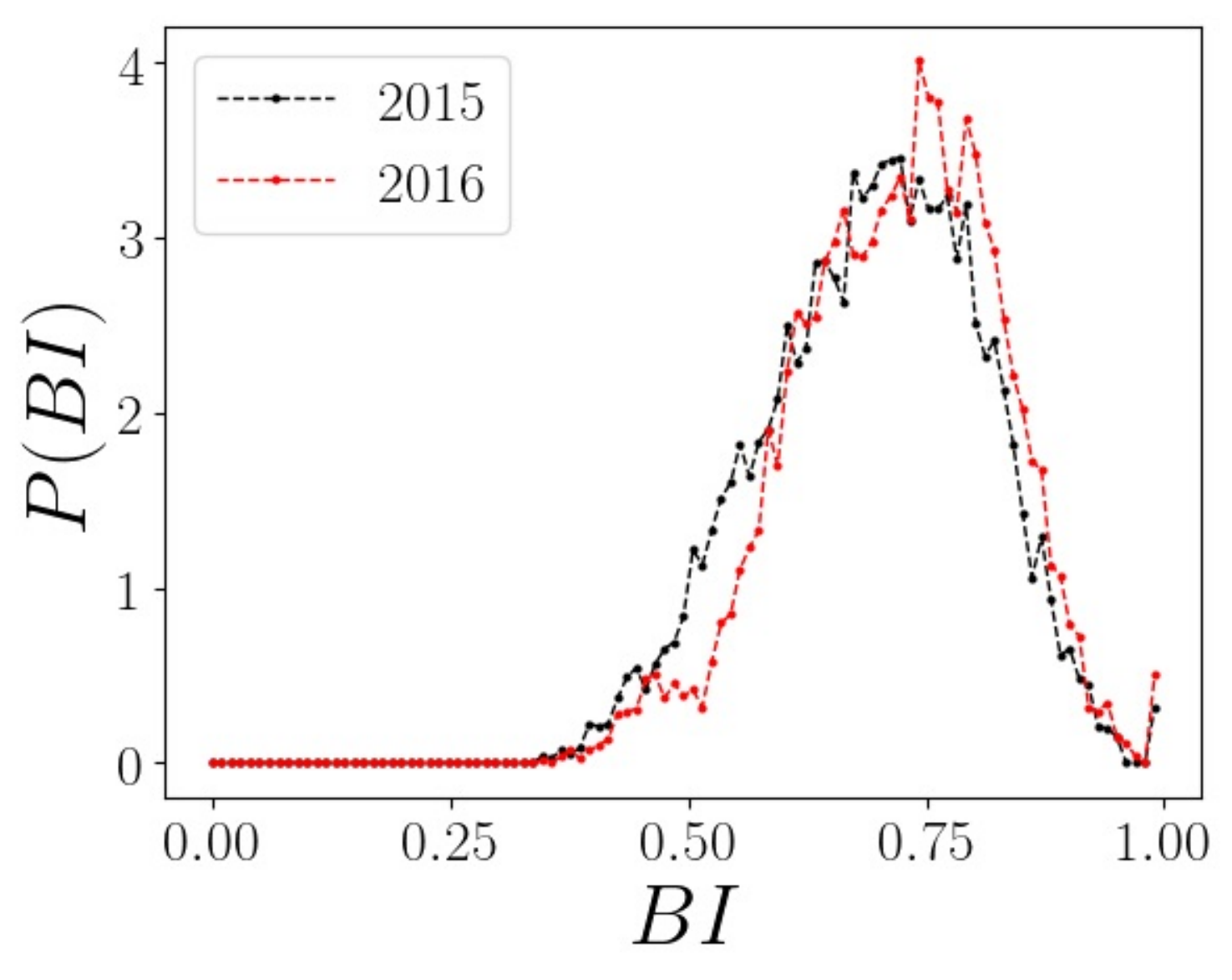}
\caption{(Left and middle panels) Heat map of the Bipartisanship index (BI, see the text) at the municipality level (the darker, the larger BI, and the more bipartisanship the system is) for 2015 elections (left panel) and 2016 elections (right panel). One can see that overall Bipartisanship decreases from 2015 to 2016, and that Bipartisanship breakdown is more acute closer to important, cosmopolite cities (e.g. Madrid, Barcelona, etc). (Right panel) Frequency histogram of BI for 2015 and 2016. In 2015 the distribution has a clear Gaussian shape and such shape is slightly perturbed in 2016.}
\label{fig:BI}
\end{figure*}

\noindent {\bf Bipartisanship index (BI). }As a crude metric, we initially define a {\it Bipartisanship Index} ($\text{BI}_i$) of a given municipality (respectively precinct, etc) $i$ as the sum of the vote percentage ratio (between 0 and 1) of the two most-voted parties in $i$. For a purely bipartisanship region $i$ ($N=2$), $\text{BI}_i$ approaches $1$, whereas in the ideal case of a multipartidist region, BI would approach its minimum value $2/N$. This metric allows us to compare the relative level of Bipartisanship across regions.\\
In Fig.\ref{fig:BI} we provide a spatial heat map of Spain where we plot $\text{BI}_i$ for each municipality $i=1,2,\dots,8215$, for the 2015 and 2016 elections data respectively (note that for illustration constraints Canary Islands are not represented here, but we shall emphasize that data and results there are qualitatively equivalent to those obtained for the Iberian peninsula and Balearic Islands). First, we can observe that there is a clear tendency towards relatively lower Bipartisanship in areas that correspond to highly-populated regions, e.g. Madrid, Catalonia. This finding is well aligned with the social observation that the process of Bipartisanship breakdown, as other social changes, initially develops close to important cosmopolite cities and then percolate to more rural areas. A second interesting finding is that from 2015 to 2016 there is a {\it stalling} in the Bipartisanship breakdown, and its {average index over all spanish municipalities even slightly increases from $\langle \text{BI}\rangle=0.70\pm 0.12$ in 2015 to $\langle \text{BI}\rangle=0.73\pm 0.11$ in 2016} (this small increase is however within error bars so one cannot rule out this being a statistical artifact). A possible sociological interpretation is the following: after the 2015 elections parliament was unable to build the necessary coalitions to make a workable majority, and this was the trigger to the new elections only six months after the previous ones, in June 2016. As society is averse to the uncertainty generated by frustrated elections, risk aversion might have stalled the overall inertia that a priori was driving the Bipartisanship breakdown, and the fear of not being able to find workable majorities might have forced some voters in the most conservative regions to turn back to a bipartisanship strategy that indeed guarantees those much-needed majorities.\\
At a regional level we can observe that this phenomenon is highly heterogeneous: whereas there is a very acute trend towards Bipartisanship breakdown consolidation in specific regions such as the autonomous community of Valencia, in other regions such as the autonomous community of Galicia the trend is pretty much the opposite.\\
In the right panel of Fig.\ref{fig:BI} we plot the frequency histogram of Bipartisanship indices, for both 2015 and 2016. For 2015 a clear Gaussian-like shape emerges, and such shape is slightly perturbed for the 2016 case. In the latter case we observe weird peaks and pits emerging in the distribution leaving a trace of wild fluctuations which are not present in the 2015 statistics. Comparing both histograms we can perceive a subtle shift towards higher values of BI. \\

\noindent {\bf Entropy index (E). }Intuitively, Bipartisanship tends to accumulate vote percentage in a few parties, whereas multipartidism tends to spread out votes across parties. Following this argument, in order to quantify more precisely the concentration of vote percentages over the whole set of parties one can define an {\it entropy index} for region $i$ as
$$E_i=-\frac{\sum_{j=1}^N v_j \log v_j}{\log N}$$
This quantity is bounded in $0\leq E_i\leq1$, reaching the minimum for unipartidism (a single party gets 100$\%$ of the votes in the region) and reaching its maximum for multipartidism (all $N$ parties get the same percentage of votes).\\
Qualitatively we have found similar results when exploring spatial distributions of the entropy index $E_i$ to those obtained with the more crude quantifier BI (see Fig. \ref{fig:E} in the appendix).\\

\begin{figure*}
\centering
\includegraphics[width=0.32\columnwidth]{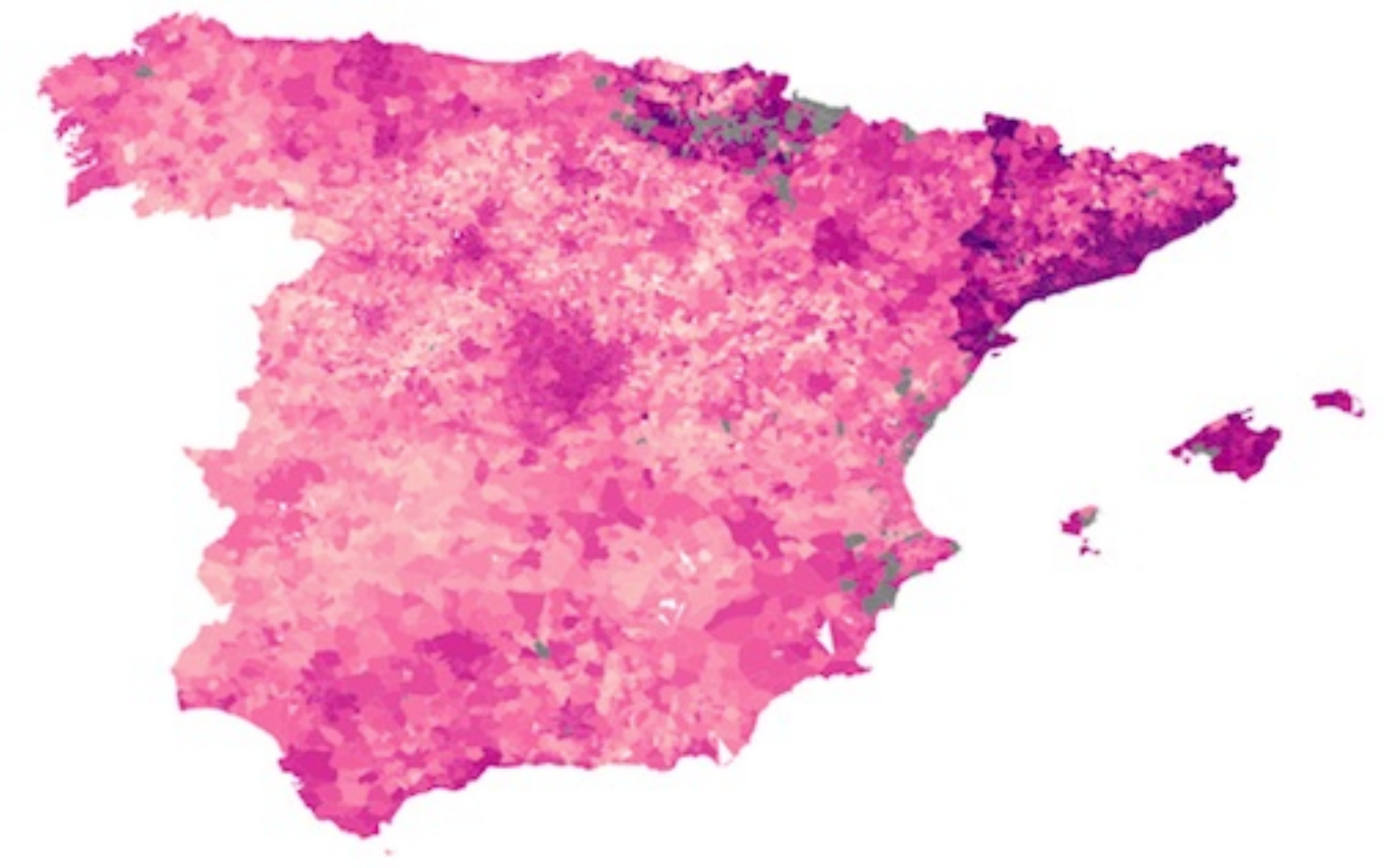}
\includegraphics[width=0.32\columnwidth]{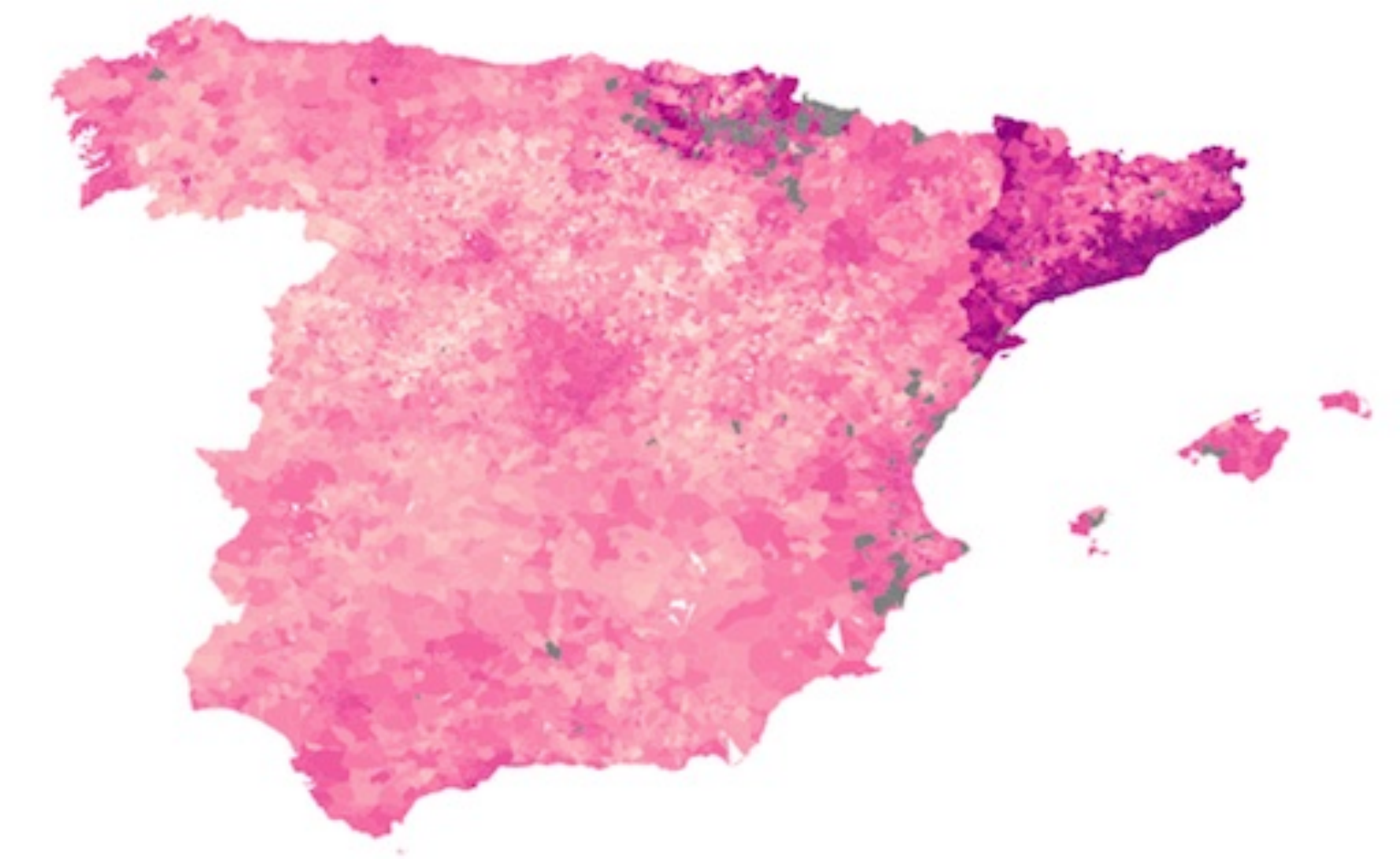}
\includegraphics[width=0.32\columnwidth]{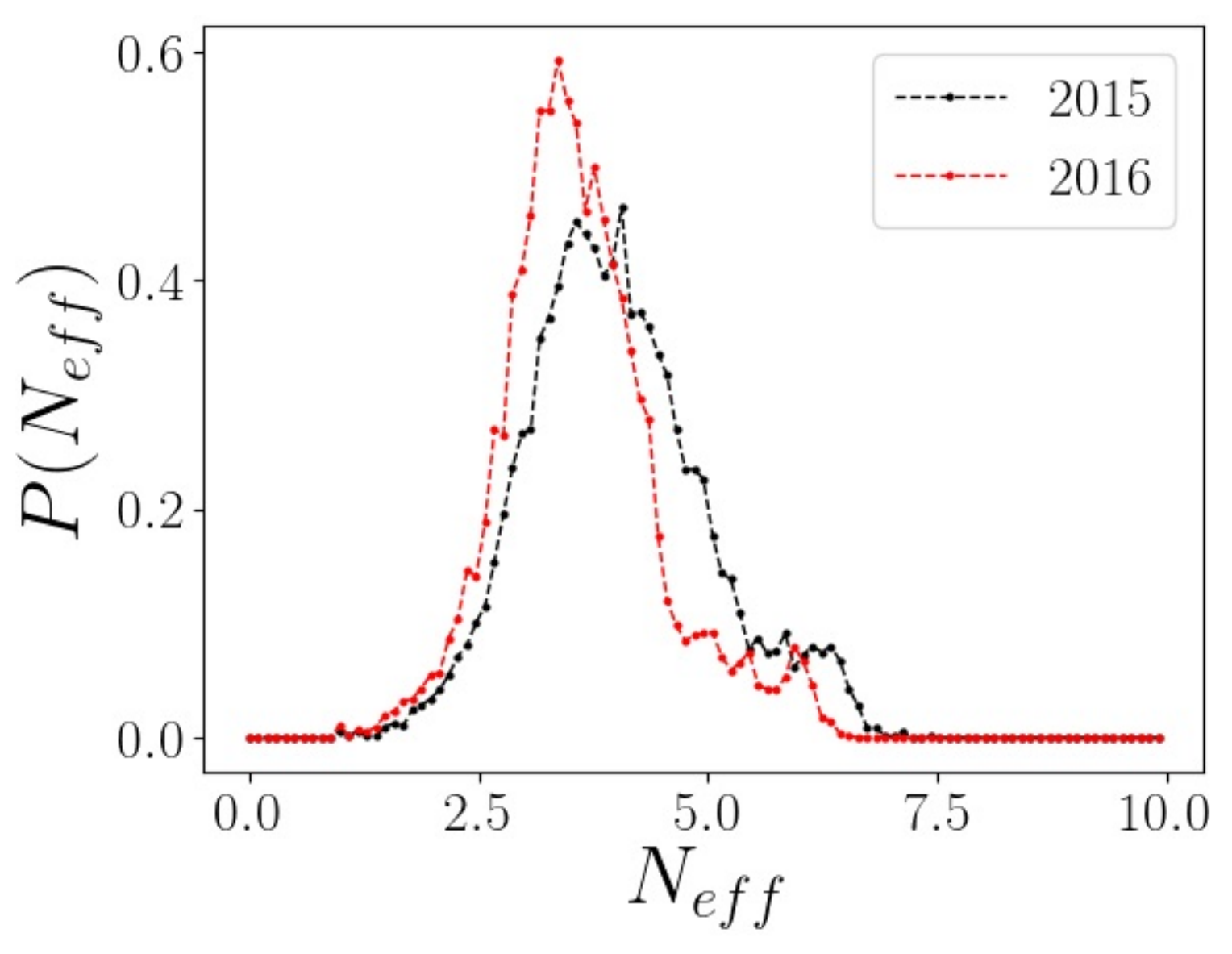}
\caption{(Left and middle panels) Heat map of the diversity index ($N_{\text{eff}}$, see the text) at the municipality level (the darker, the larger $N_{\text{eff}}$, and the less bipartisanship the system is) for 2015 elections (left panel) and 2016 elections (right panel). One can see that overall the system is more diverse in 2015 and its diversity decreases in 2016. This result is in consonance with Bipartisanship breakdown stalling, probably due to risk aversion that generates the undertainty associated to not obtaining workable majorities in parliament. (Right panel) Frequency histogram of $N_{\text{eff}}$ for 2015 and 2016, highlighting an initially diverse voting ecosystem with $\langle N_{\text{eff}} \rangle= 4.1$ (over four relevant parties on average) in 2015, and a clear shift towards a smaller diversity in 2016.}
\label{fig:Neff}
\end{figure*}

\noindent {\bf Diversity index. }Inspired by ecological metrics \cite{eco}, here we further define the diversity index $N_{\text{eff}}$ of a given region as the effective number of political parties. This refers to the number of `equally voted parties' needed to obtain the same mean proportional parties vote percentage as that observed in the dataset (where all parties may not be equally abundant):
$$N_{\text{eff}}=\exp(E \log N),$$
where $E$ is the entropy index as defined above. In other words, $N_{\text{eff}}$ counts, assuming that effective parties all get the same number of votes, the number of such effective parties one would need to find the same entropy index as found by computing $E$ to the vote statistics. In the left and middle panels of figure \ref{fig:Neff} we plot such index for 2015 (left) and 2016 (middle) elections. We clearly observe two important stylized facts, namely (i) as already found in BI and $E$, there is a clear separation between regions close to cosmopolite and largely populated cities, whose diversity index tends to be large, entailing a high number of effective parties at play, and regions which are typically less populated (rural areas) with lower diversity index. (ii) At odds with BI, the overall index evidences a marked decrease between 2015 and 2016, visually observed by a drift in heat map towards lighter color (i.e. lower values of diversity) and a change in the diversity distribution (right panel of figure \ref{fig:Neff}).

\section{Functional network analysis}

In this section we make use of tools from Network Theory \cite{network} to explore the vote similarity across regions. Attached to each municipality we consider the vote percentage vector $\bf v$ defined above. We measure the similarity between the voting statistics of two municipalities $i$ and $j$ via the so-called cosine similarity
$$S_{ij}=\frac{\langle {\bf v}^{(i)},{\bf v}^{(j)}\rangle}{||{\bf v}^{(i)}|| \cdot ||{\bf v}^{(i)}||},$$
where $\langle\cdot,\cdot\rangle$ is the standard scalar product and $||\cdot||$ is the ${\ell}_2$ norm. By construction, $0\leq S_{ij}\leq 1$, where complete similarity ($S_{ij}=1$) is reached when the vote statistics are identical and null similarity is reached when ${\bf v}^{(i)}\perp {\bf v}^{(j)}$, i.e. when a non-null percentage to a given party in one municipality is always matched to a null percentage in the other municipality.\\

\begin{figure*}
\centering
\includegraphics[width=0.38\columnwidth]{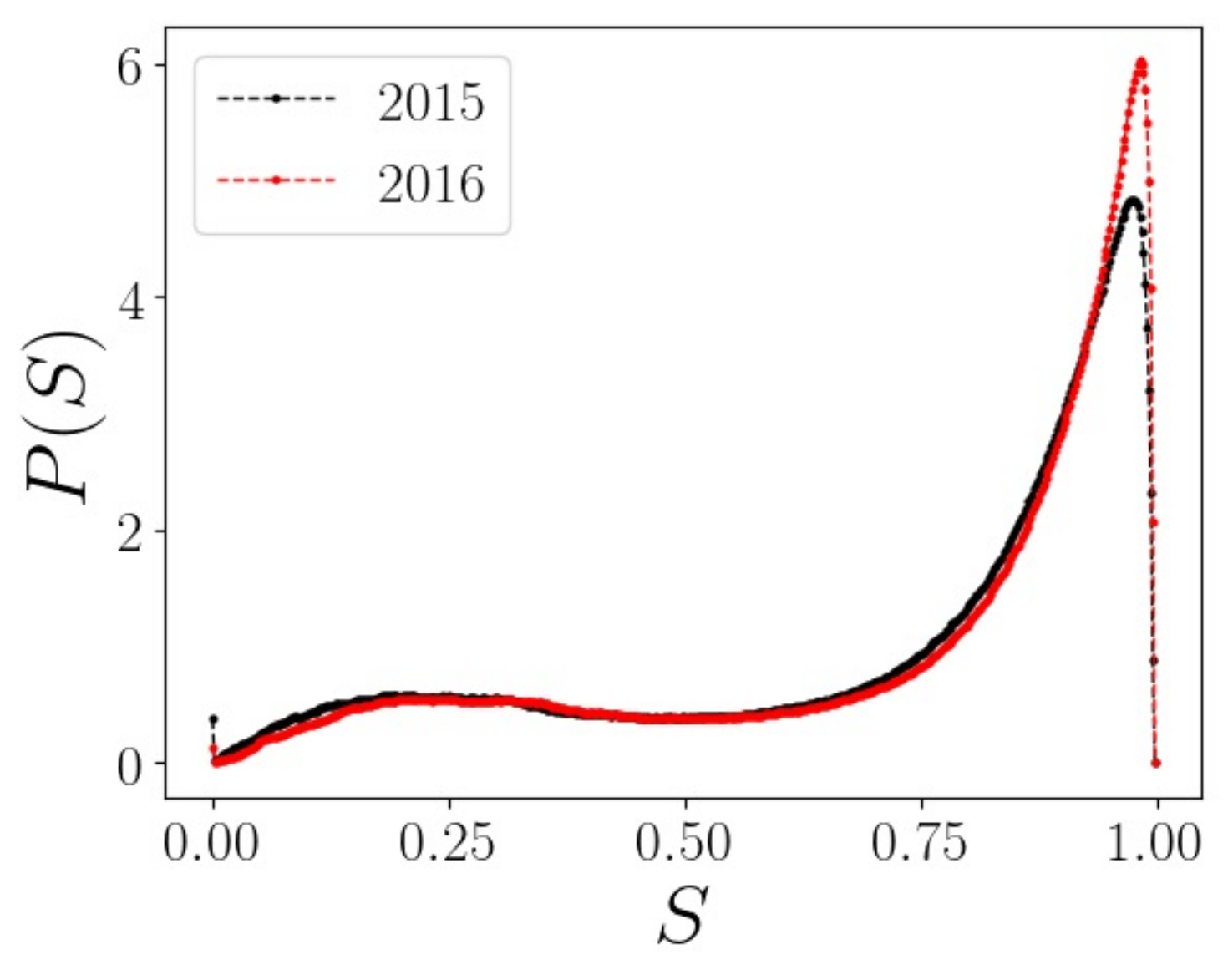}
\caption{Estimated probability density of similarity values $P(S)$ for the fully connected, weighted similarity matrix {\bf S}.}
\label{fig:P_S}
\end{figure*}

\begin{figure*}
\centering
\includegraphics[width=0.45\columnwidth]{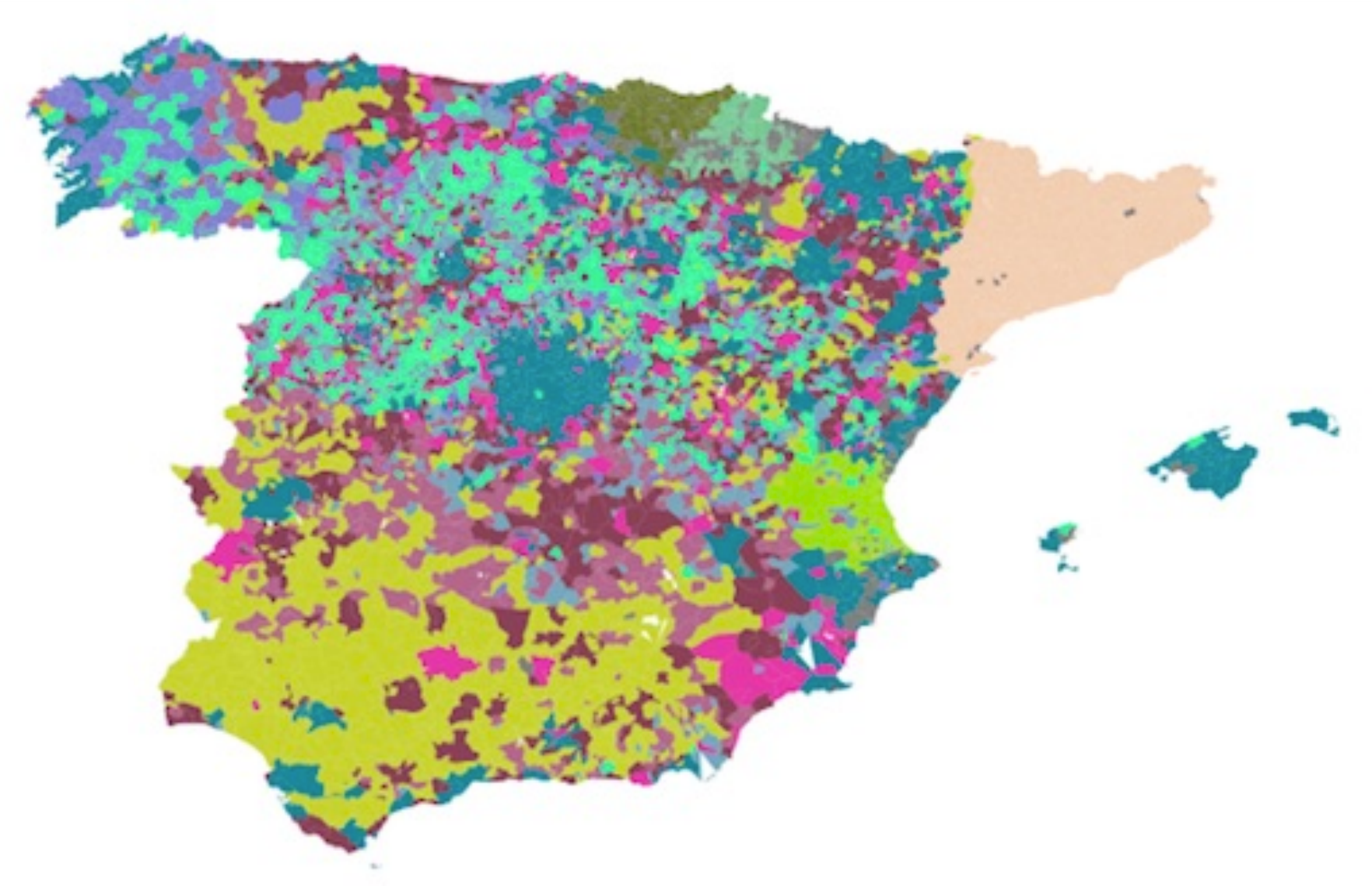}
\includegraphics[width=0.45\columnwidth]{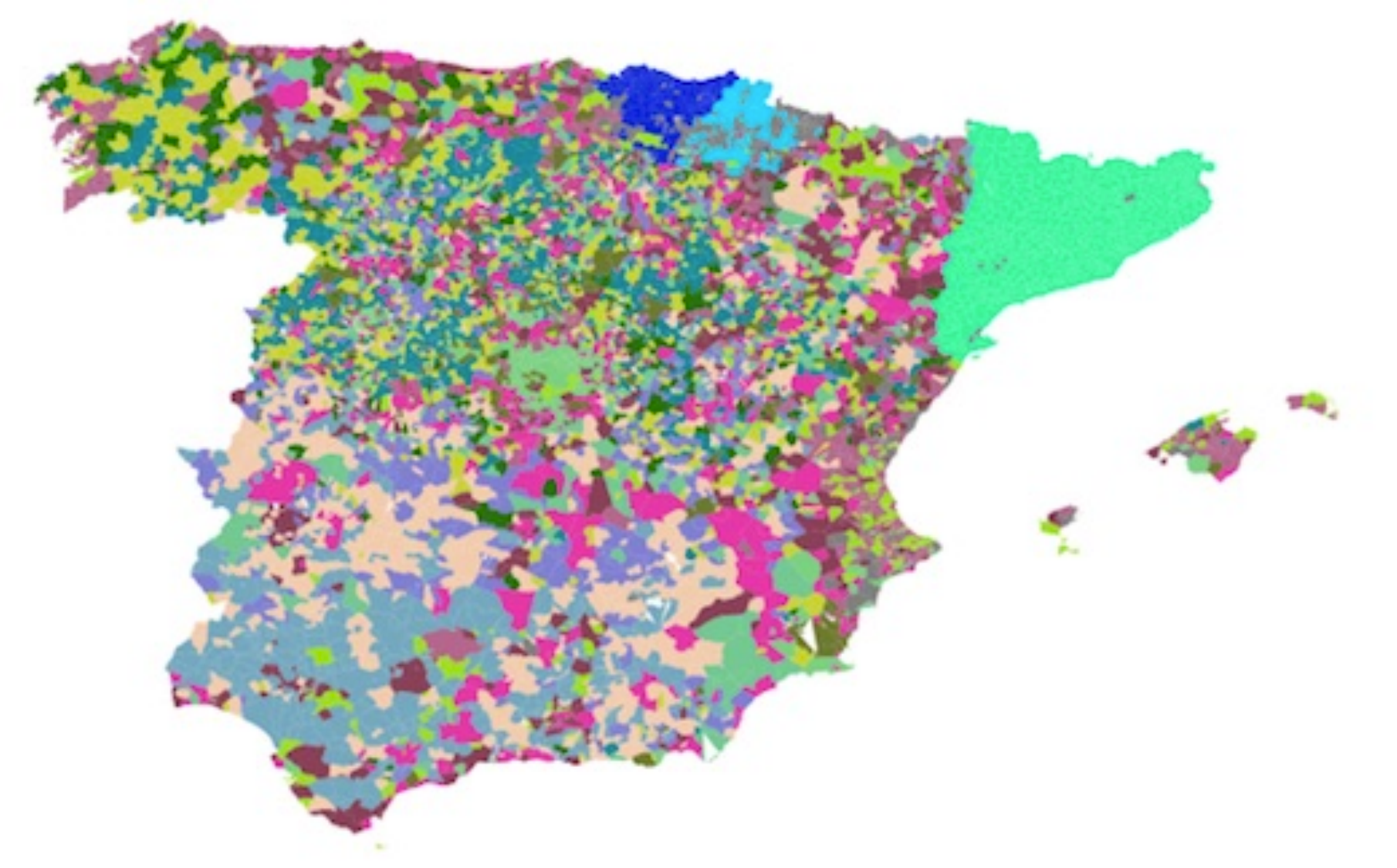}
\caption{First level communities obtained via community detection algorithm Infomap, performed on the thresholded, undirected and weighted network based on the similarity matrix {\bf S} (see the text). Here we color nodes belonging to the same community in the same colour. There is an appreciable matching between functional network communities in the network of municipalities and actual Spanish autonomous communities, although this correspondence is much more evident for the 2015 round, where Bipartisanship breakdown was at its zenith.}
\label{fig:net}
\end{figure*}

The matrix ${\bf S}=\{S_{ij}\}$ naturally defines a fully-connected weighted network where nodes are municipalities, and every pair of nodes $i$ and $j$ are linked with an edge with weight $S_{ij}$. In our database we have a total number of about $8\cdot 10^3$ municipalities, hence a fully connected network of about $8\cdot 10^3$ nodes and $64\cdot 10^6$ edges.  {In fig \ref{fig:P_S} we plot the edge's weight probability density $P(S)$. As expected, weights are concentrated in a region relatively close to the maximum value, something that can be justified already by noting that, in most of the municipalities, all four parties PP, PSOE, Cs and PODEMOS} have a non-null vote percentage. To get some insight beyond this simple statistc, we now run an algorithm to detect communities --i.e. large groups of nodes which are similar to each other and less similar to the nodes in the other groups--. To do this we run Infomap \cite{infomap_th, infomap} on the undirected and weighted network after a simple thresholding is performed on ${\bf S}$: for a given node $i$, we only conserve the largest similarity weights $S_{ij}$ such that all nodes at least have degree $k=10$. In other words, we perform a parallel pruning on the edges, starting from those with smaller weights, and we prune the network up to the point where we cannot prune anymore if we want to make sure every node has a degree $k\geq 10$. The first level of Infomap reveals a total of $14$ communities in 2015 and $16$ communities in 2016. In fig. \ref{fig:net} we plot a spatial projection of the network, where nodes are municipalities. For exposition reasons, edges are not plotted in the figure as otherwise the image wouldn't carry much information. Nodes belonging to the same network community are colored equally. Resemblance between network communities obtained via Infomap in our functional network and actual Spanish autonomous communities is remarkable, and such similitude is more acute in 2015 --where Bipartisanship breakdown is slightly more pronounced-- than in 2016, where Bipartisanship slightly reduced --as reported by the diversity index-- and this has the effect of fuzzing up the relation between network communities and autonomous community divisions. Interestingly, one finds what we could call stable and compact autonomous communities (those which have a well-defined, stable over time and cohesive functional community counterpart): Catalonia, Madrid, Basque Country, Navarra, whereas other set of communities have a clear counterpart in 2015 but a fuzzier one in 2016 (e.g. Comunitat Valenciana, Andalucia, Murcia, etc). Another set of autonomous communities present a highly heterogeneous voting profiles and don't present any clear functional community counterpart.
Theoretical digressions and insights that could give a socio-political justification for this classification are left as an open question for future work.

\section{Forensic analysis}
\subsection{Benford's law}
The first significant digit (or leading digit) of a number is defined as its non-zero leftmost digit (for instance the leading digit of 123 is 1 whereas the leading digit of 0.025 is 2). The so-called Benford's law is an empirical statistical law stating that in particular types of numeric datasets the probality of finding an entry whose first significant digit is $d$ decays logarithmically as
\begin{equation}
P(d)=\log_{10}(1+1/d),
\label{BL}
\end{equation}
where $\log_{10}$ stands here for the decimal logarithm (note that trivially $\sum_{d=1}^9 P(d)=1)$. Perhaps counterintuitively, this law is quite different from the expected distribution arising from an uncorrelated random process (e.g. coin tossing or extracting numbers at random from an urn) which would yield a uniform distribution where every leading digit would be equally likely to appear.
The logarithmically decaying shape given in eq.\ref{BL} was empirically found first in 1881 by astronomer Simon Newcomb and later popularized and exhaustively studied by Frank Benford \cite{benford}. Empirical datasets that comply to Benford's law emerge in as disparate places as for stock prizes or physical constants, and some mathematical sequences such as binomial arrays or some geometric sequences have been shown to conform to Benford. A possible origin of this law has been rigorously explained by Hill \cite{hill}, who proved a central limit-type theorem by which random entries picked from random distributions form a sequence whose leading digit distribution converges to Benford's law. Another explanation comes from the theory of multiplicative processes, as it is well known that power-law distributed stochastic processes follow Benford's law for the specific case of a density $1/x$ (see \cite{tolo} and references therein for details). In practice, this law is expected to emerge in a range of empirical datasets where part or all of the following criteria hold: (i) the data ranges a broad interval encompassing several orders of magnitude rather uniformly, (ii) the data are the outcome of different random processes with different probability densities, (iii) the data are the result of one or several multiplicative processes. \\

\noindent Mainly advocated by Nigrini \cite{nigrini}, the application of Benford's law to detect fraud and irregularities -by observing anomalous and statistically significant deviations from eq.\ref{BL} for datasets which otherwise should conform to that distribution- has become popular in recent years, and from now on we quote this a 1BL test. Mansilla \cite{mansilla} and Roukema \cite{roukema} applied this methodology to assess Mexican and Iranian vote count results respectively. On the other hand, Mebane \cite{mebane} advocates instead to look at the second significant digit (which follows an extended version of Benford's law \cite{hill2}) and argues that the frequencies of election vote counts at precinct level approximate a Benford distribution for the second digit, and accordingly mismanaged or fraudulent manipulation of vote counts would induce a statistically significant deviation in the distribution of the second leading digit, detected by a simple Pearson $\chi^2$ goodness of fit test. Mebane applied this so-called 2BL test to assess the cases of Florida 2004 and Mexico 2006, and other authors have subsequently applied this in many other occasions (see \cite{germany} and references therein). In this case the theoretical distribution takes a more convoluted shape than eq.\ref{BL}, namely
\begin{equation}
P_2(d)=\sum_{k=1}^{9}\log_{10}\left(1+\frac{1}{10k+d}\right),
\label{2BL}
\end{equation}
and a good numerical approximation \cite{hill2,germany} is given by
$$P_2(d)\simeq(0.11968, 0.11389, 0.10882, 0.10433, 0.10031, 0.09668, 0.09337, 0.09035, 0.08757, 0.08500).$$\\ 

\begin{figure*}
\centering
\includegraphics[width=0.4\columnwidth]{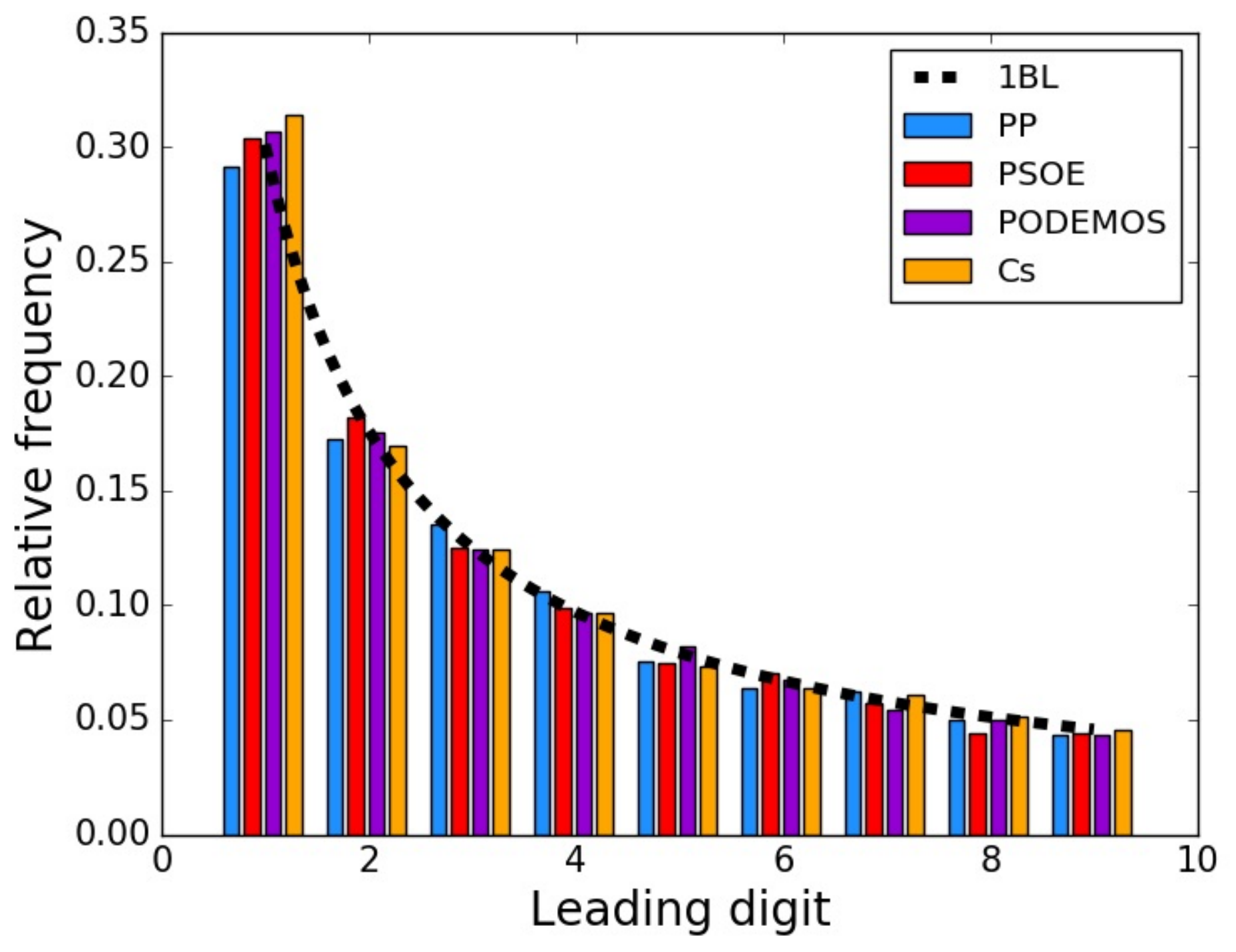}
\includegraphics[width=0.4\columnwidth]{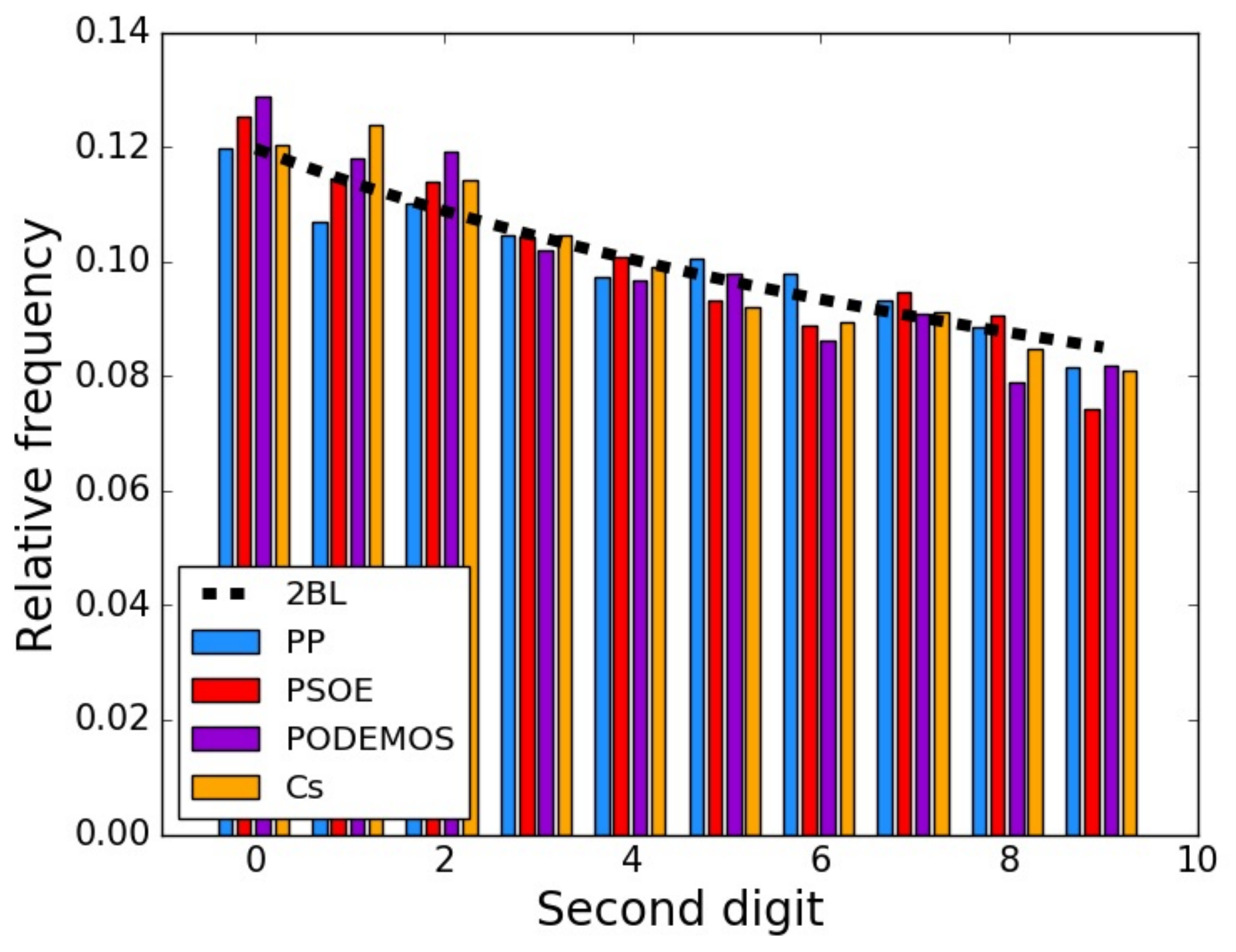}
\includegraphics[width=0.4\columnwidth]{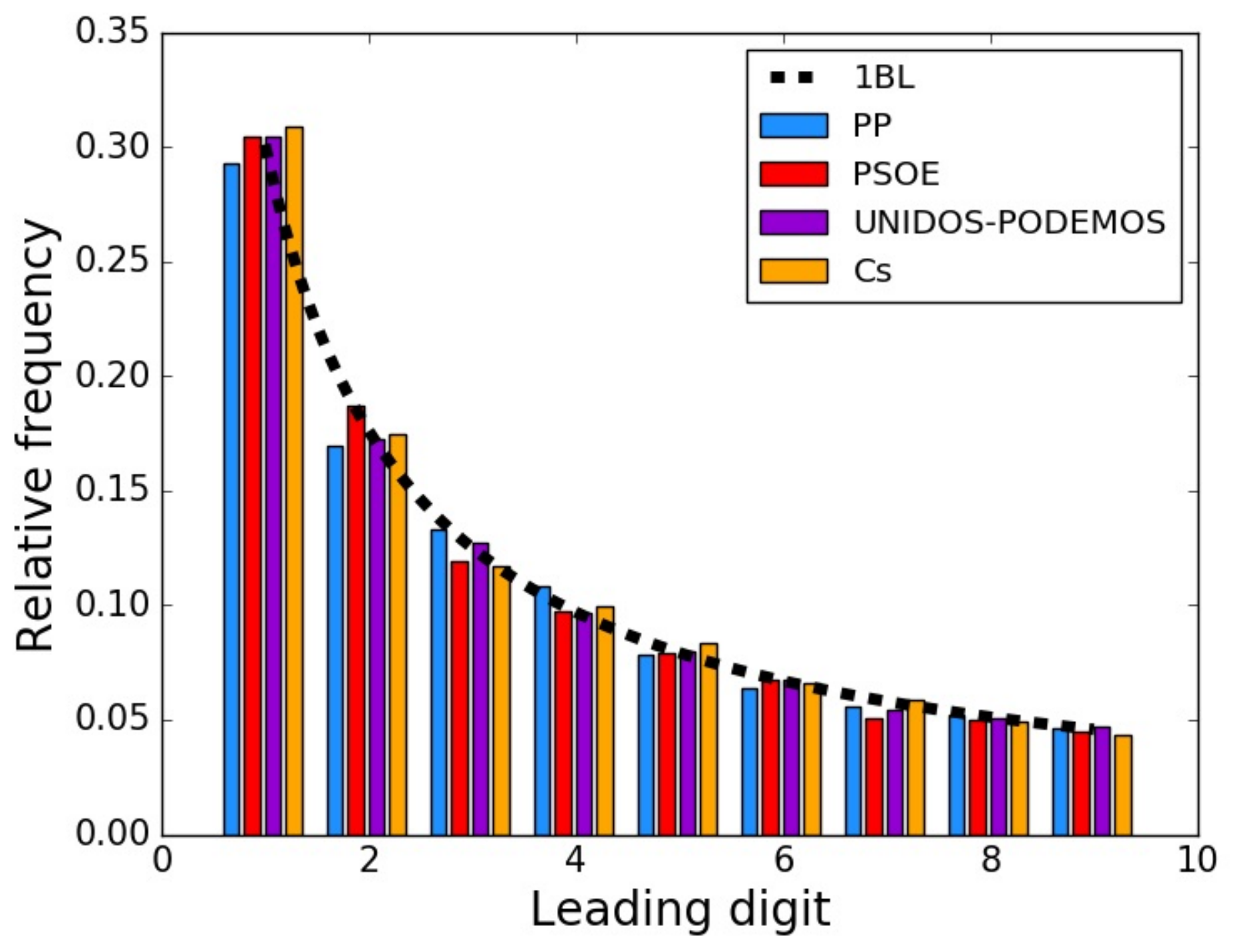}
\includegraphics[width=0.4\columnwidth]{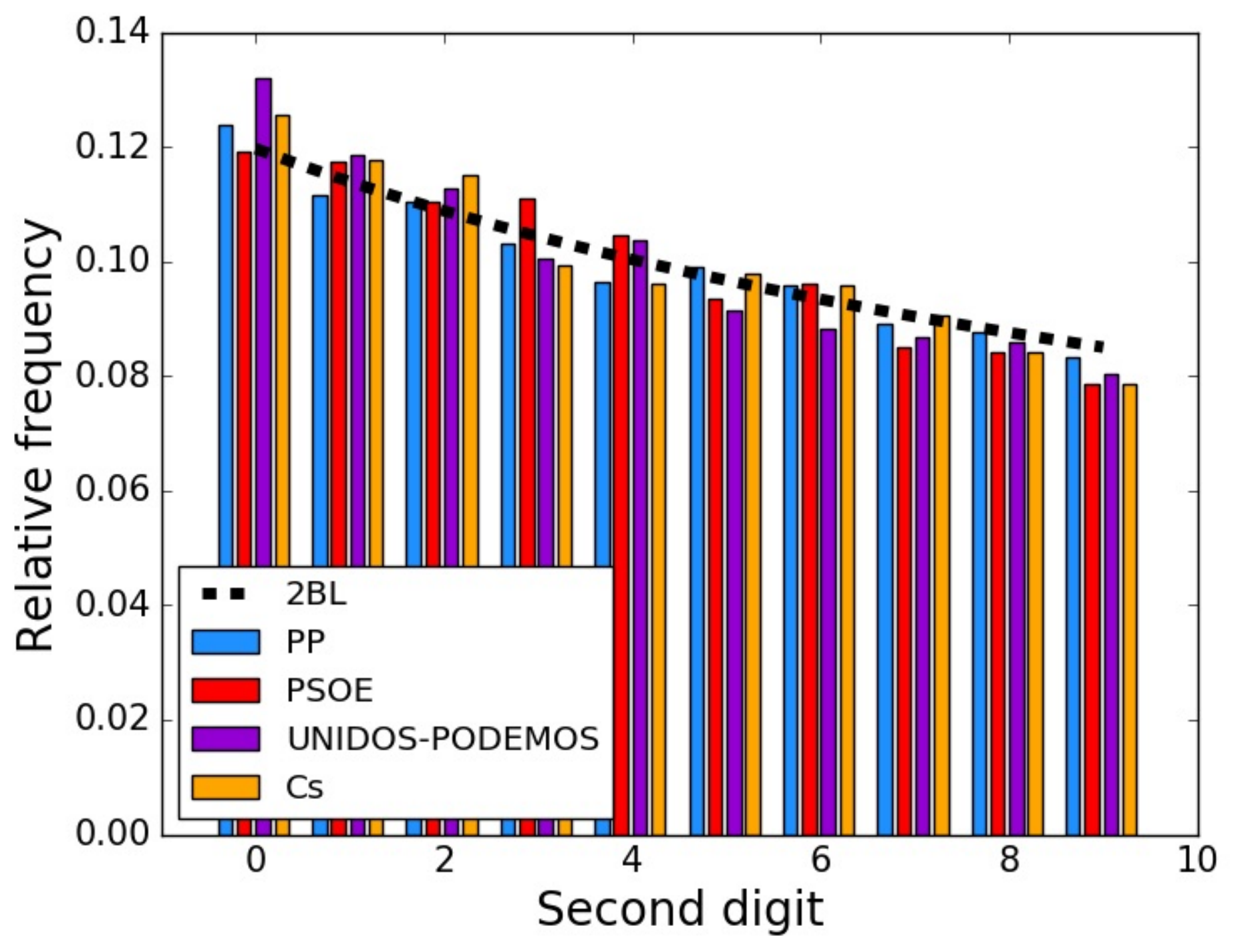}
\caption{Histograms of relative frequencies for the first (left panels) and second (right panels) significant digits of the four most important political parties vote counts over municipalities (more than 8000 in each case) for the 2015 (upper panels) and 2016 (bottom panels) elections}
\label{fig:BL_todos}
\end{figure*}

\noindent We start by exploring 1BL and 2BL tests applied to vote count statistics nationally using the fine-grained data given by splitting vote counts at the level of municipalities (with over 8000 samples, vote counts ranging in about five orders of magnitude). With socio-political impact in mind, from now on we focus on the vote statistics to the main, national-wide parties PP, PSOE, Cs and PODEMOS.\\
 Results for the 1BL are shown in the left panels of Fig.~\ref{fig:BL_todos} (left panel depicts results for the 2015 elections while the right panel does the same for the 2016 case). As expected the distributions seem to be close to Benford's law for all political parties, at least visually, and there are no obvious differences between 2015 and 2016. To have a better quantitative understanding, we have made use of two statistics: (i) the classical Pearson's $\chi^2$ and (ii) the mean absolute deviation (MAD) test as proposed by Nigrini \cite{nigrini}. In both cases the null hypothesis $H_0$ is that data conform to Benford's law. The former statistic reads
$$\chi^2=N\sum_{d=m}^9\frac{[P_{\text{obs}}(d)-P_{\text{th}}(d)]^2}{P_{\text{th}}(d)},$$
where $P_{\text{th}}(d)$ and $P_{\text{obs}}(d)$ are the theoretical and observed relative frequencies of each digit and $m=1$ for 1BL and $m=0$ for 2BL. This statistic has 8 degrees of freedom for 1BL and 9 for 2BL (as in this latter case the digit zero has to be incorporated as a candidate) and is to be compared to certain critical values, such that if $\chi^2>\chi^2_{n,a}$ then $H_0$ is rejected with the selected level of confidence level $a$. For $n=8$ degrees of freedom, the critical values at the $95\%$ and $99\%$ are 15.507 and 20.090 respectively, whereas for $n=9$ degrees of freedom the critical values at the $95\%$ and $99\%$ are 16.919 and 21.666 respectively.\\
The mean absolute deviation is defined as
$$\text{MAD}=\frac{1}{10-m}\sum_{d=m}^9 |P_{\text{obs}}(d)-P_{\text{th}}(d)|,$$
where $m$ is the initial digit (1 for 1BL, 0 for 2BL). Whereas this statistic lacks clear cut-off values, Nigrini provides the following rule of thumb for 1BL: MAD between $0$ and $0.004$ implies close conformity; from
$0.004$ to $0.008$ acceptable conformity; from $0.008$ to $0.012$
marginally acceptable conformity; and, finally, greater than $0.012$, nonconformity.  To the best of our knowledge, the critical values for MAD have not yet been established for 2BL so all over this work we will assume the same ones as for 1BL.\\
\noindent Results for 1BL can be found in table \ref{table:chi2}. We conclude that for the Pearson $\chi^2$ test, $H_0$ cannot be rejected with sufficiently high confidence in three out of the four main political parties but the $\chi^2$ result for PP is consistently large and suggests rejection of the null hypothesis with a confidence of $99\%$. These results are in contrast with those found using the MAD statistic, where according to Nigrini all political parties conform to Benford's law (PP only showing acceptable conformity and the rest showing close conformity).\\
\noindent The results on 2BL are shown in the right panels of Fig.~\ref{fig:BL_todos} and test statistics are summarized again in table \ref{table:chi2}. These suggest an overall conformance to the second digit law, with exception flagging nonconformace raised by $\chi^2$ that rejects $H_0$ at 95$\%$ for Podemos (2015), Unidos-Podemos (2016) and PSOE (2015). 

\begin{table*}[t]
\centering
\begin{tabular}{l|l|l|l|l|l|l}
  Year&Political party &$\#$ observations&$\chi^2$ 1BL & MAD 1BL&$\chi^2$ 2BL&MAD 2BL\\
  \hline
  \hline
2015&PP&8182&{\color{red}23.079}&0.0052&8.737&0.0027\\
2016&PP&8186&{\color{red}21.408}&0.0046&4.142&0.0021\\
 \hline
2015&PSOE&8135&13.486&0.0030&{\color{blue}17.648}&0.0038\\
2016&PSOE&8121&15.040&0.0033&13.065&0.0038\\
 \hline
2015&Podemos $\&$ Co.&7927&4.845&0.0020&{\color{blue}21.329}&0.0050\\
2016&Unidos Podemos &8056&3.537&0.0019&{\color{blue}18.314}&0.0048\\
 \hline
2015&C's&8037&11.933&0.0036&10.934&0.0034\\
2016&C's&8001&9.671&0.0033& 10.951&0.0039\\
\hline      
\hline
\end{tabular}
\caption{Statistical tests of conformance to Benford's law for the first (1BL) and second (2BL) significant digit distribution for the vote counts of each political party (at the level of municipalities), along with $\chi^2$ and MAD statistics. In blue we highlight the datasets where the null hypothesis can be rejected with 95$\%$ confidence but not with 99$\%$ and in red cases for which where the null hypothesis can be rejected with more than 99$\%$ confidence according to $\chi^2$. On the basis of MAD statistic the null hypothesis of conformance to Benford's laws cannot be rejected for any case.}
\label{table:chi2}
\end{table*}

\subsubsection{Individual analysis at the precincts level}
\begin{figure*}
\centering
\includegraphics[width=0.45\columnwidth]{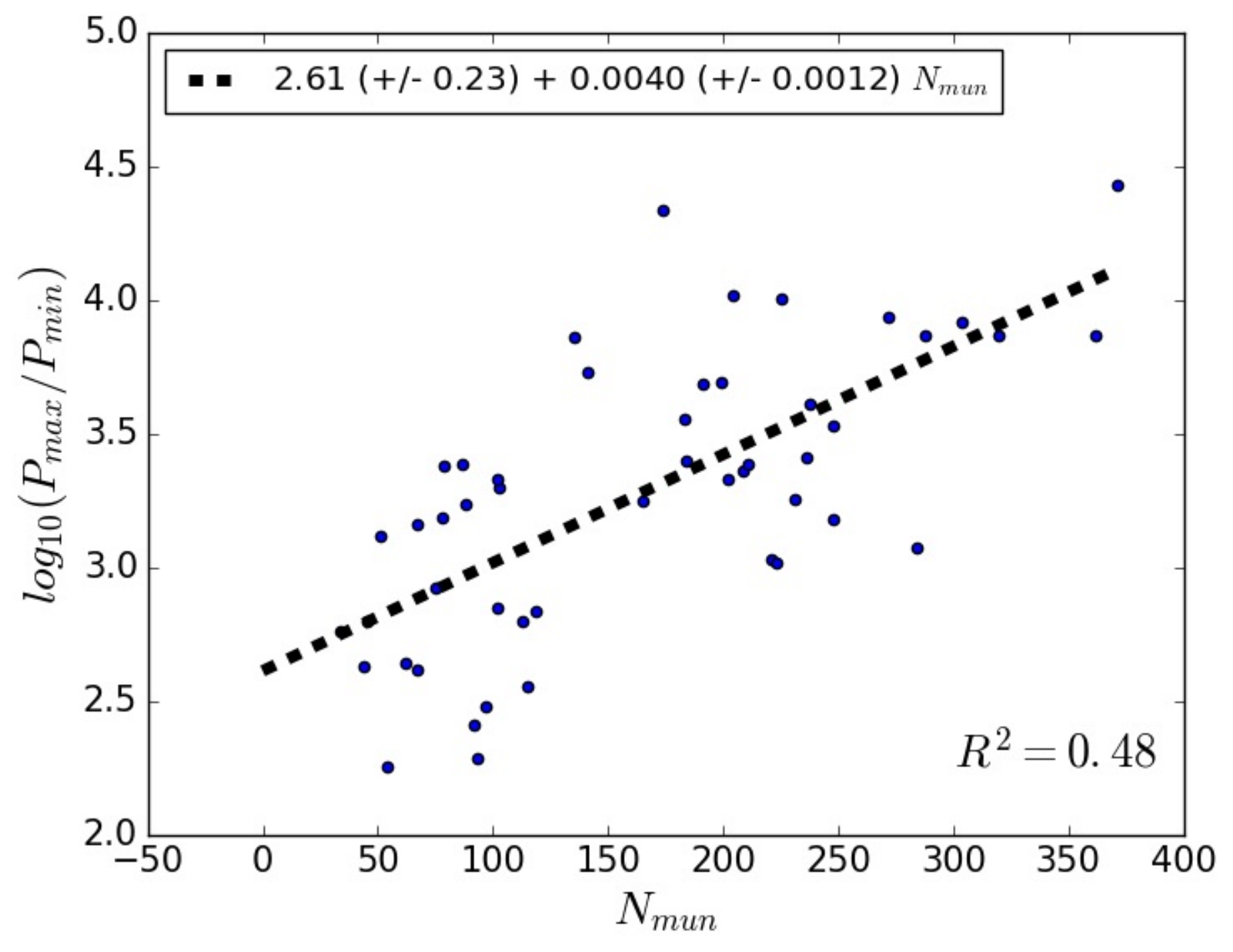}
\includegraphics[width=0.45\columnwidth]{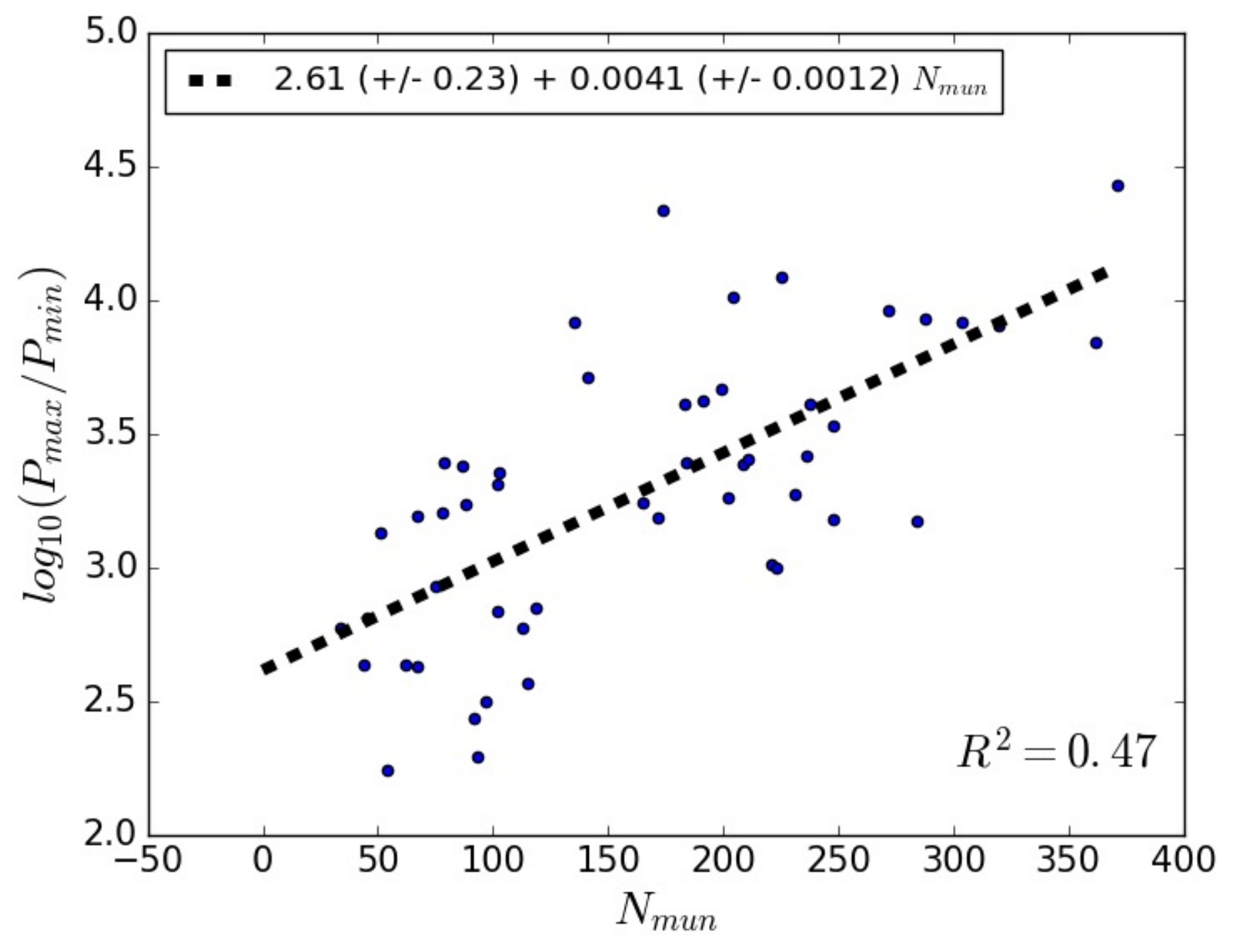}
\caption{Scatter plot of the number of orders of magnitude spaned by the voting populations of a precinct as a function of the number of municipalities in each precinct, for 2015 (left) and 2016 (right). We find a positive correlation with $R^2\approx 0.47$ suggesting that the "size" of a precinct in terms of the number of municipalities explains with $47\%$ of the variation in the support (in terms of orders of magnitude) of the number of votes (the larger the number of municipalities, the more likely that the number of votes take values from a larger number of orders of magnitude. The coefficient of $0.004$ indicates that, on average, when we move from a precinct with $x$ municipalities to one with $x+100$, the ratio of the biggest population to the smallest is $2.5$ times bigger.}
\label{corr}
\end{figure*}

In order to give a closer look to the vote count distributions we now explore the statistics taking place at each separate precinct. At this point we need to recall that among other criteria Benford's law is expected to emerge in datasets where data range several orders of magnitude. This hypothesis was fulfilled at the national scale as the population of municipalities ranges several orders of magnitude ($\mathcal{O}(1)$ -- $\mathcal{O}(10^5)$, see Table~\ref{table:precincts}) if we consider all of them. However this is not straightforward at the precinct level, where the number of municipalities is highly heterogeneous from precinct to precinct. In Fig.~\ref{corr} (appendix) we have checked that the number of orders of magnitude that vote counts span at the precinct level is indeed linearly correlated with the number of municipalities the precinct contains ($R^2=0.47$). This means that the larger the number of municipalities considered in a single analysis, the more we should expect data to conform to Benford's law.\\
That being said, for each and every precinct in Spain we proceed to extract the frequencies of the first and second significant digits found for all the municipalities inside that precinct, and make a goodness of fit test between these empirical distributions and 1BL and 2BL using both $\chi^2$ and MAD statistics. Results on 1BL are summarized for the case of $\chi^2$ in the left panels of \ref{fig:circu_chi2} and in the appendix Fig.~\ref{fig:circu_chi2_unoporuno_1BL}, finding an overall good conformance to 1BL at the precinct level. Conversely, MAD statistics (Fig.~\ref{fig:circu_mad} in the appendix) say just the opposite, suggesting systematic non-conformance. As for the 2BL test, there exists a strong deviation from the expected distribution (right panels of Fig.~\ref{fig:circu_chi2} for $\chi^2$ and appendix Fig.\ref{fig:circu_mad} for MAD), and both statistics consistently reject the null hypothesis of conformance to Benford's law for all political parties.\\ 

\begin{figure*}
\centering
\includegraphics[width=0.4\columnwidth]{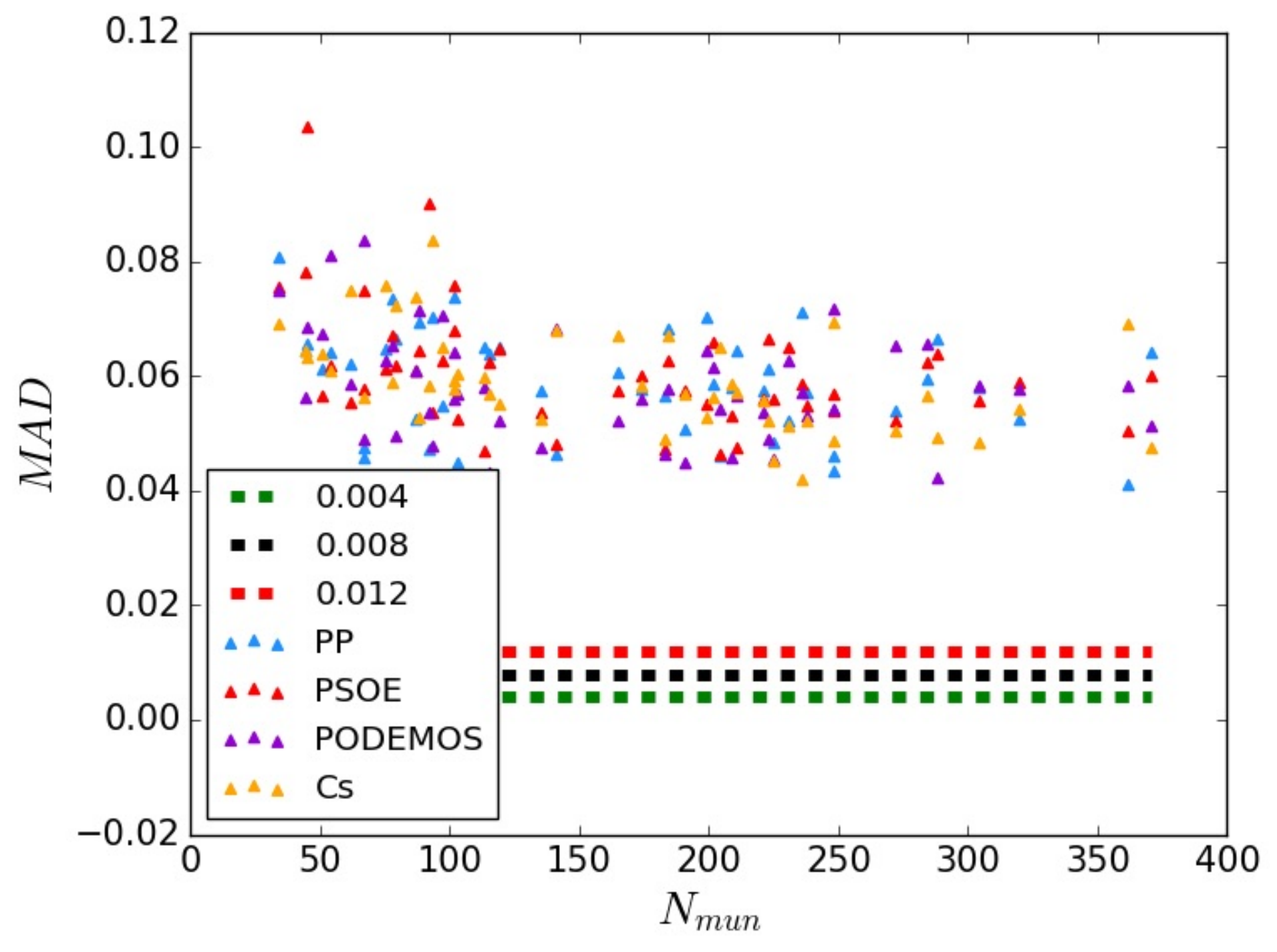}
\includegraphics[width=0.4\columnwidth]{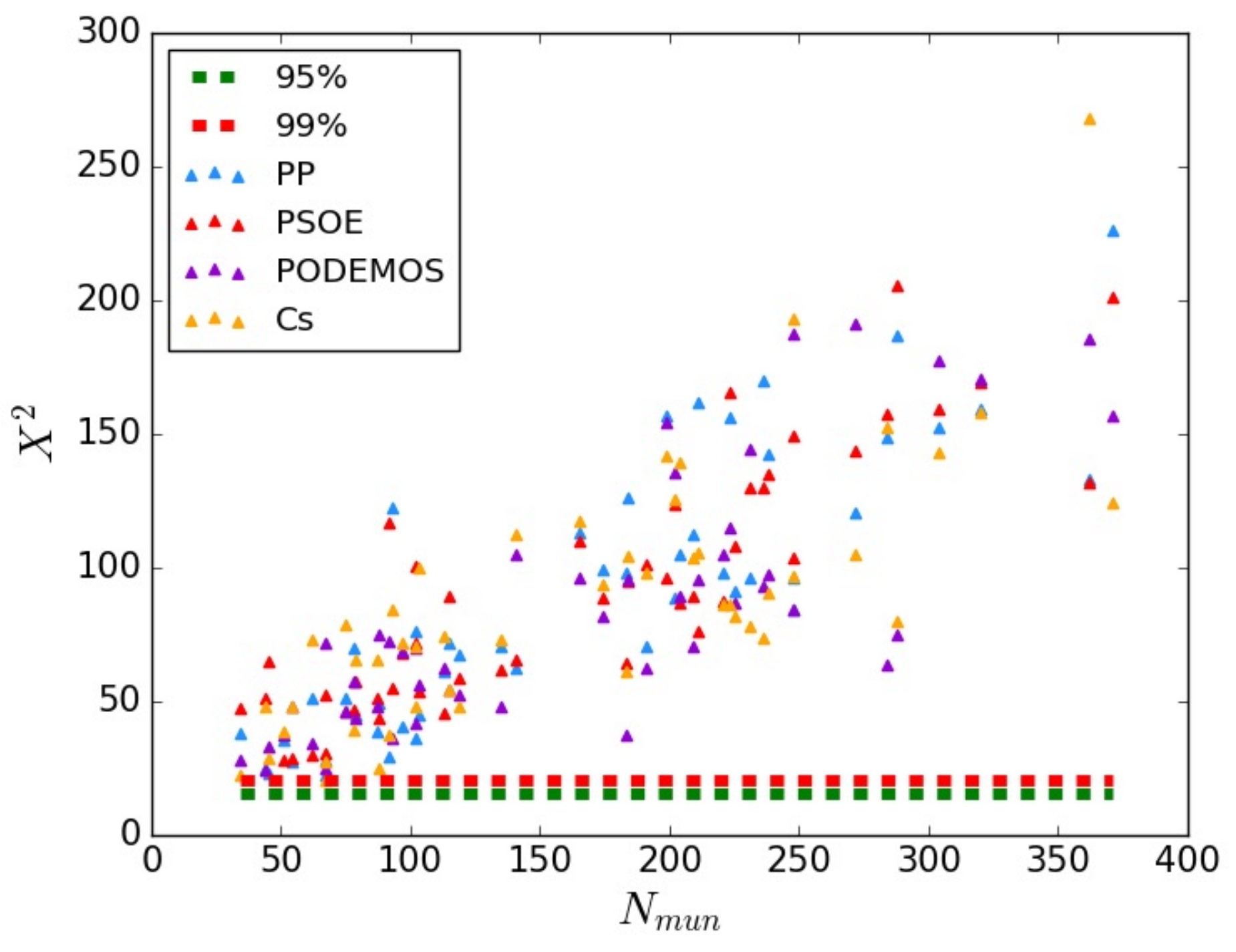}
\includegraphics[width=0.4\columnwidth]{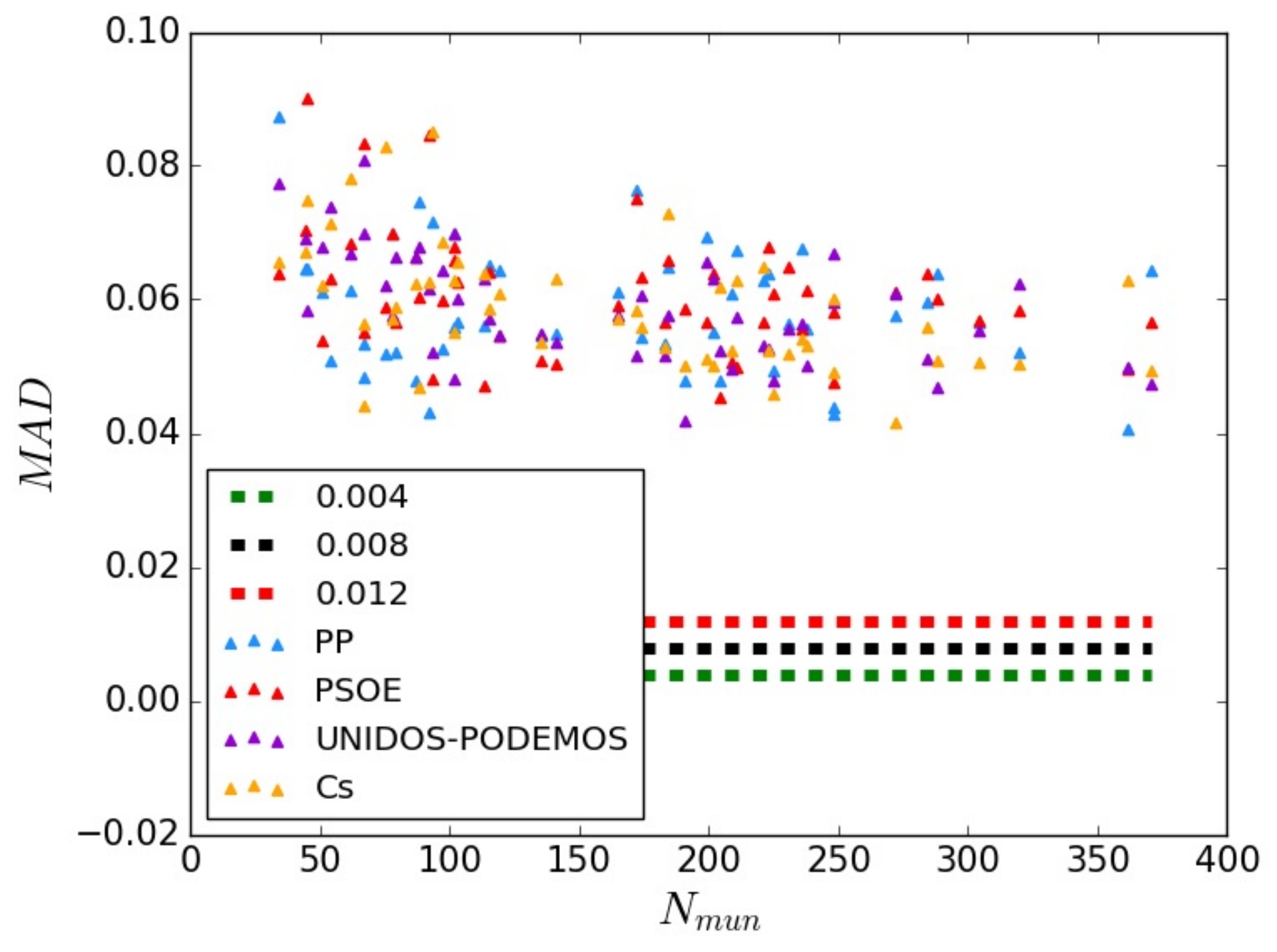}
\includegraphics[width=0.4\columnwidth]{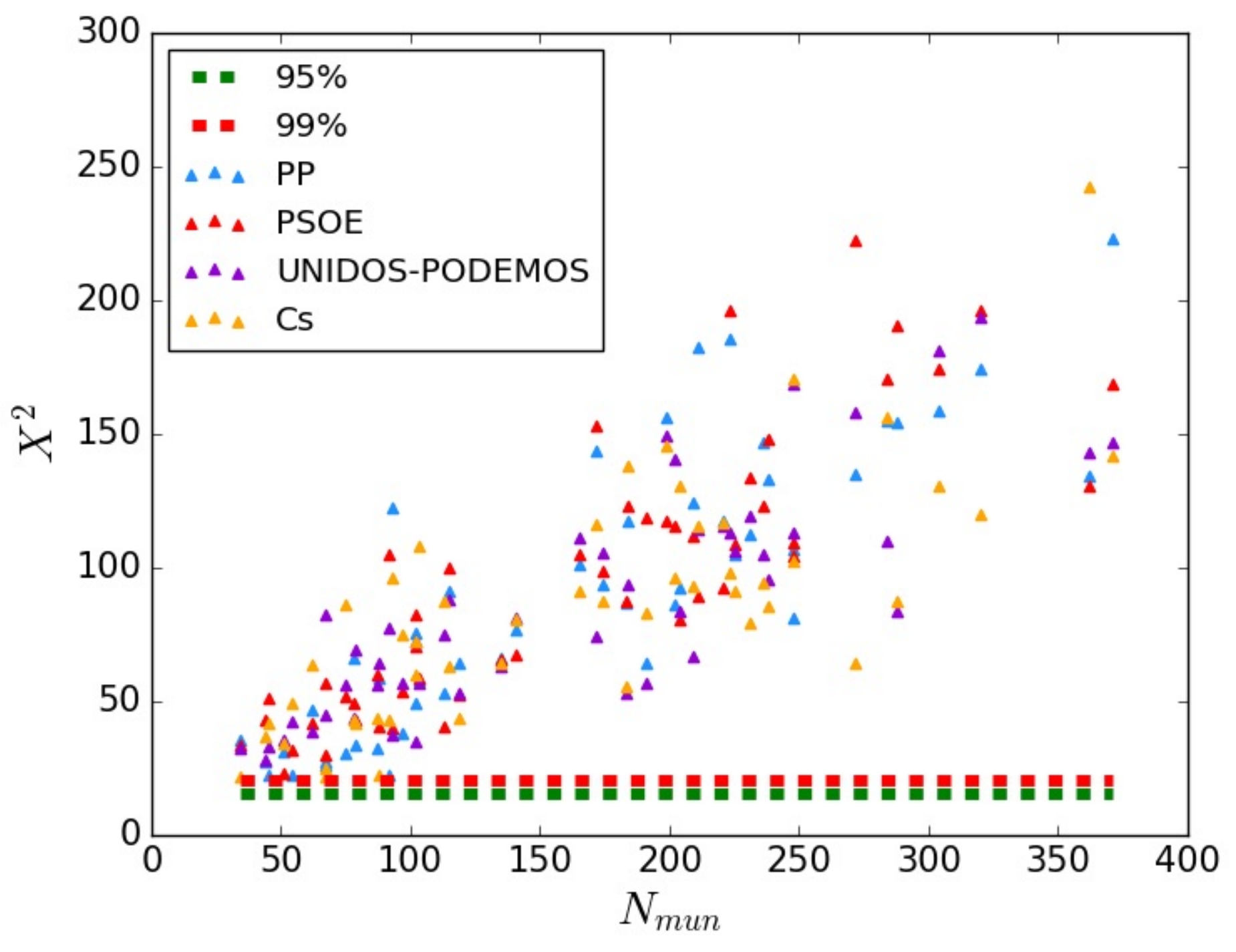}
\caption{Scatter plot of the MAD (left panels) and $\chi^2$ (right panels) statistics extracted from the 2BL test of each precinct as a function of the number of municipalities in each precinct (2015 results are shown in the top panels and 2016 ones are shown in the bottom panels, with no obvious differences). In the case of MAD, we find a weak negative correlation as expected, but this correlation is not enough to explain the systematic nonconformance to 2BL. In the case of $\chi^2$, the effect is quite the opposite, and nonconformance is stronger as the size of the precinct increases, thereby suggesting that nonconformance to 2BL at this level of aggregation is a genuine result and not a spurious effect of finite size statistics.}
\label{fig:scatter_bl2}
\end{figure*}

\noindent Now, note that at each individual precinct we expect statistics to be a priori poorer than at the national scale, as the average number of municipalities per precinct is of the order of $\mathcal{O}(10^2)$ (see table \ref{table:precincts} for details), that is, one order of magnitude smaller. As MAD does not include any correction term that depends on the sample set, one should therefore take the results associated to MAD with a pinch of salt. This is not necessarily the case for the $\chi^2$ as this latter statistic takes into account in its definition the number of samples. In any case, in order to assess whether the strong nonconformance to 2BL at this level of aggregation is just due to finite size effects we explore the dependence of both the $\chi^2$ and MAD results on the precinct's size. Accordingly, in Fig.~\ref{fig:scatter_bl2} we plot for each precinct its $\chi^2$ and MAD result as a function of the number of municipalities present in that precinct. As expected, we find  that MAD suffers from finite size effects and is over conservative for small sample sizes, however this effect is rather weak and not enough to explain the systematic nonconformance to 2BL. In the case of the $\chi^2$ statistic we observe quite the opposite effect: the larger the number of municipalities in a given precinct, the more likely the null hypothesis to be rejected. An equivalent size dependence analysis for the 1BL test is reported in the appendix Fig.~\ref{fig:scatter_bl}.
\begin{figure*}
\centering
\includegraphics[width=0.4\columnwidth]{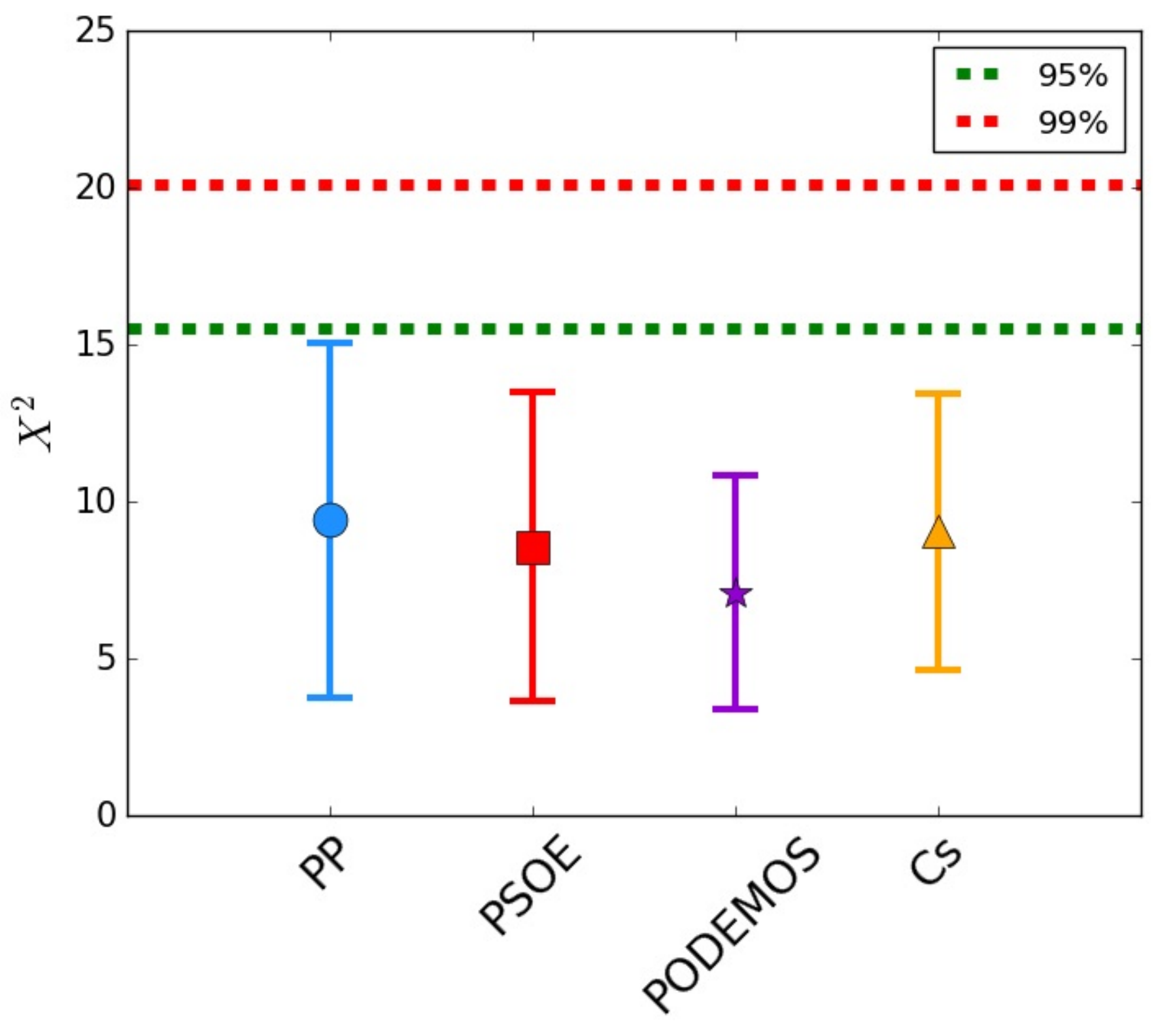}
\includegraphics[width=0.4\columnwidth]{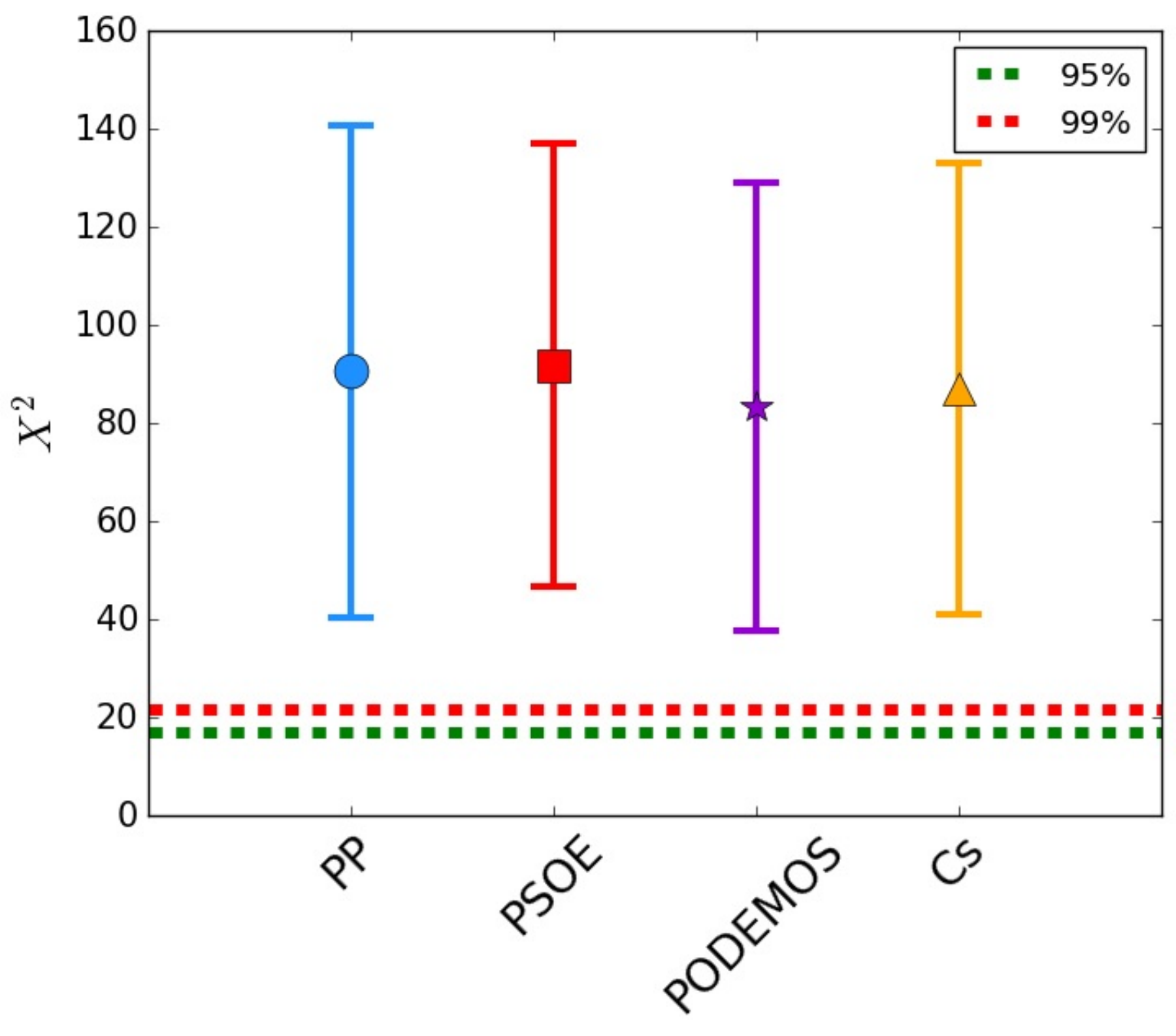}
\includegraphics[width=0.4\columnwidth]{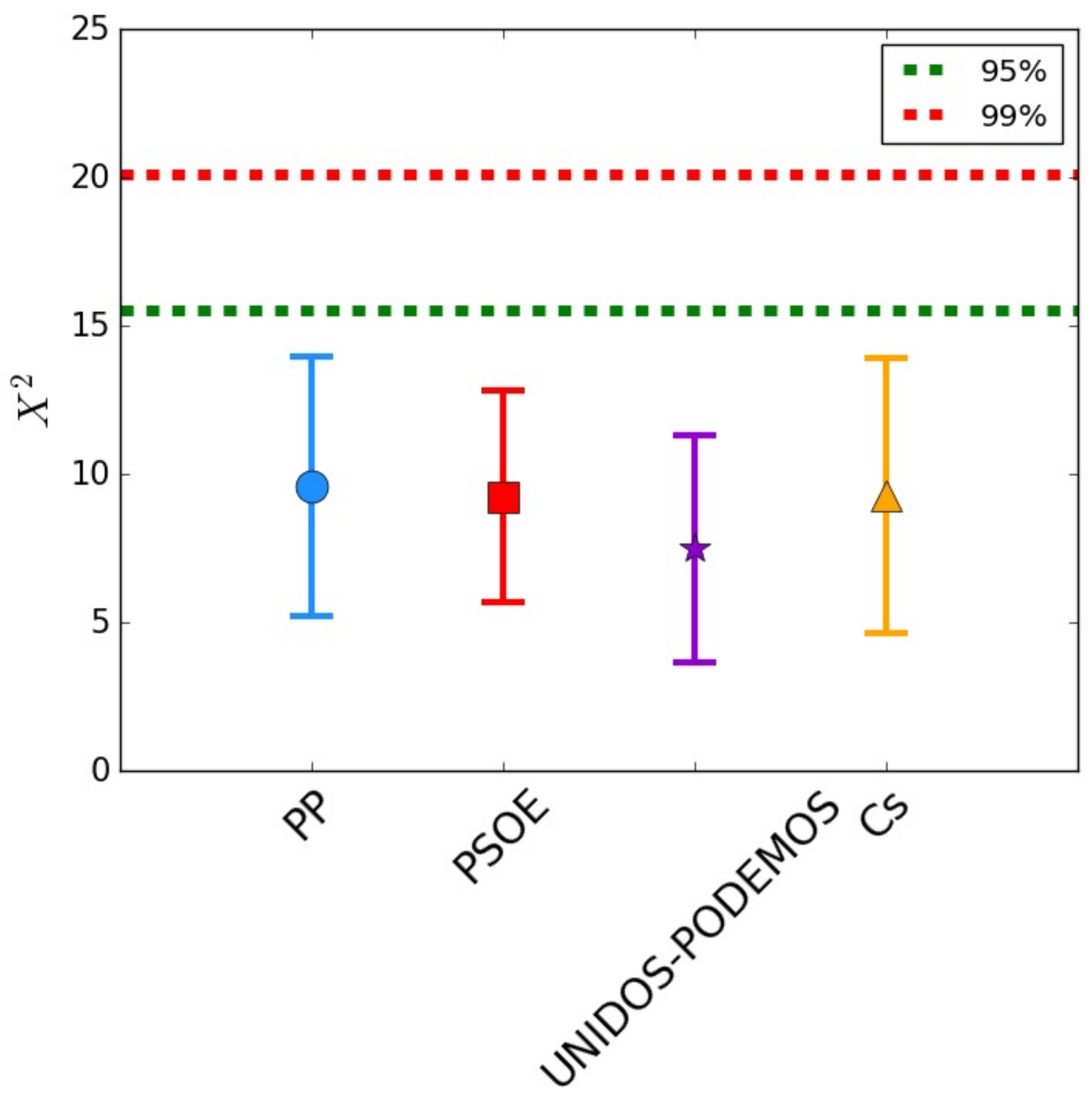}
\includegraphics[width=0.4\columnwidth]{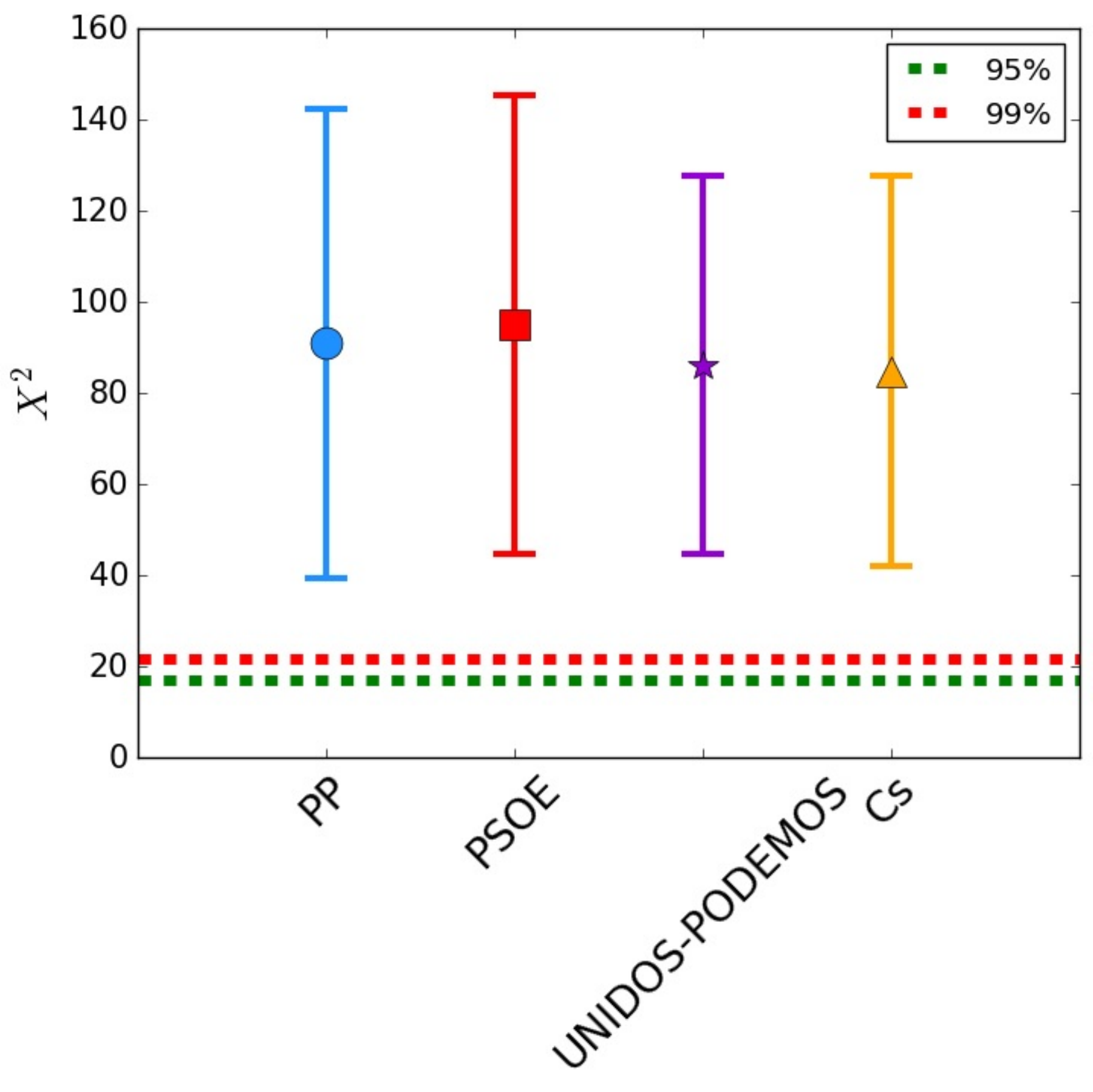}
\caption{Summary of Pearson $\chi^2$ goodness of fit to 1BL (left panels) and 2BL (right panels) for 2015 (upper panels) and 2016 (bottom panels) extracted from analysis of each individual precinct (each precinct contains a different number of municipalities and shows a precise distribution and an associated $\chi^2$, so here we plot the mean $\pm$ standard deviation over all Spanish precincts (excluding Ceuta and Melilla, precincts with a single municipality) for the main parties. In every case the critical values for rejection at the 95 and 99$\%$ confidence level are shown.
Interestingly, in the case of 1BL for a large majority we accept conformance to Benford's law, whereas in the case of 2BL for a large majority the null hypothesis is rejected. Results based on MAD suggest that neither 1BL nor 2BL is accepted (Fig.~\ref{fig:circu_mad} in the appendix).}
\label{fig:circu_chi2}
\end{figure*}

\subsubsection{Aggregate analysis at the precincts level}
\begin{figure*}
\centering
\includegraphics[width=0.4\columnwidth]{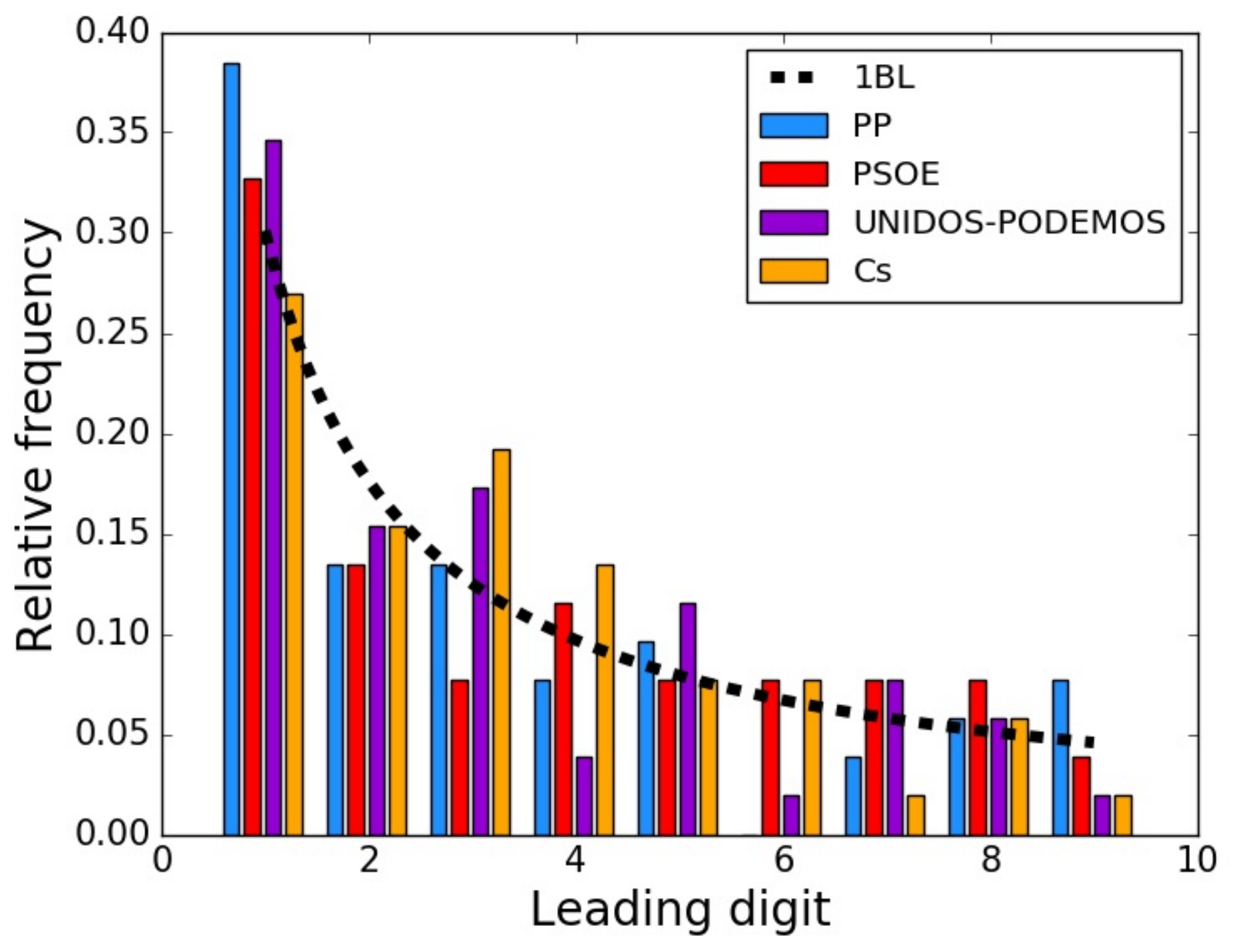}
\includegraphics[width=0.4\columnwidth]{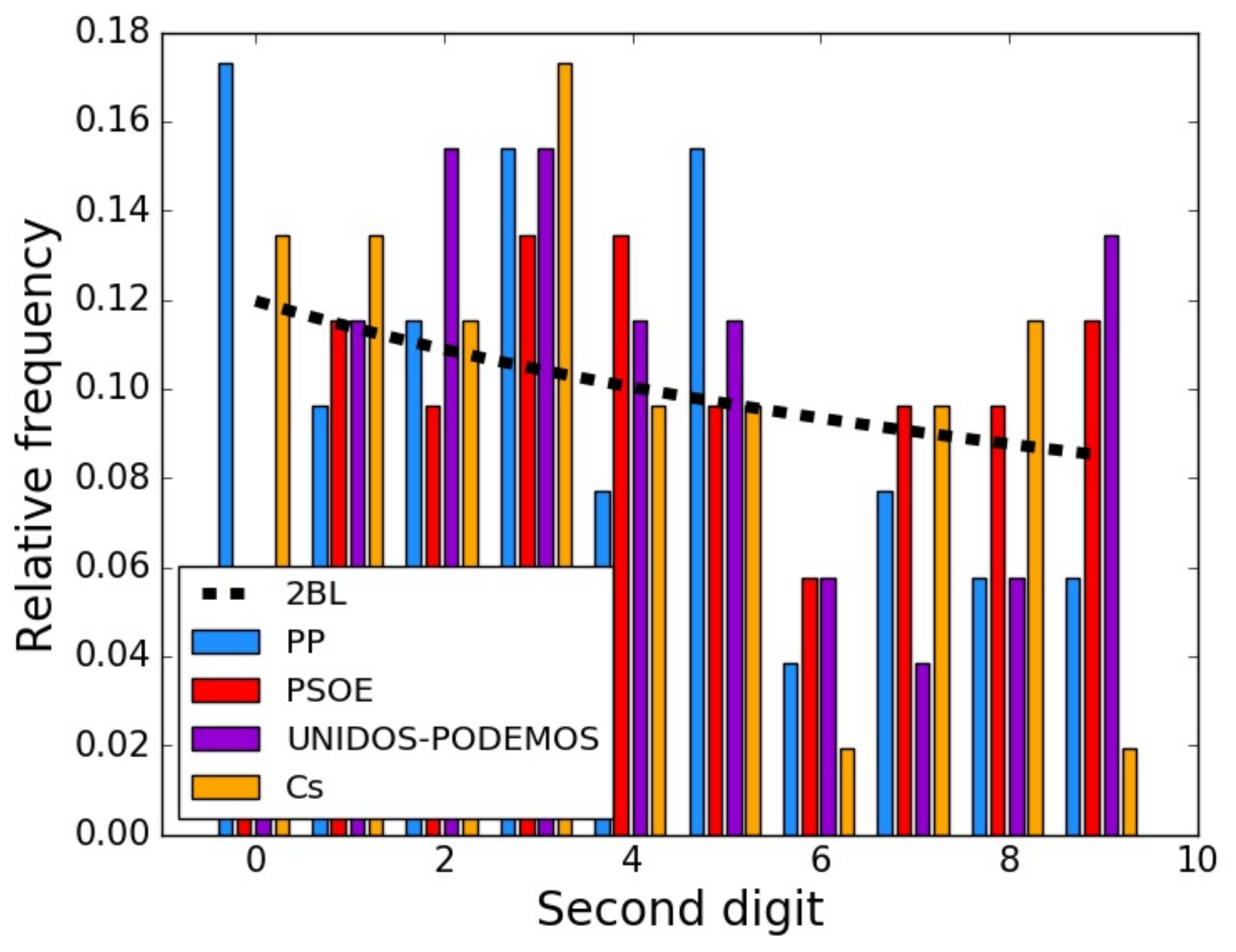}
\caption{Histograms of relative frequencies for the first (left panel) and second (right panel) significant digits for the main political parties vote counts aggregated over precincts, for the case of 2016 (2015 is shown in appendix Fig.~\ref{fig:BL_aggregate2015})}
\label{fig:BL_aggregate}
\end{figure*}

\begin{figure*}
\centering
\includegraphics[width=0.4\columnwidth]{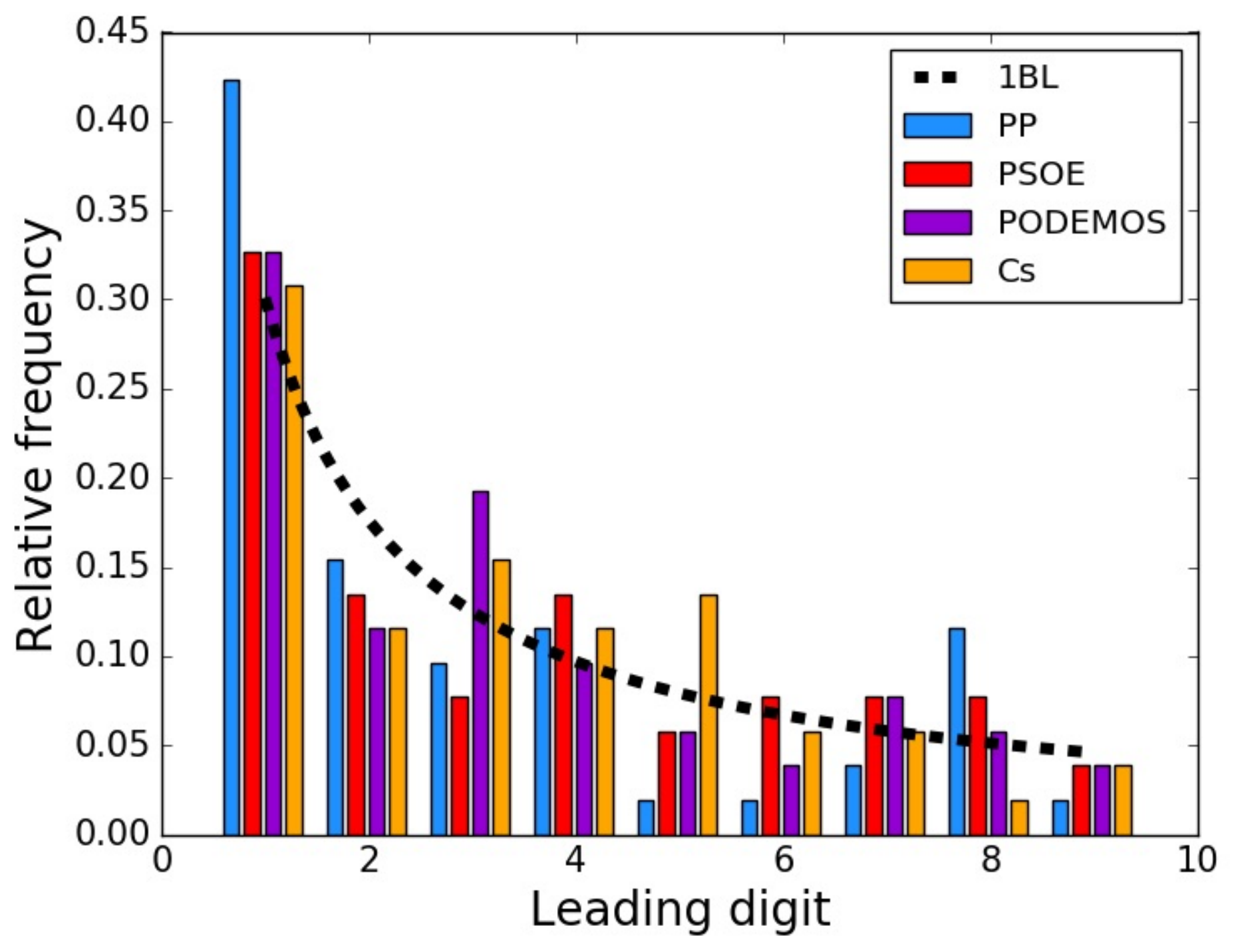}
\includegraphics[width=0.4\columnwidth]{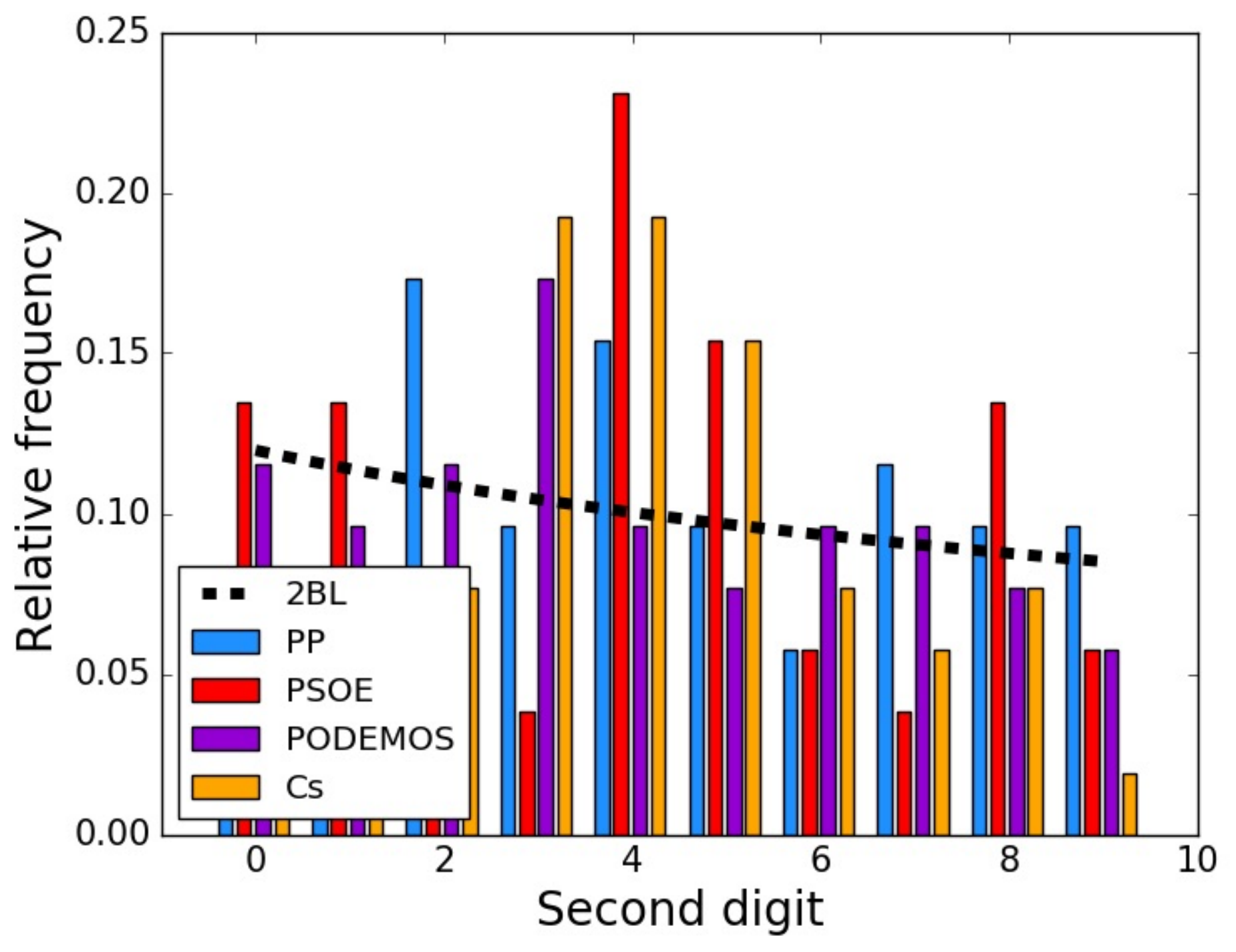}
\caption{Histograms of relative frequencies for the first (left panel) and second (right panel) significant digits for the main political parties vote counts aggregated over precincts, for the case of 2015.}
\label{fig:BL_aggregate2015}
\end{figure*}

To round off our analysis with a third level of aggregation, we explore conformance to 1BL and 2BL when vote counts are aggregated per precinct. In this case we only have 52 samples (52 precincts) so we expect the distributions to be more noisy. From the previous analysis, we learned that MAD suffers from finite size effects so we expect MAD to be more conservative than $\chi^2$  at this level of aggregation. In Fig.~\ref{fig:BL_aggregate} we show the results for 2016 and we refer the readers to the appendix Fig.~\ref{fig:BL_aggregate2015} to find analogous results for 2015, which don't show substantial differences at the qualitative level. As expected the distributions show larger fluctuations and, in absolute terms, deviate more from the theoretical laws (depicted in dashed lines). In table \ref{table:aggregate} we depict the $\chi^2$ and MAD statistics, which, again as expected, show inconsistent results: while $\chi^2$ systematically cannot be rejected with above 95$\%$ confidence level, in turn MAD systematically suggests nonconformity. We conclude that this level of aggregation is less informative than previous ones.

\begin{table}[]
\centering
\begin{tabular}{l|l|l|l|l}
  Year&Political party&test&$\chi^2$&MAD\\
  \hline
  \hline
2015&PP&1BL&12.71&0.0455\\
2016&PP&1BL&7.13&0.0329\\
2015&PP&2BL&7.79&0.0325\\
2016&PP&2BL&7.43&0.0333\\
\hline
2015&PSOE&1BL&3.78&0.02628\\
2016&PSOE&1BL&2.90&0.0220\\
2015&PSOE&2BL&20.90&0.0541\\
2016&PSOE&2BL&4.15&0.0222\\
\hline
2015&Podemos $\&$ Co. &1BL&4.45&0.02638\\
2016&Unidos Podemos&1BL&7.09&0.0344\\
2015&Podemos $\&$ Co. &2BL&3.29&0.0168\\
2016&Unidos Podemos&2BL&8.46&0.0359\\
\hline
2015&C's&1BL&4.81&0.02433\\
2016&C's&1BL&5.24&0.0270\\
2015&C's&2BL&15.39&0.0474\\
2016&C's&2BL&8.86&0.0289\\
\hline
\hline      
\hline
\end{tabular}
\caption{$\chi^2$ values of conformance to 1BL and 2BL for each political party extracted from the analysis performed when we aggregate votes at the precinct level.}
\label{table:aggregate}
\end{table}

\begin{figure*}
\centering
\includegraphics[width=0.4\columnwidth]{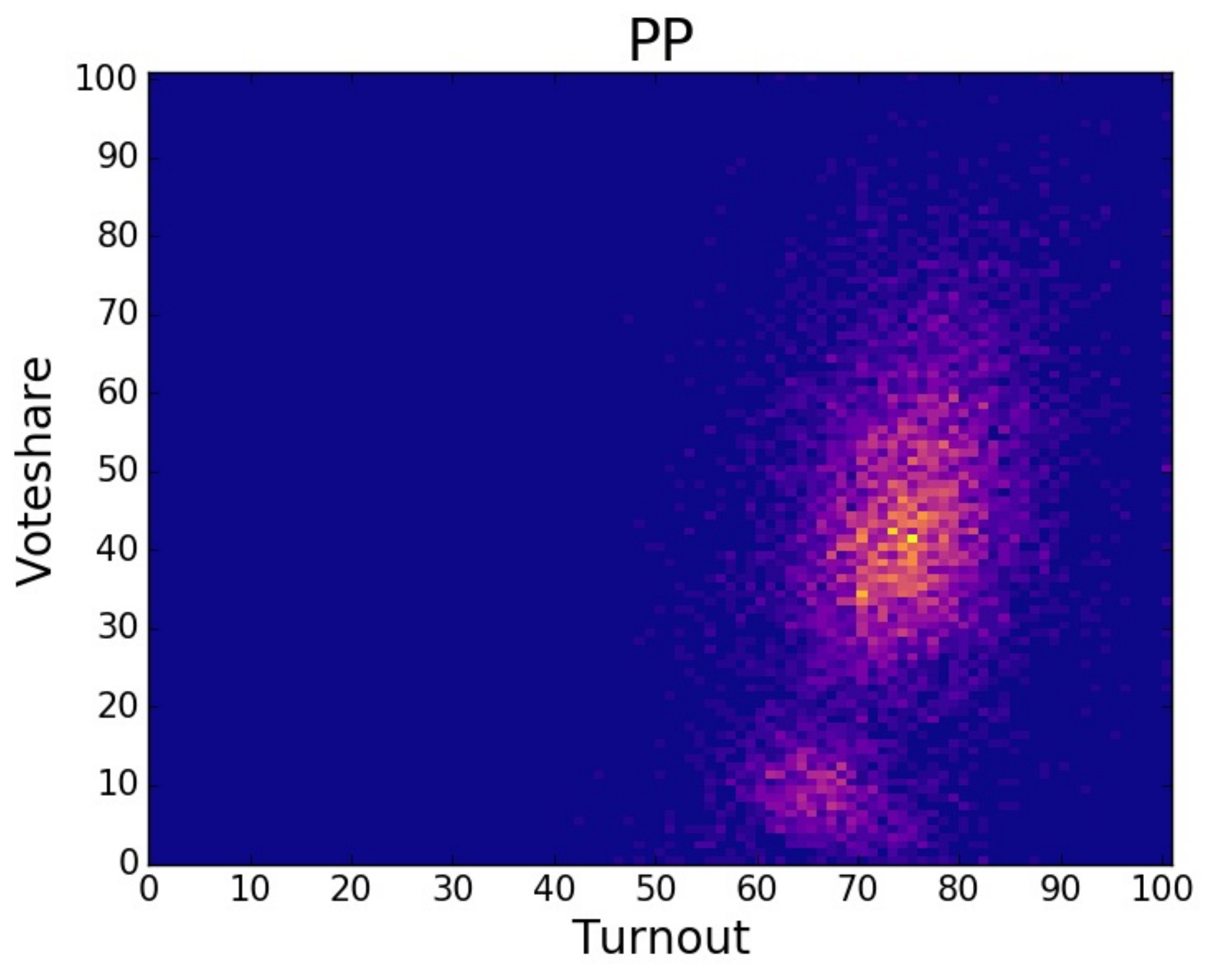}
\includegraphics[width=0.4\columnwidth]{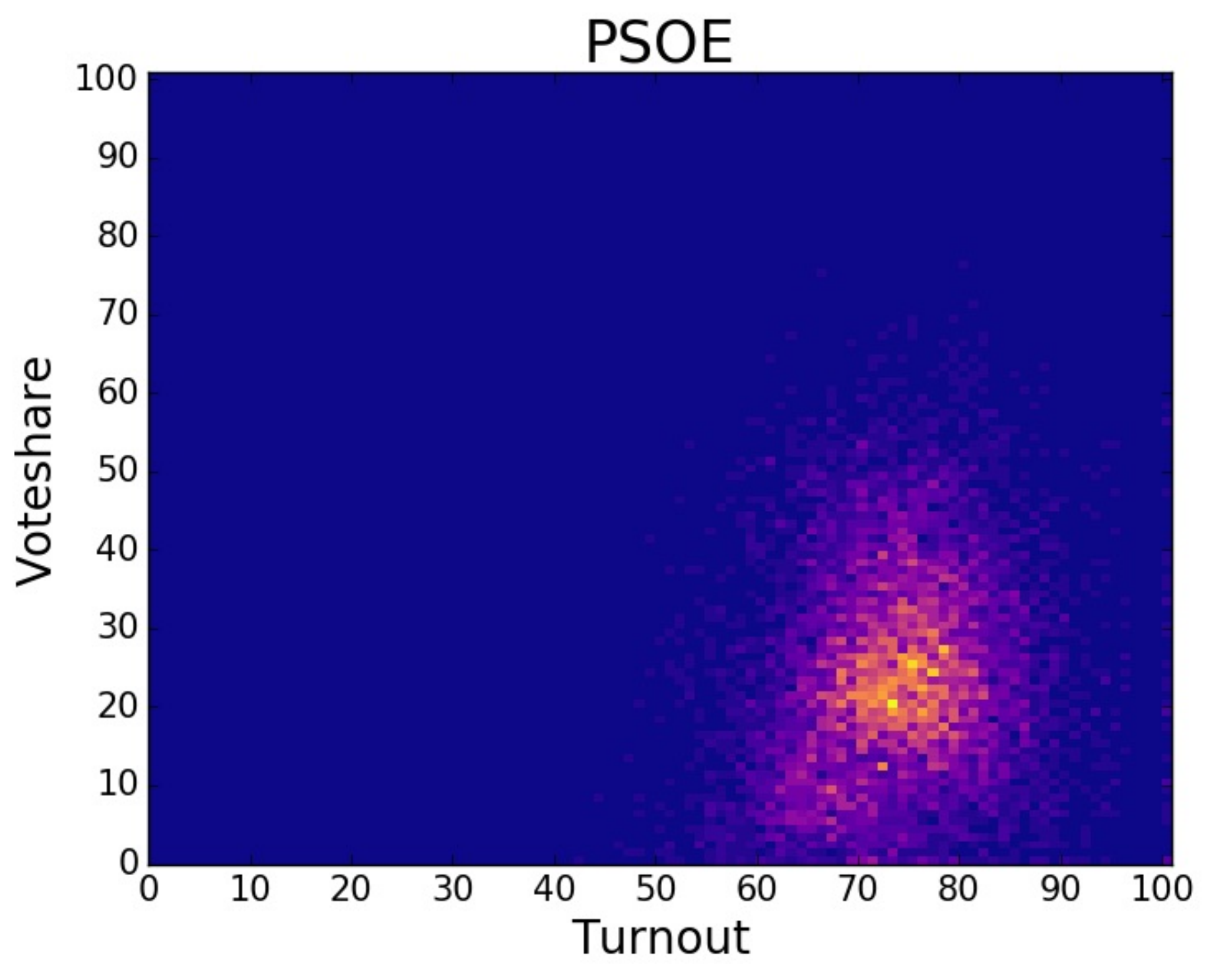}
\includegraphics[width=0.4\columnwidth]{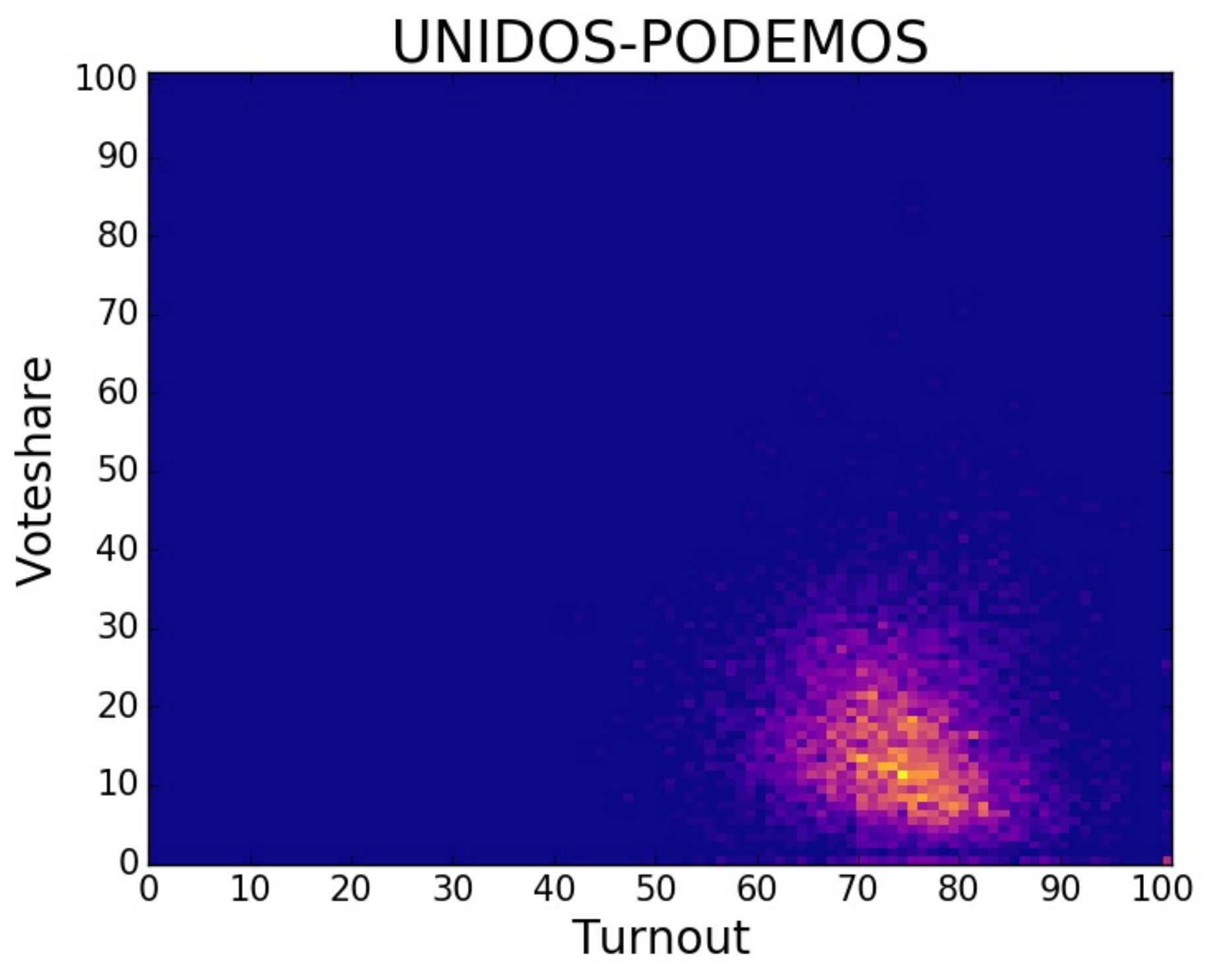}
\includegraphics[width=0.4\columnwidth]{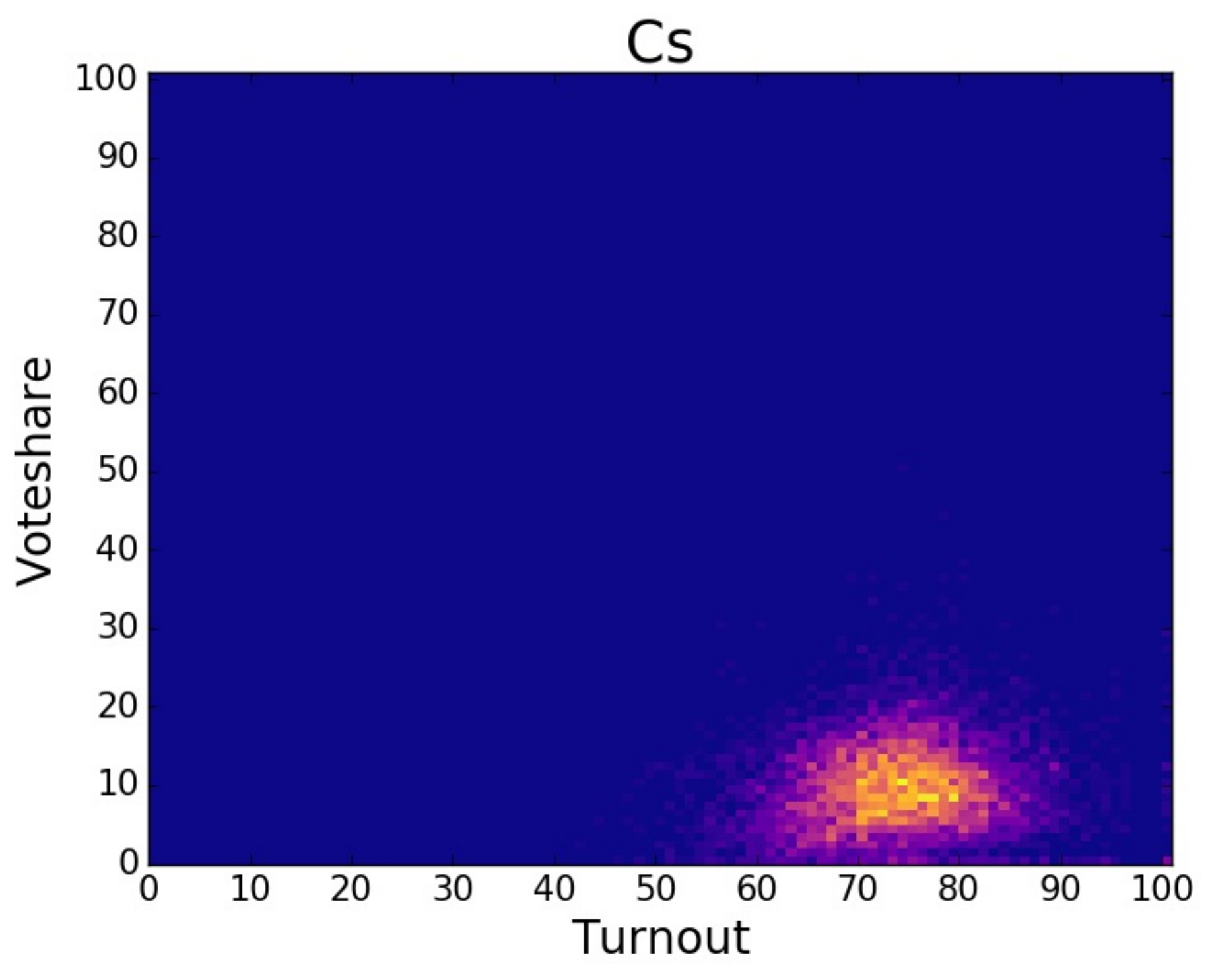}
\caption{Heatmaps plotting the percentage (in color scale) of municipalities where a given political party has received a certain percentage of votes, as a function of the relative participation. These are results from the 2016 elections, the 2015 case is reported in Fig.\ref{fig:heatmap_todos_2015}. According to Klimek et al. \cite{klimek} a smear out of the cluster towards the top-right corner of the heat map is a sign of incremental fraud, whereas extreme fraud would occur for bimodal distributions where a cluster emerges at the top-right corner.}
\label{fig:heatmap_todos}
\end{figure*}
\begin{figure*}
\centering
\includegraphics[width=0.4\columnwidth]{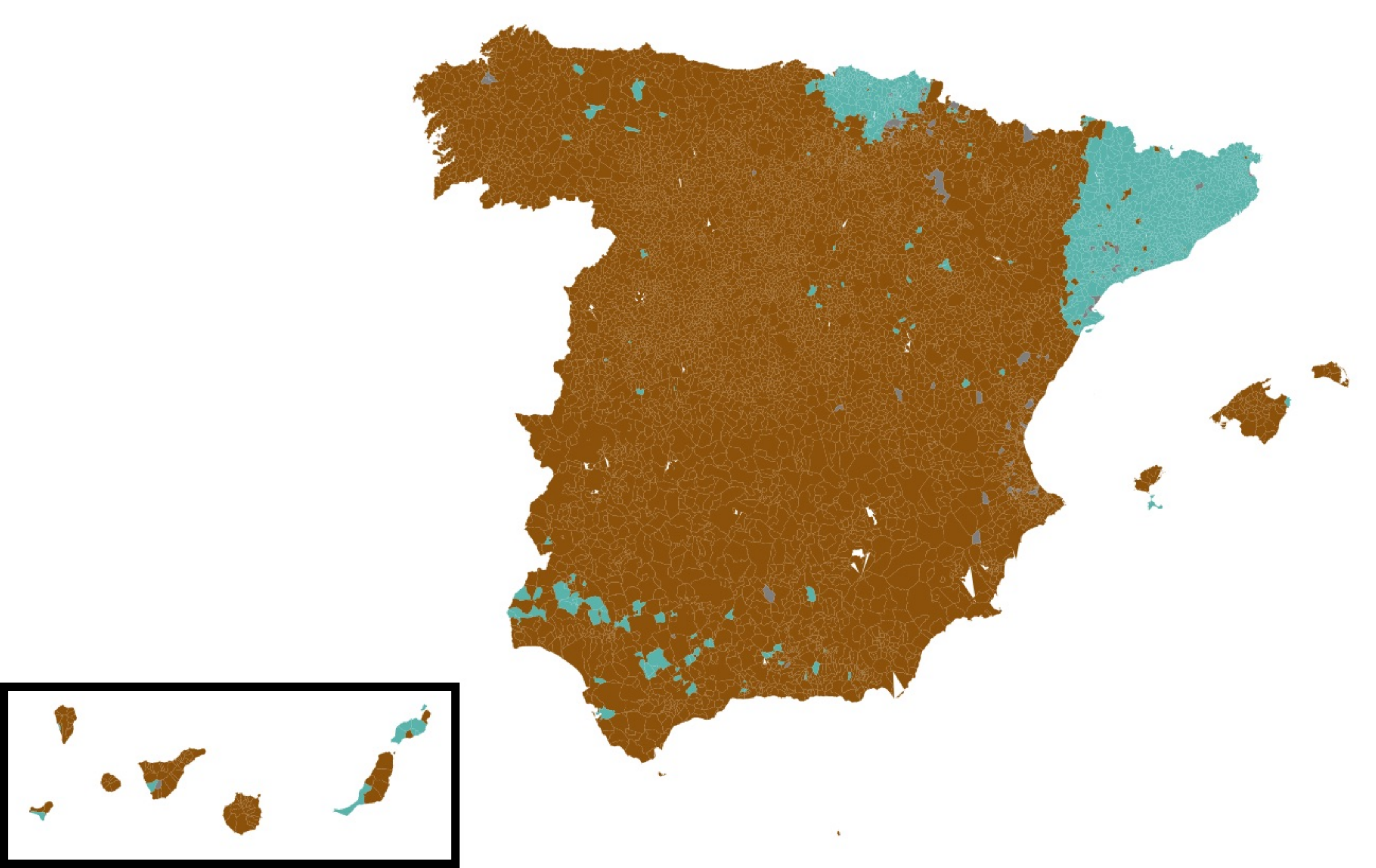}
\includegraphics[width=0.4\columnwidth]{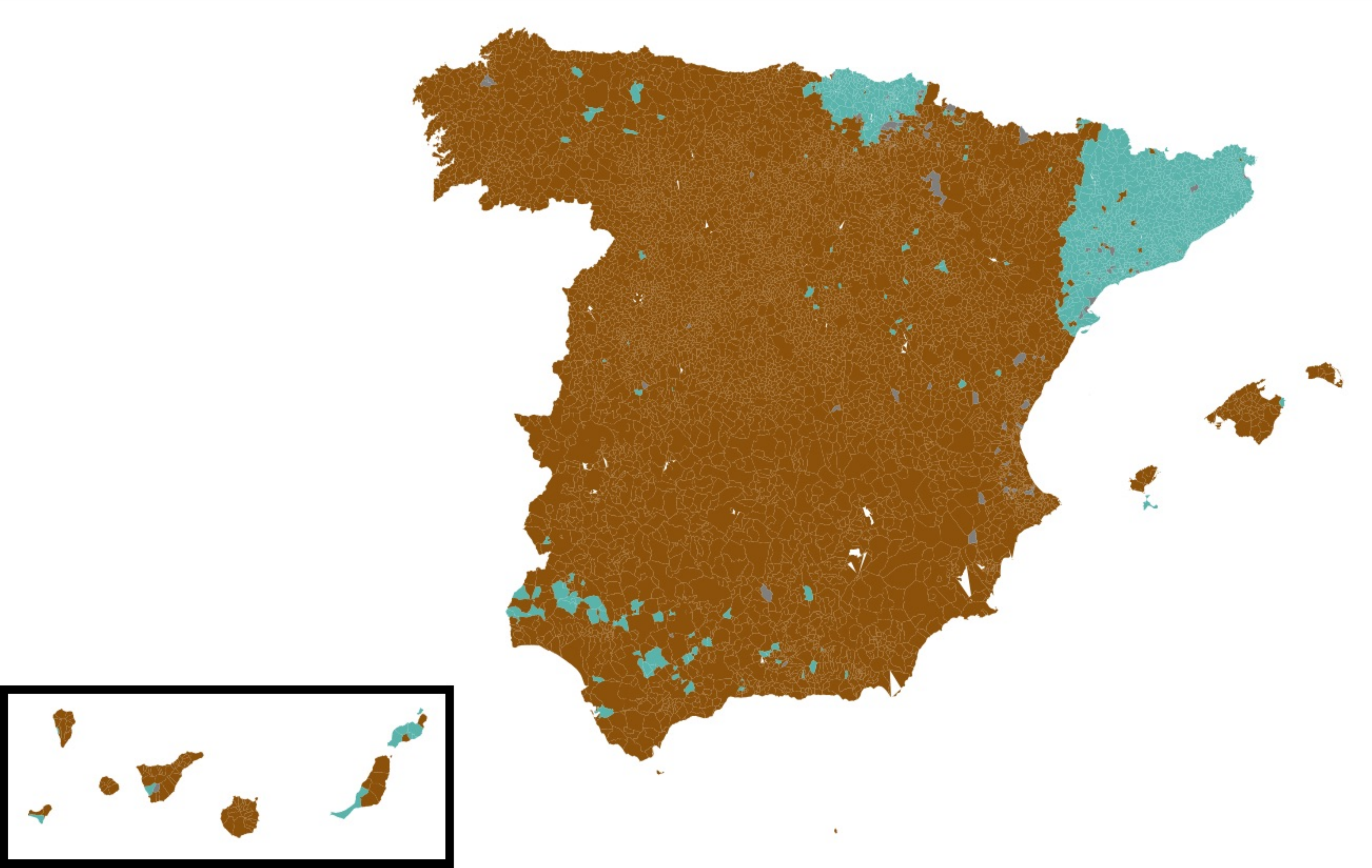}
\caption{Focusing on the bimodal distribution for the conservative party that emerges in the heat map of Fig.~\ref{fig:heatmap_todos}, here we show in a spatial map of Spain where we assign a brown color to those municipalities that belong to the larger cluster (high voteshare) and a turquoise color to those that belong to the smaller cluster (low voteshare). We find that the low voteshare cluster are predominantly linked with Catalonya and the Basque Country, the two areas of Spain with some pro-independence aspirations. No obvious change is perceived between 2015 and 2016}
\label{fig:map2015}
\end{figure*}
\subsection{Co-occurence heat maps}
Our second analysis is inspired by a recent study \cite{klimek} that explore the co-occurring statistics of vote and turnout numbers and the associated double mechanism of incremental and extreme fraud by plotting two dimensional histograms (heat maps) reporting, for a given political party, the percentage of vote (voteshare) it got as compared to the percentage of participation. According to Klimek and co-authors, incremental fraud occurs when with a given rate, ballots for one party are added to the urn and/or votes for other parties are taken away, and this mechanism is revealed when the histograms smear out towards the top right corner of the histograms. On the other hand, extreme fraud (which corresponds to reporting close to complete turnout and almost all votes for a single party) emerges when the distribution transitions from unimodal to bimodal and one of the modes corresponding to a cluster that concentrates close to that corner of $100 \%$ participation (complete turnout) and very large vote percentage. They applied these statistical principles to several national elections, concluding that in the cases of Russia and Uganda fraudulent manipulation was the most likely underlying mechanism.
In Fig.~\ref{fig:heatmap_todos} we plot such heat maps for the 2016 case for all four political parties. Data for PSOE, Unidos-Podemos and C's do not show any sign of fraudulent manipulation. In the case of PP results are less clear, as there indeed exists a (rather weak) tendency of the data to smear out towards the top right corner (results for 2015 are very similar and have been reporter in appendix Fig.~\ref{fig:heatmap_todos_2015}). We don't find any sign of systematic extreme fraud, although it is worth stating that we have found a small subset of municipalities where just one party received 100\% of the voteshare (see table \ref{table:raros}). Without exception, this is the popular party (PP), something that is in principle suspicious. Nevertheless a closer inspection reveals that these municipalities are extremely small and thus consensus in one political option cannot be ruled out statistically.\\
A further interesting peculiarity for the case of the conservative party PP) is the existence of two clusters of municipalities  (bimodal distribution) that gathers two different voting strategies: one relatively small, located at small voteshare and the other one at high voteshare, which is more spread out (we don't observe bimodality for the rest of political parties). We have labeled municipalities according to which cluster they belong (assigning a brown label for the larger cluster and a turquoise label for the smaller one) and plotted them in Fig.~\ref{fig:map2015}. Just by visual inspection we can appreciate that the category linked with the smaller cluster is mainly formed by Catalonia and the Basque Country (regions with pro-independence aspirations and a strong nationalist tradition), something that was recently pointed out independently \cite{new2}, and some further municipalities in regions that have been considered PSOE strongholds historically.

\begin{table}[]
\centering
\begin{tabular}{l|l|l|l|l}
  Year&Municipality&Population&Turnout&Party receiving100\% of the voteshare \\
  \hline
  \hline
2015&Castilnuevo (Guadalajara)&8&88\%&PP\\
2015&Valdemadera (La Rioja)&7&100\%&PP\\
2016&Castilnuevo (Guadalajara)&7&100\%&PP\\
2016&Rebollosa de Jadraque (Guadalajara)&10&90\%&PP\\
2016&Congostrina (Guadalajara)&16&62\%&PP\\
2016&La Vid de Bureba (Burgos)&16&65\%&PP\\
2016&Portillo de Soria (Soria)&16&75\%&PP\\
2016&Valdemadera (La Rioja)&8&88\%&PP\\
\hline      
\hline
\end{tabular}
\caption{List of municipalities where a single party receives $100\%$ of the voteshare.}
\label{table:raros}
\end{table}

\section{Discussion}
In this work we have studied the statistical properties of vote counts in the Spanish national elections that took place in December 2015 and June 2016, focusing on three separate questions: (i) breakdown of Bipartisanship, (ii) region-to-region similarity in vote percentage, and (iii) election forensics for fraud detection.\\

\noindent On relation to (i), our results highlight that the bipartisanship system has suffered a clear breakdown in 2015 --at least in regions associated to a more widespread cosmopolite society--. Such breakdown consolidate over time but doesn't increase, probably due to the risk aversion of not finding workable majorities in the second election round, and even evidences a subtle decrease as captured by the diversity index $N_{\text{eff}}$. Bipartisanship breakdown is actually a quite complex phenomenon with a high degree of heterogeneity at a regional level, probably due to regional political particularities.\\

\noindent Second, on relation to (ii) we have constructed a functional network of municipalities via cosine similarity of voting profiles. Interestingly, there is a very good matching between network communities which emerge by a community detection algorithm on the functional network and the actual Spanish autonomous communities. In particular, a classification of autonomous communities emerges naturally: we find that some autonomous communities whose functional network community counterpart is more cohesive and stable over time (e.g. Catalonia, Basque Country, Madrid, Navarra), some whose counterpart is only well defined in 2015 when Bipartisanship breakdown was more acute (e.g. Murcia, Valencia, etc), and the rest where there is no clear matching. Beyond the probably amplified role played by regionalist parties in 2015, we don't have clear socio-political explanations for such emergent classification and we leave this as an open problem. Other aspects left for future work include network pruning using different criteria to the ones applied in this work, such as using a fixed similarity threshold.\\

\noindent On relation to fraud detection, for the 2016 elections, the unusually high discrepancy found between electoral surveys preceding and on the day of the elections (26th June) and the actual electoral results have been a source of debate and controversy in Spanish media.
To the best of our knowledge, this work is among the first systematic analysis of its kind for Spanish elections (see however \cite{new, new2}). The first and general conclusion on relation to question (iii) we have extracted is that the voting distributions don't show any systematic and significant change between the 2015 and the 2016 elections, as all statistical results are qualitatively identical. This is in line with the original analysts thesis that were discussed soon after it was learned that Spain had to go into a second election given the inability of the parliament to find a suitable coalition, but at odds with most of the polls and surveys of vote intention which were predicting a much different scenario as 26th June approached.\\
The first analysis is based on the hypothesis that under clean conditions, vote count data should conform to Benford's law. At the national scale we have found a general good qualitative and quantitative conformance to Benford's law for the first (1BL) and second (2BL) digits, with small deviations only occurring for 1BL in the conservative party, where the null hypothesis can be rejected at $99\%$ confidence in both years according to the standard Pearson $\chi^2$ hypothesis test, a result which is not confirmed using an alternative test (mean absolute deviation) proposed by Nigrini. For 2BL only $\chi^2$ flags up some concerns at the $95\%$ confidence level for Podemos / Unidos-Podemos, but the null hypothesis cannot be rejected at $99\%$ and again in this case, MAD statistic is less conservative and accepts the null hypothesis for every party.\\ 
If we change the resolution and explore results for each individual precinct, results show a completely different story: conformance to 1BL is accepted according to $\chi^2$ but systematically rejected according to MAD, and conformance to 2BL is consistently rejected according to both $\chi^2$ and MAD statistics for every precinct and every political party. We have also shown that these are genuine results that cannot be associated to a lack of statistics. Finally, by aggregating vote counts per precinct and analyzing conformance to 1BL and 2BL at this level of aggregation, we obtain inconsistent and therefore inconclusive results, as $\chi^2$ cannot reject the null hypothesis above 95$\%$ confidence level systematically but conversely MAD suggests systematic nonconformance. This lack of consistency raises the question about what level of aggregation might be better-suited for BL-type analysis and which statistic is more reliable when assessing the goodness of fit, issues that certainly deserve further investigation.\\
Given the somewhat mixed results and acknowledging that the applicability of Benford's law tests to election forensic is not completely free from controversy \cite{contro1, contro2}, as a complementary analysis we further explored the correlations between percentage of participation and percentage of votes for each municipality, plotting two-dimensional histograms to detect the presence of so-called incremental and/or extreme fraud as described by Klimek et al. \cite{klimek}. Our results suggest that the results for PSOE, Unidos-Podemos and C's are apparently free from these mechanisms whereas in the case of PP we find a weak evidence of cluster smearing out similarly to what Klimek et al. refer to incremental fraud, an evidence which needs to be studied in more detail. The heat map of the conservative party also shows two clusters instead of a single one hence bimodality in the voteshare tendency: there exist two different groups of municipalities, including a small one where the tendency is to give a small voteshare to PP and a larger one where the voteshare takes larger values. Interestingly, according to a spatial analysis we have been able to confirm that the low voteshare cluster typically corresponds to regions which are considered nationalist (Catalonia, Basque Country) where the strength of regional options outperforms those that prevail at a nationwide scale.\\ 
\noindent All in all, these results suggest that further investigations and enquiries should be conducted in order to confirm and clarify the presence or absence of some of these apparent irregularities, to elucidate their source and quantify their impact in election results. On this respect, systematic comparative studies with historical SBipartisanshippanish data and analogous data (analysis at different levels of aggregation) from other similar democratic countries are needed.\\



\section{Appendix}
In this appendix we depict several additional figures and tables that complement the main study (see the main text for references to each of these figures).\\

\begin{figure*}
\centering
\includegraphics[width=0.32\columnwidth]{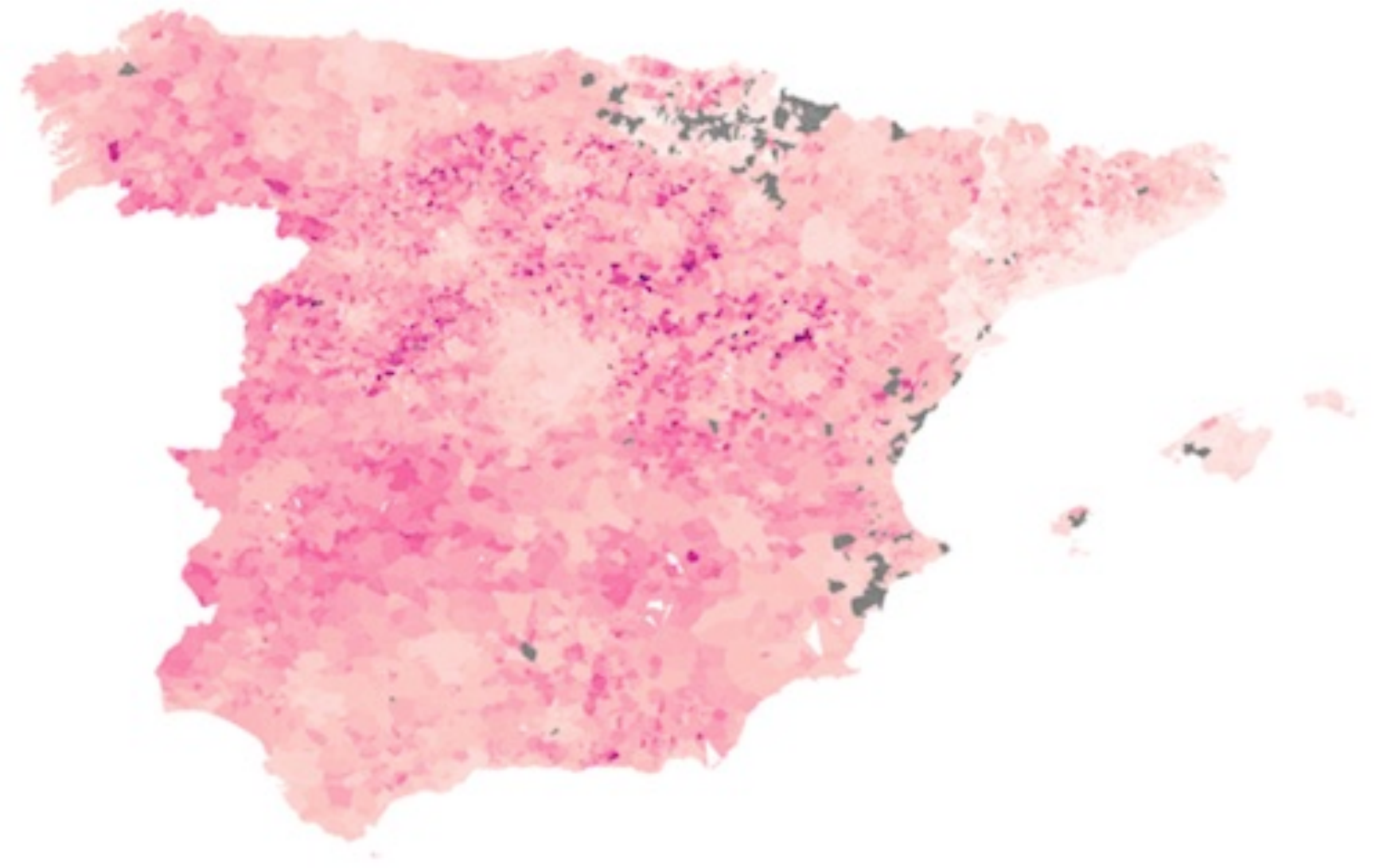}
\includegraphics[width=0.32\columnwidth]{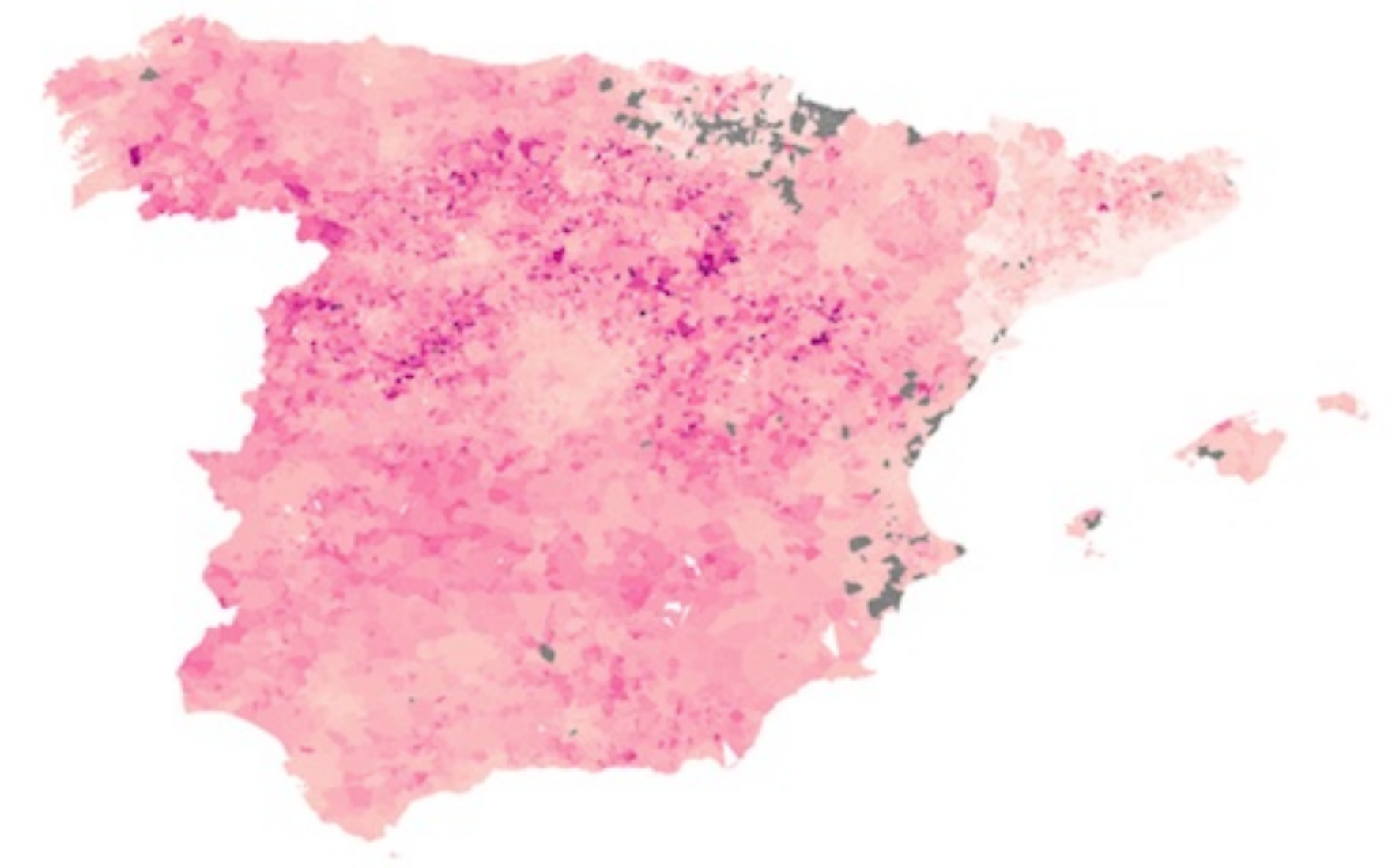}
\includegraphics[width=0.32\columnwidth]{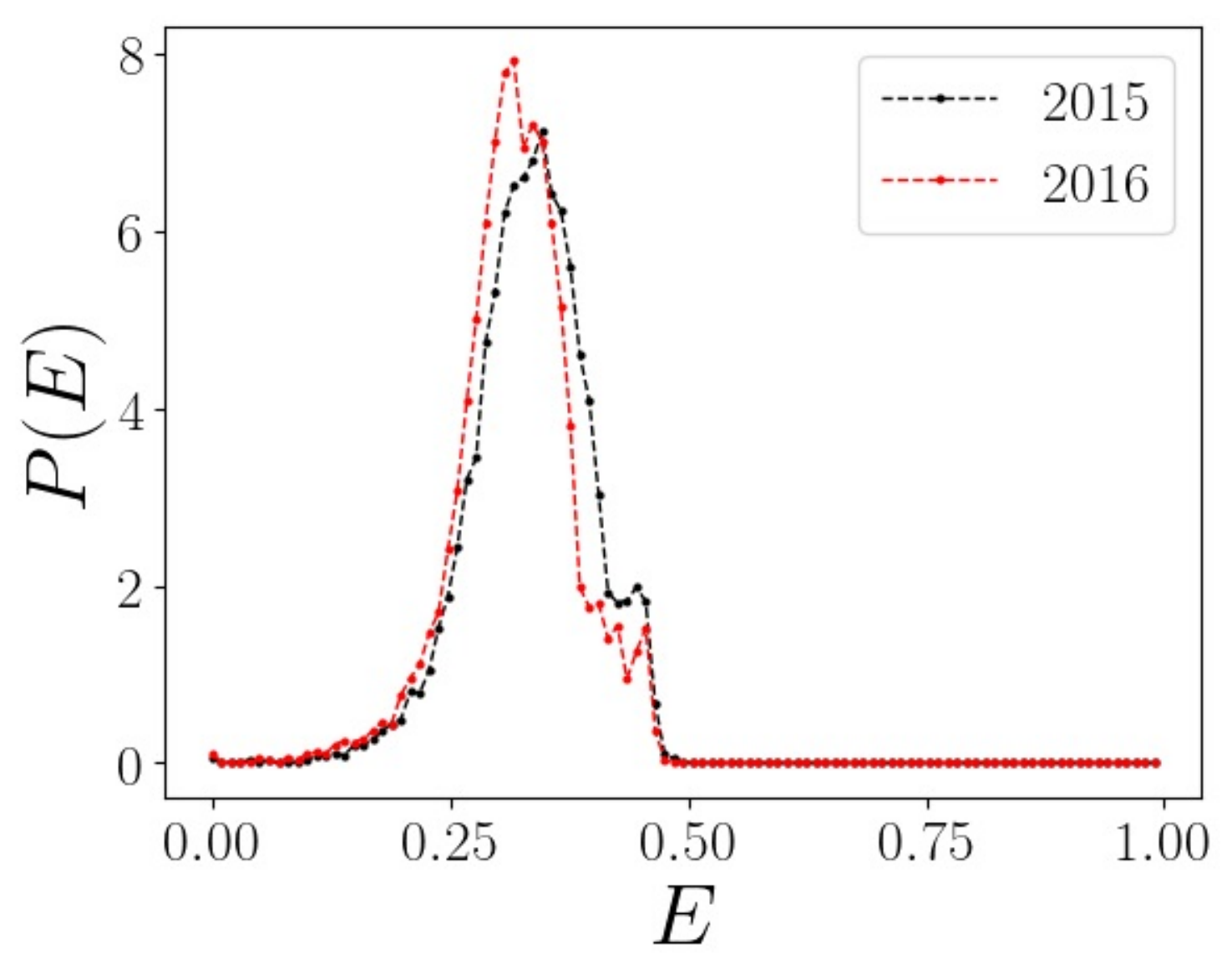}
\caption{(Left and middle panels) Heat maps of the Entropy index ($E$, see the text) at the municipality level (the darker, the lower entropy, and the more bipartisanship the system is) for 2015 elections (left panel) and 2016 elections (middle panel). Entropy is larger (leading to more heterogeneous vote and less bipartisanship) closer to cosmopolite cities (e.g. Madrid, Barcelona, etc). Overall this index is approximately constant over time ($\langle E\rangle=0.34\pm 0.06$ in 2015 and $\langle E\rangle=0.33\pm 0.06$ in 2016), similarly to the results obtained using the Bipartisanship index.}
\label{fig:E}
\end{figure*}

\begin{figure*}[h]
\centering
\includegraphics[width=0.45\columnwidth]{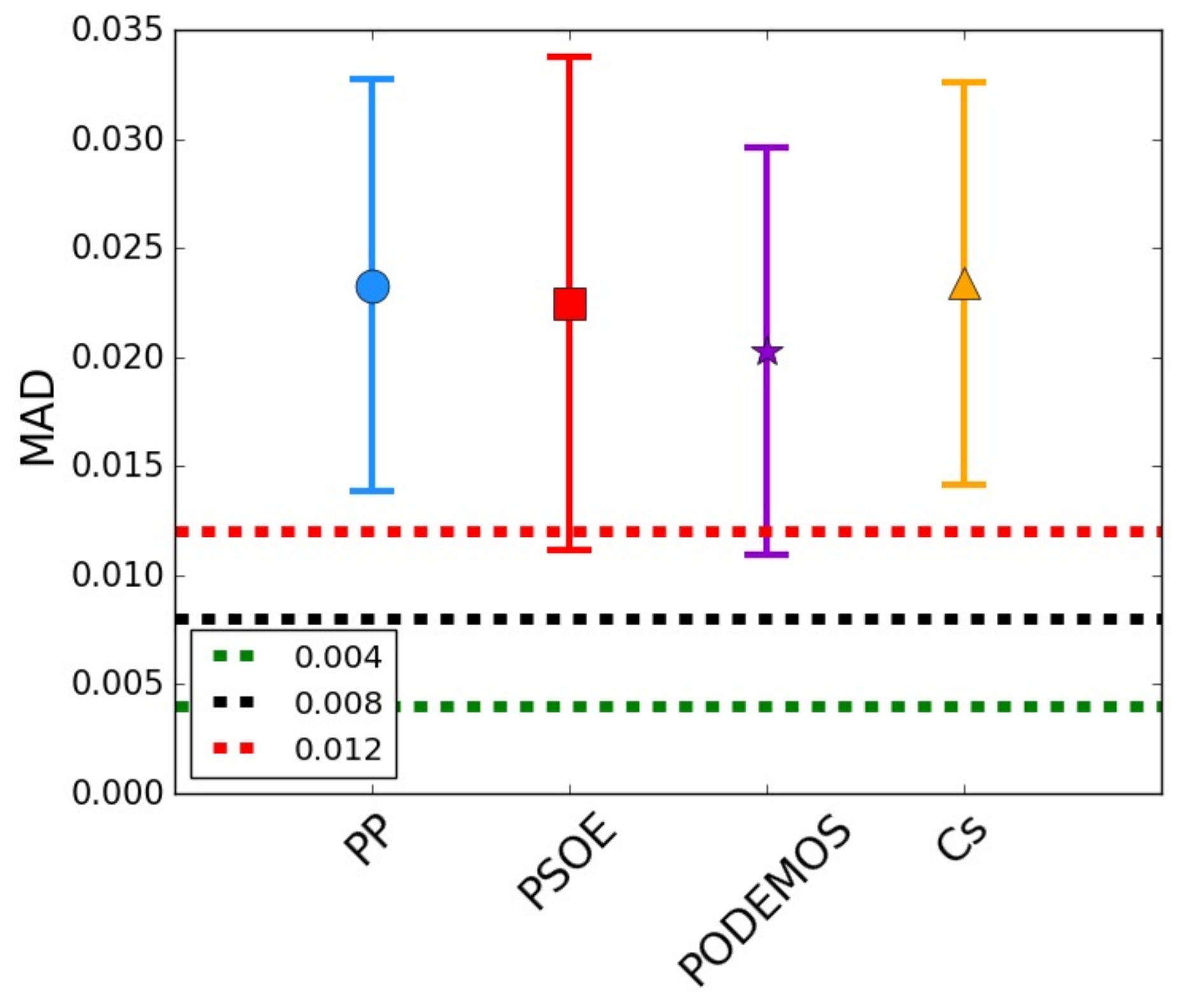}
\includegraphics[width=0.45\columnwidth]{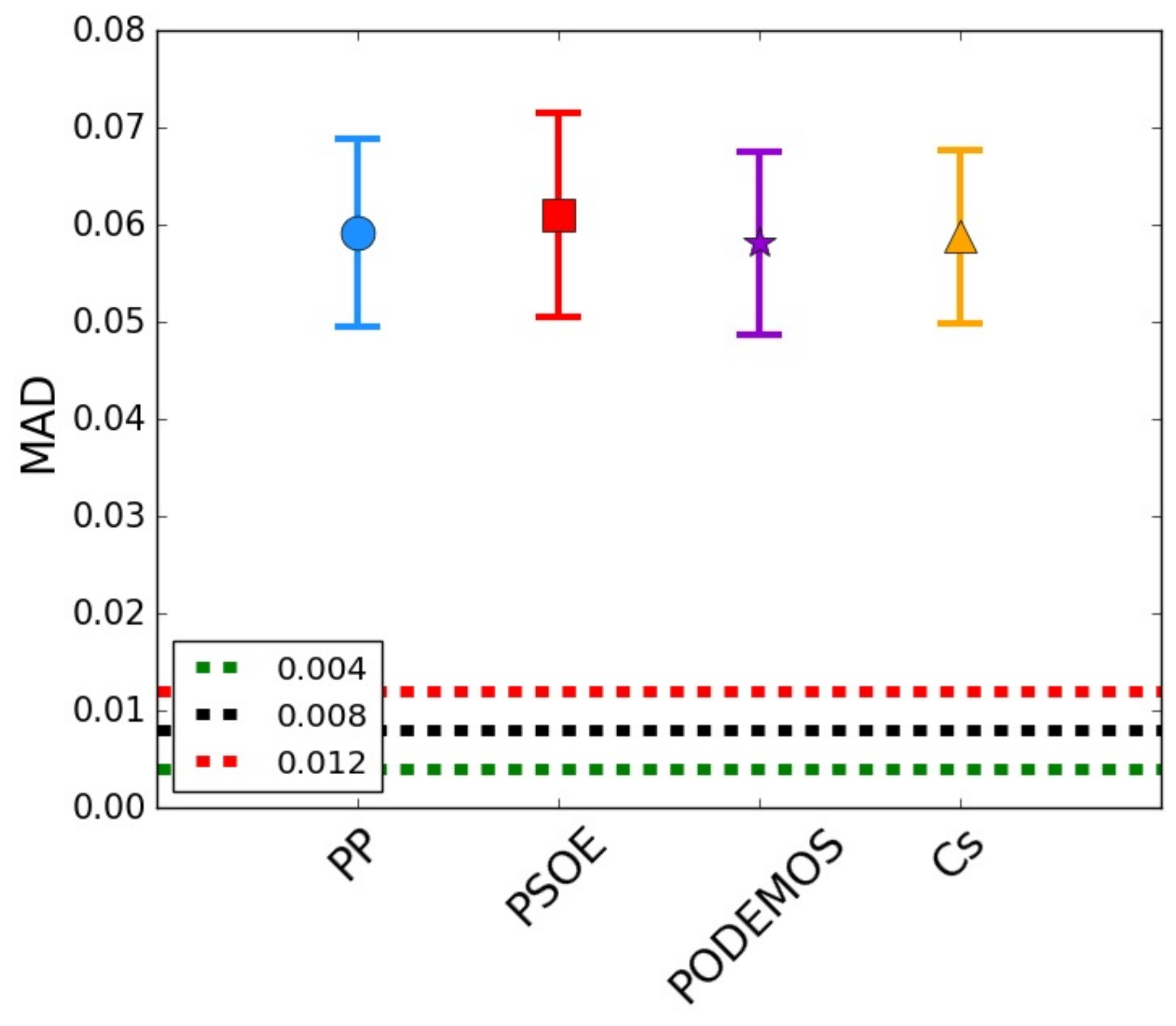}
\includegraphics[width=0.45\columnwidth]{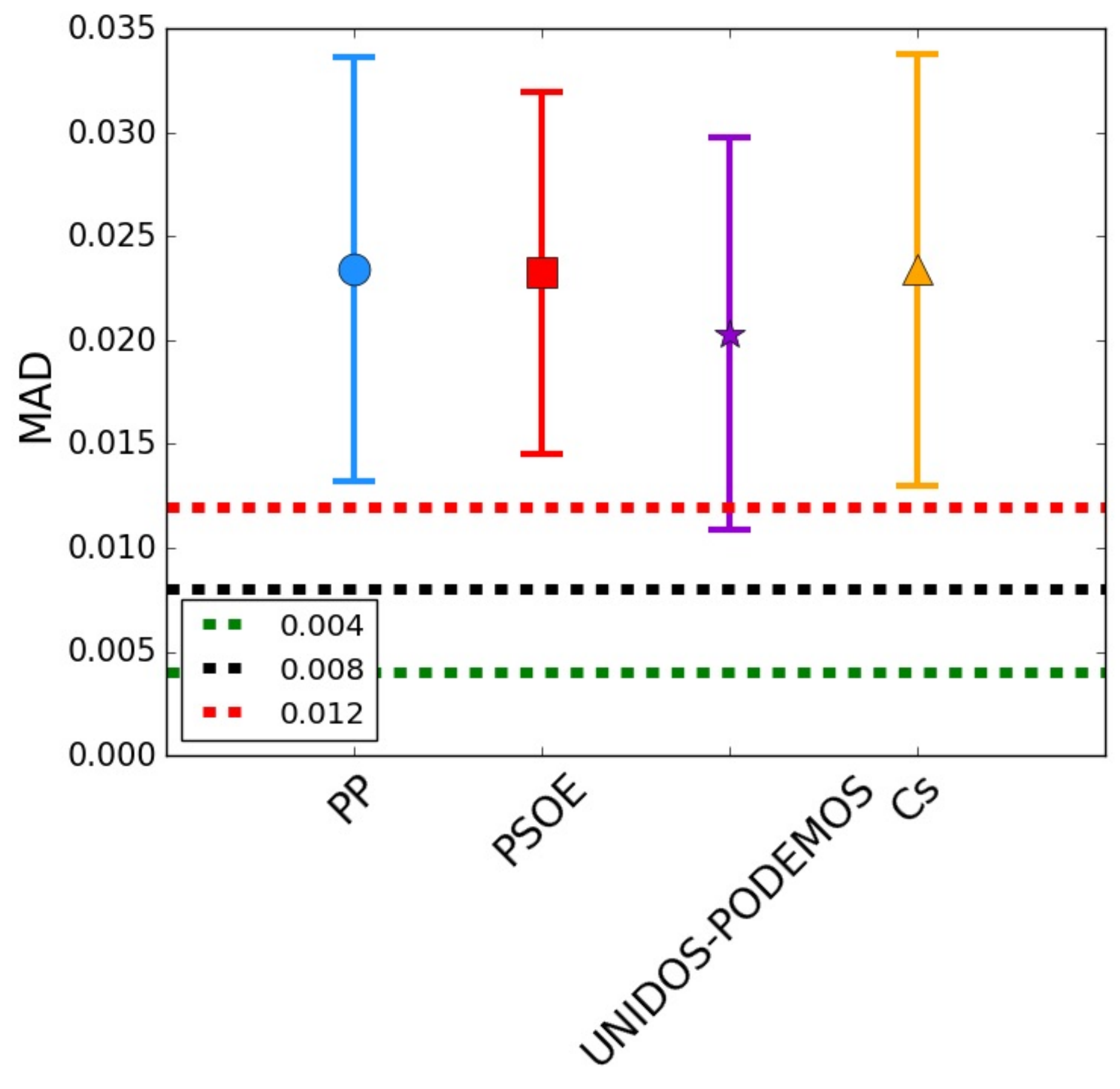}
\includegraphics[width=0.45\columnwidth]{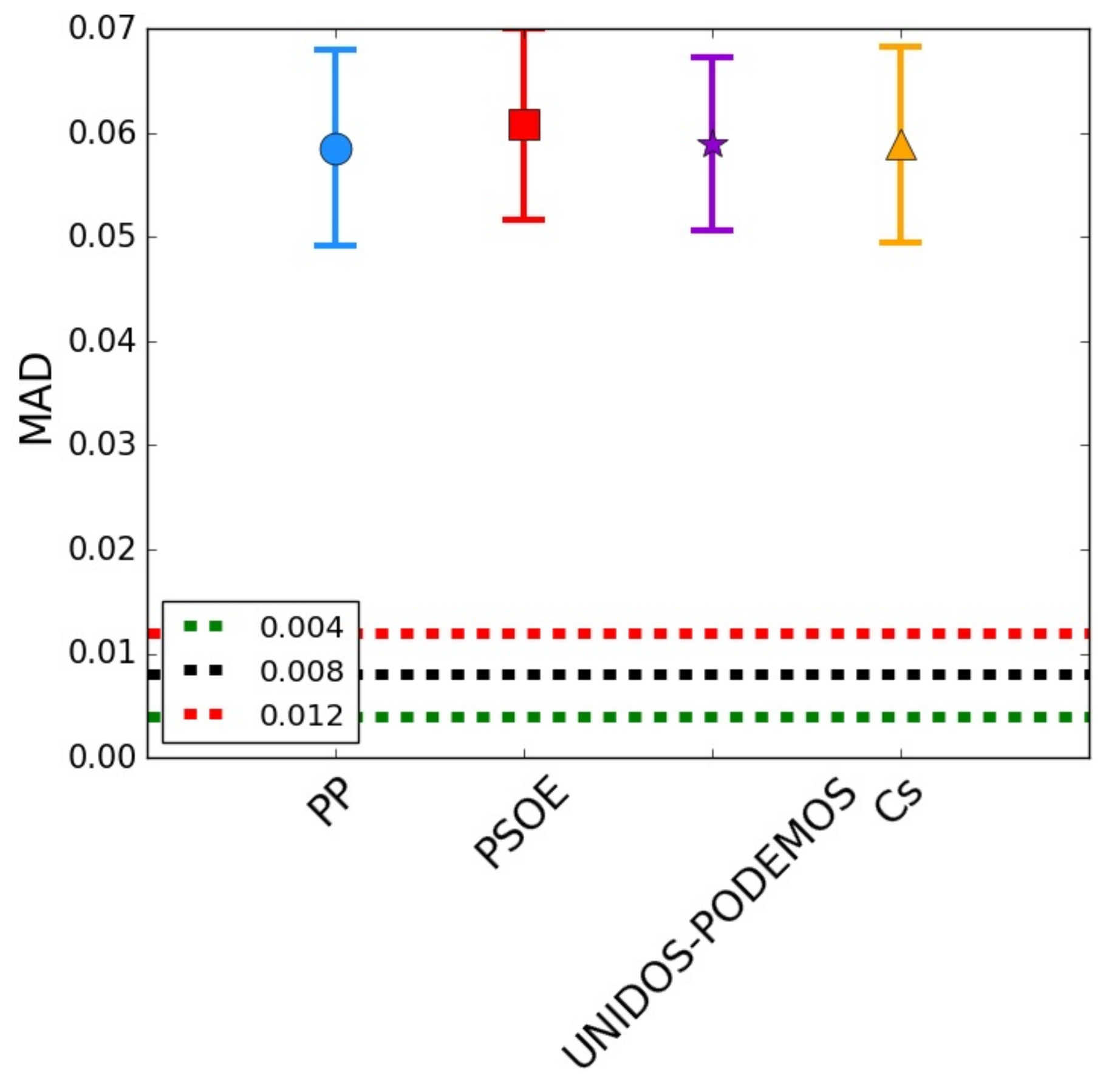}
\caption{Summary of MAD goodness of fit to 1BL (left panels) and 2BL (right panels) for 2015 (top panels) and 2016 (bottom panels), performed individually at each precinct (each precinct shows a precise distribution and an associated MAD, so here we plot the mean $\pm$ standard deviation over all spanish precincts, excluding Ceuta and Melilla, precincts with a single municipality). In every case the critical values for rejection at the 95 and 99$\%$ confidence level are shown.
Interestingly, in the case of 1BL for a large majority we accept conformance to Benford's law, whereas in the case of 2BL for a large majority the null hypothesis is rejected. All these results are consistent with the hypothesis test based on MAD (supplementary information)}
\label{fig:circu_mad}
\end{figure*}

\begin{figure*}[h]
\centering
\includegraphics[width=0.45\columnwidth]{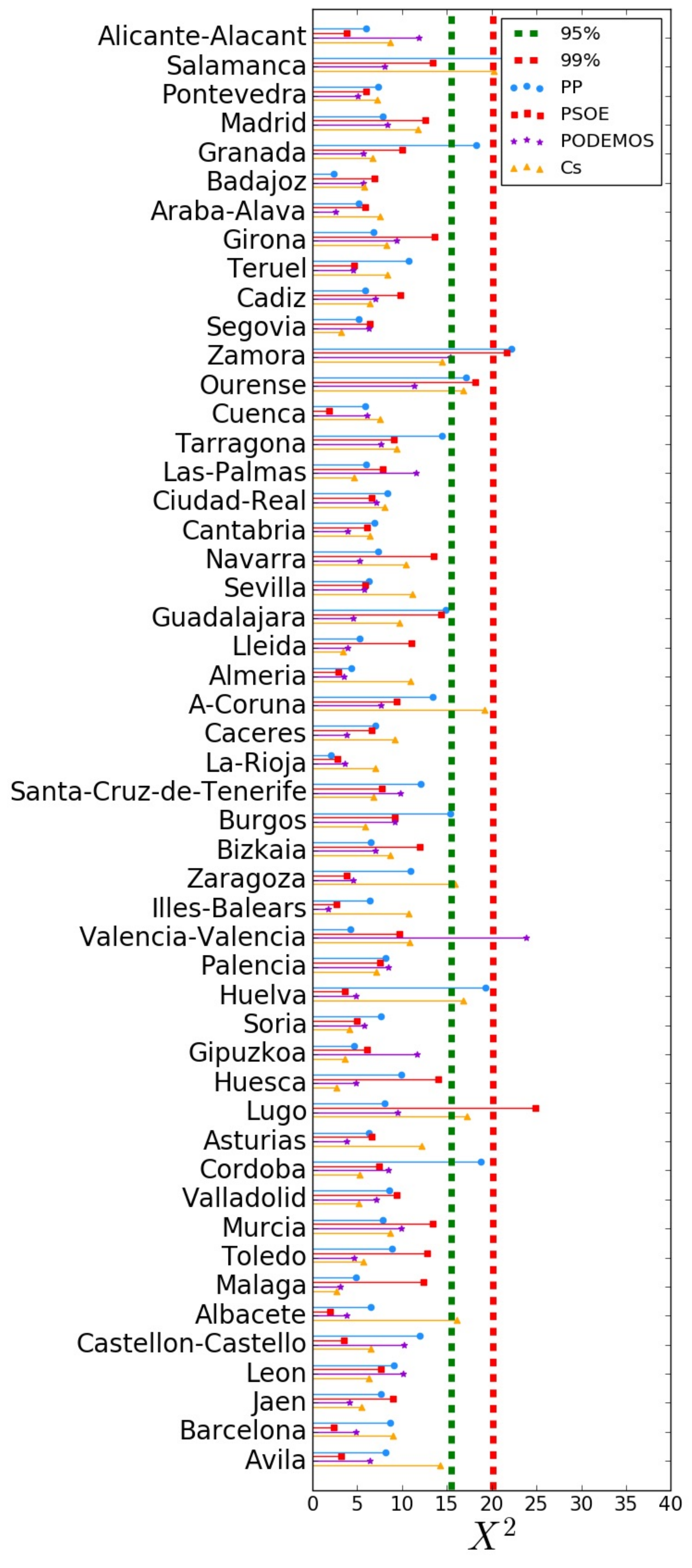}
\includegraphics[width=0.45\columnwidth]{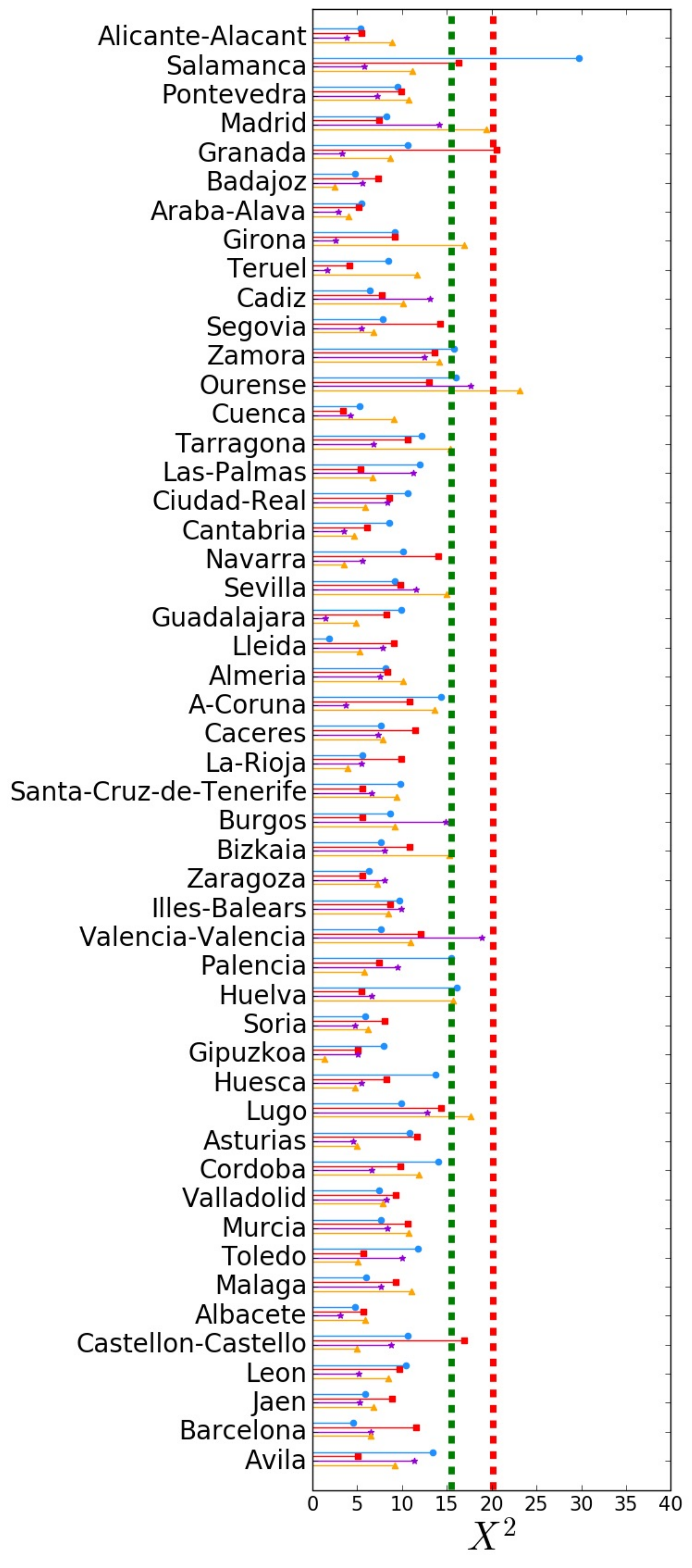}
\caption{$\chi^2$ values of the goodness of fit to 1BL for 2015 (left panel) and 2016 (right panel) at the aggregation level of precincts. In every case the critical values for rejection at the 95 and 99$\%$ confidence level are shown.
For a large majority we accept conformance to Benford's law. Note that results are inconsistent with the hypothesis test based on MAD as reported in Fig.~\ref{fig:circu_mad_unoporuno_1BL}.}
\label{fig:circu_chi2_unoporuno_1BL}
\end{figure*}

\begin{figure*}[h]
\centering
\includegraphics[width=0.45\columnwidth]{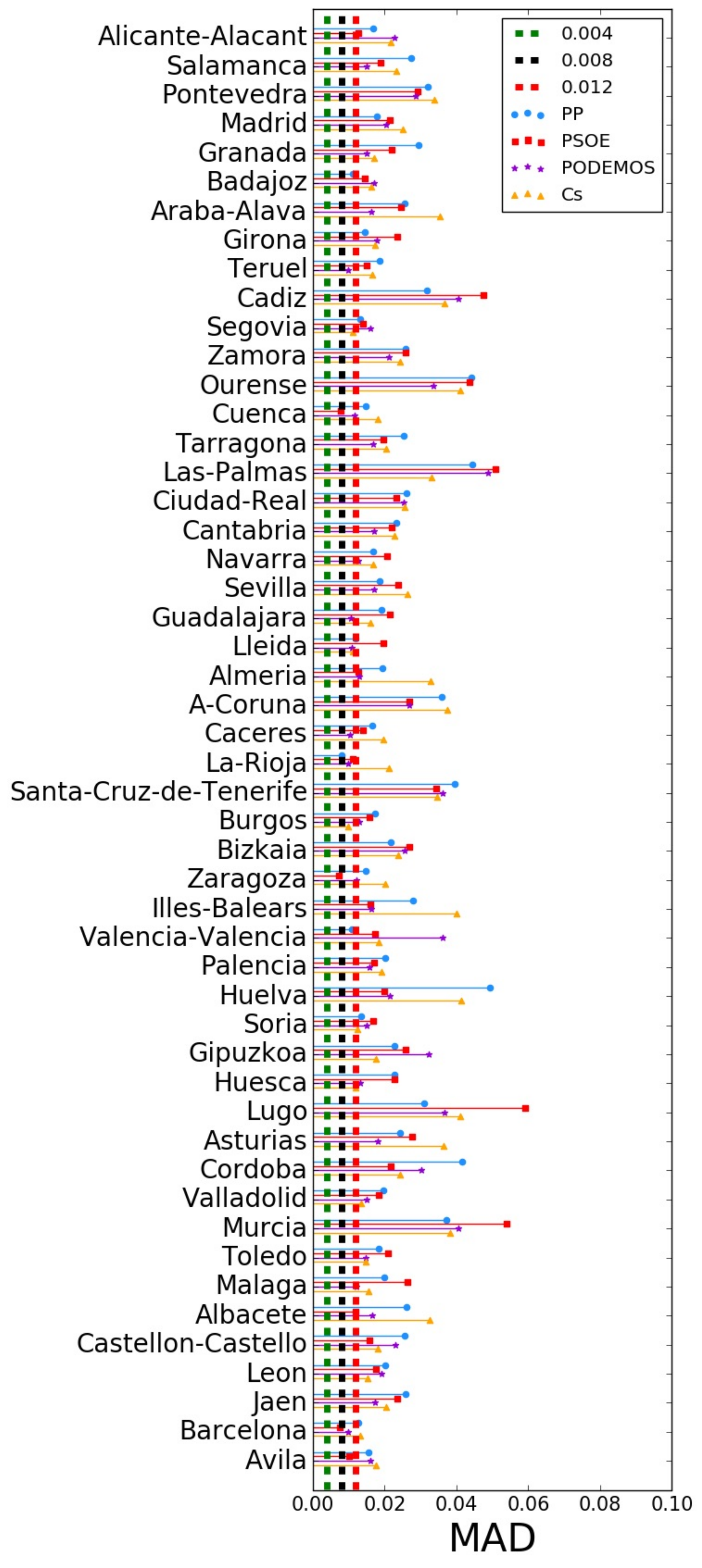}
\includegraphics[width=0.45\columnwidth]{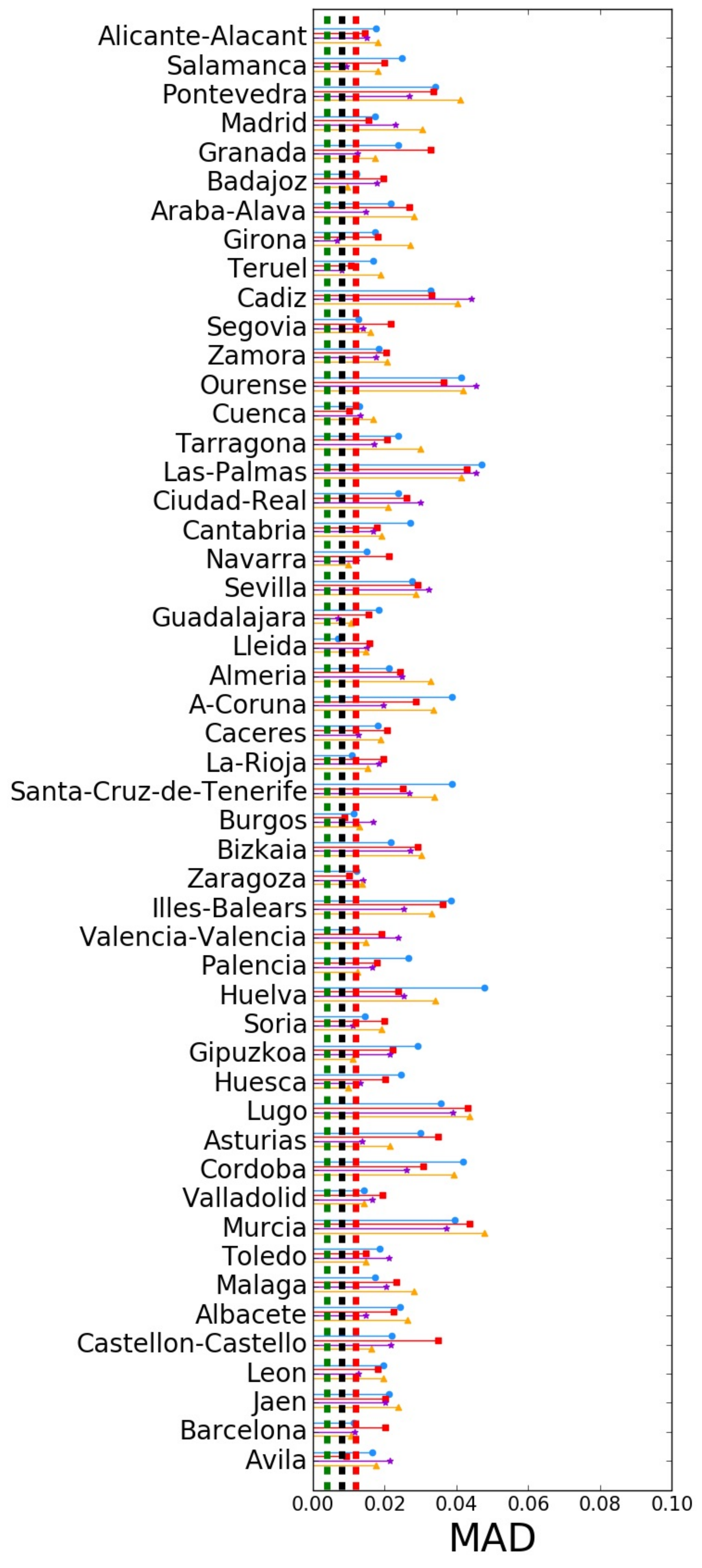}
\caption{MAD values of the goodness of fit to 1BL for 2015 (left panel) and 2016 (right panel) at the aggregation level of precincts. In every case the critical values for rejection at the 95 and 99$\%$ confidence level are shown. For a large majority we reject  conformance to Benford's law at the precinct level according to MAD, this result being inconsistent with the one found for Pearson's $\chi^2$ statistic.}
\label{fig:circu_mad_unoporuno_1BL}
\end{figure*}

\begin{figure*}[h]
\centering
\includegraphics[width=0.45\columnwidth]{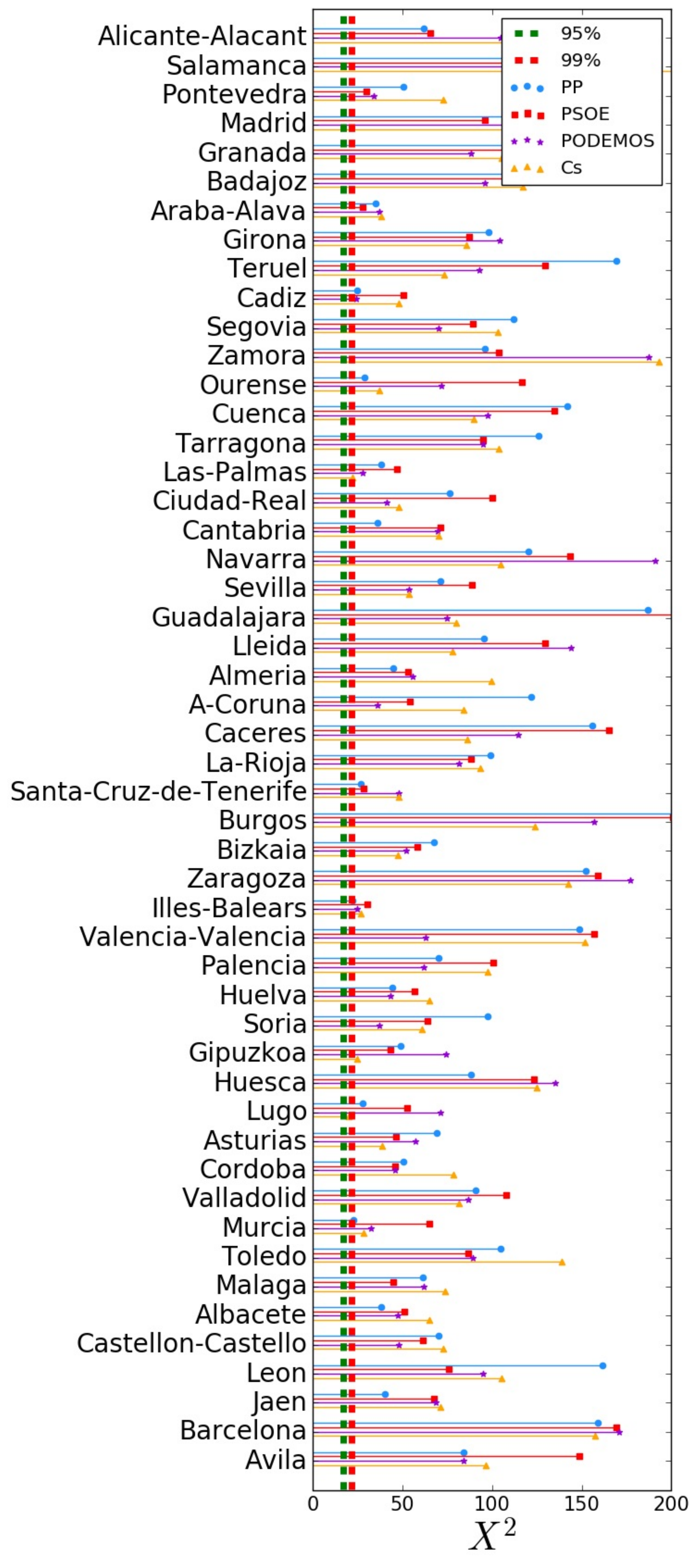}
\includegraphics[width=0.45\columnwidth]{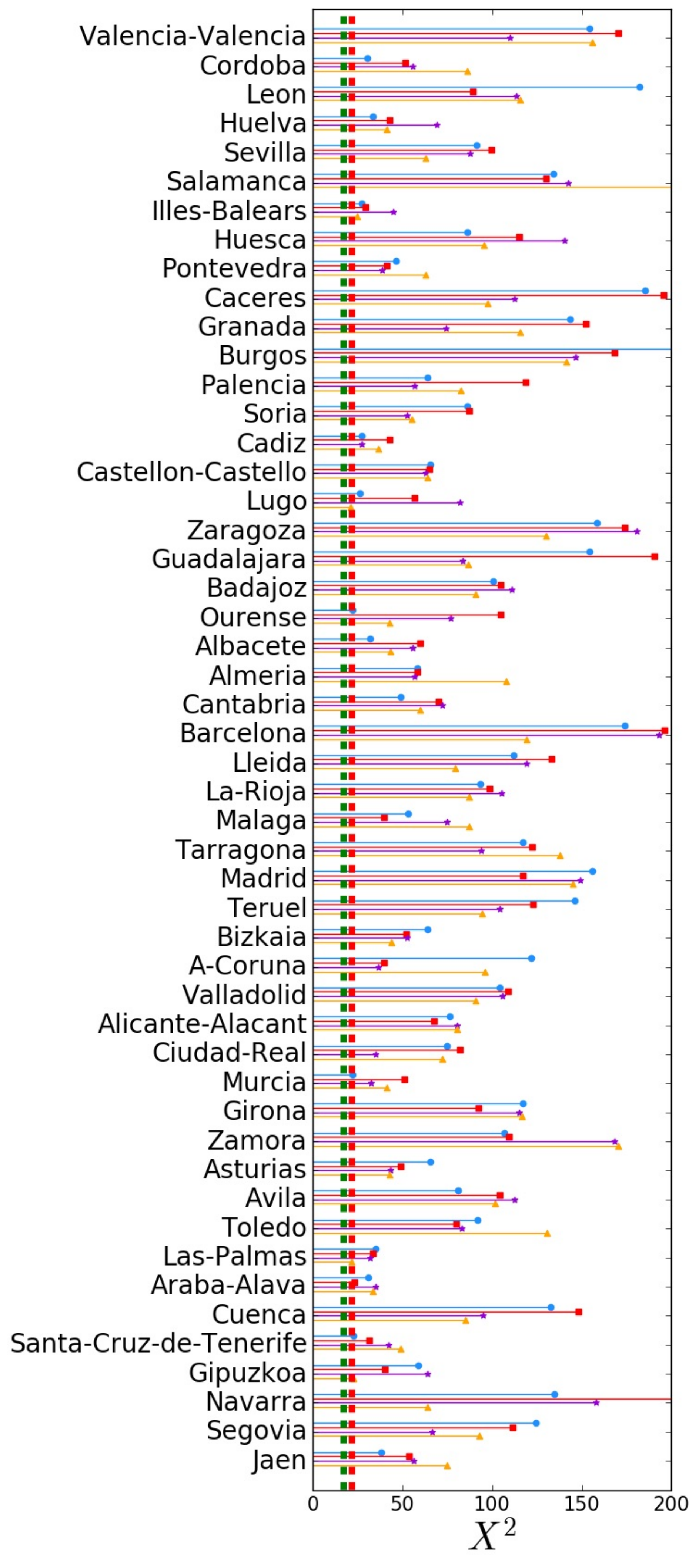}
\caption{$\chi^2$ values of the goodness of fit to 2BL for 2015 (left panel) and 2016 (right panel) at the aggregation level of precincts. In every case the critical values for rejection at the 95 and 99$\%$ confidence level are shown.
Virtually in all cases the null hypothesis is rejected. All these results are consistent with the hypothesis test based on MAD reported in Fig.~\ref{fig:circu_mad_unoporuno_2BL}.}
\label{fig:circu_chi2_unoporuno_2BL}
\end{figure*}

\begin{figure*}[h]
\centering
\includegraphics[width=0.45\columnwidth]{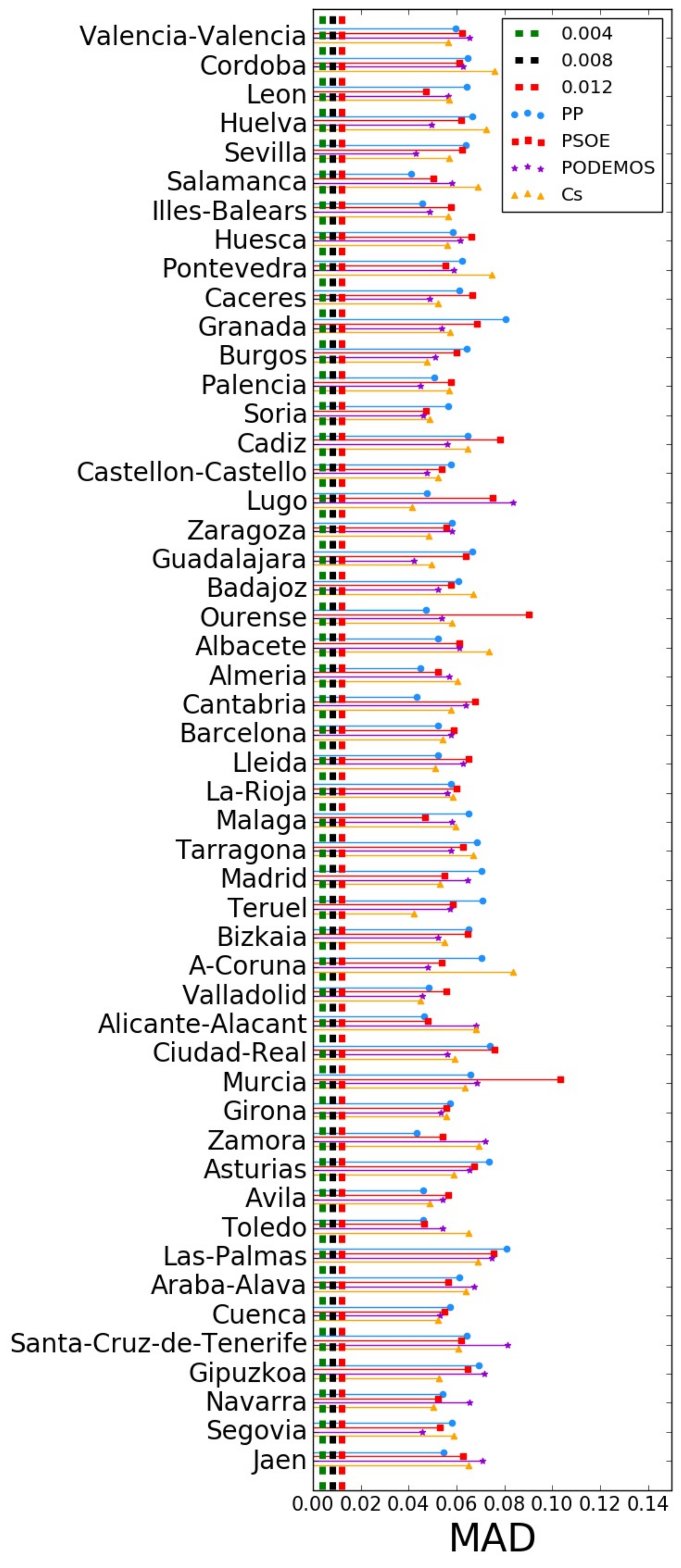}
\includegraphics[width=0.45\columnwidth]{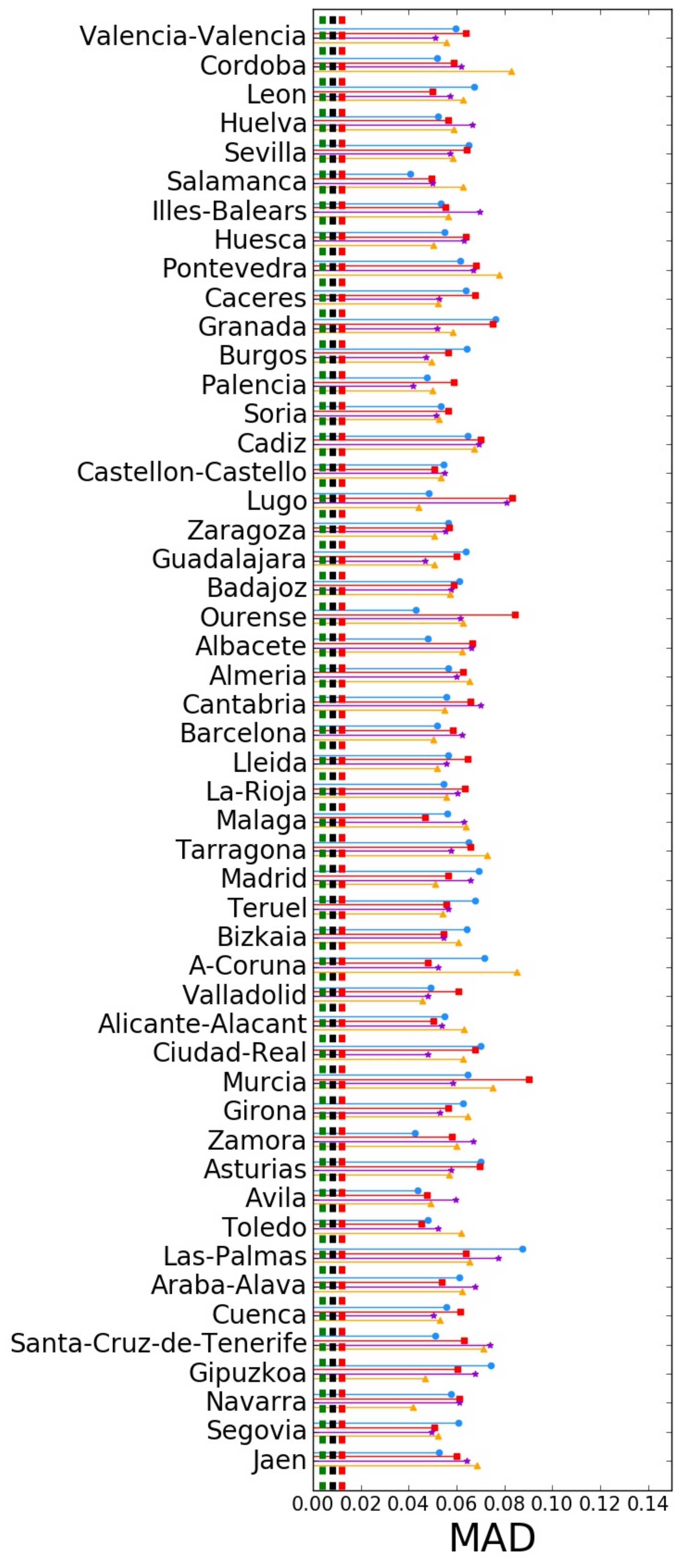}
\caption{MAD values of the goodness of fit to 2BL for 2015 (left panel) and 2016 (right panel) at the aggregation level of precincts. In every case the critical values for rejection at the 95 and 99$\%$ confidence level are shown.
Virtually in all cases the null hypothesis is rejected.}
\label{fig:circu_mad_unoporuno_2BL}
\end{figure*}

\begin{figure*}
\centering
\includegraphics[width=0.45\columnwidth]{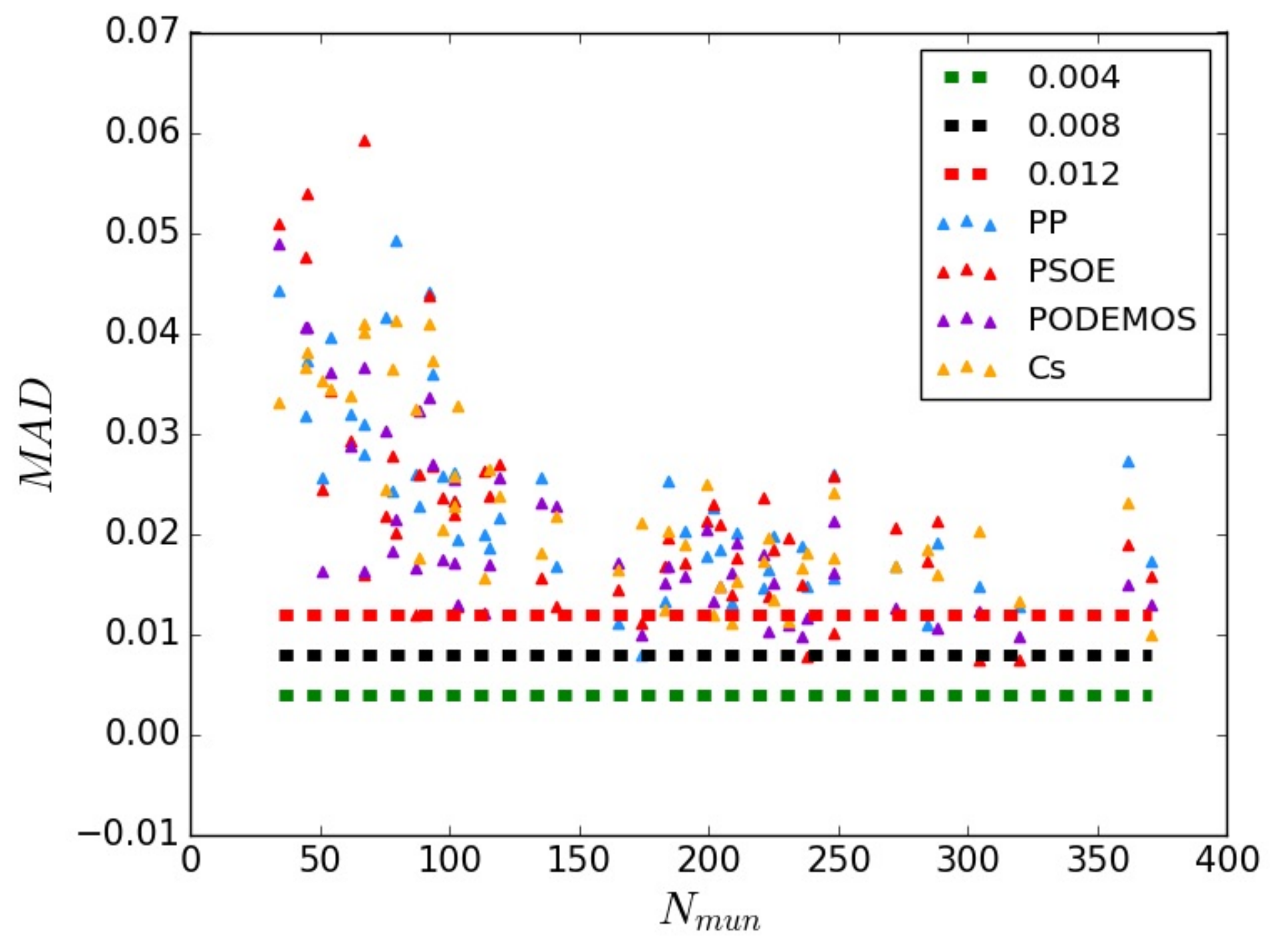}
\includegraphics[width=0.45\columnwidth]{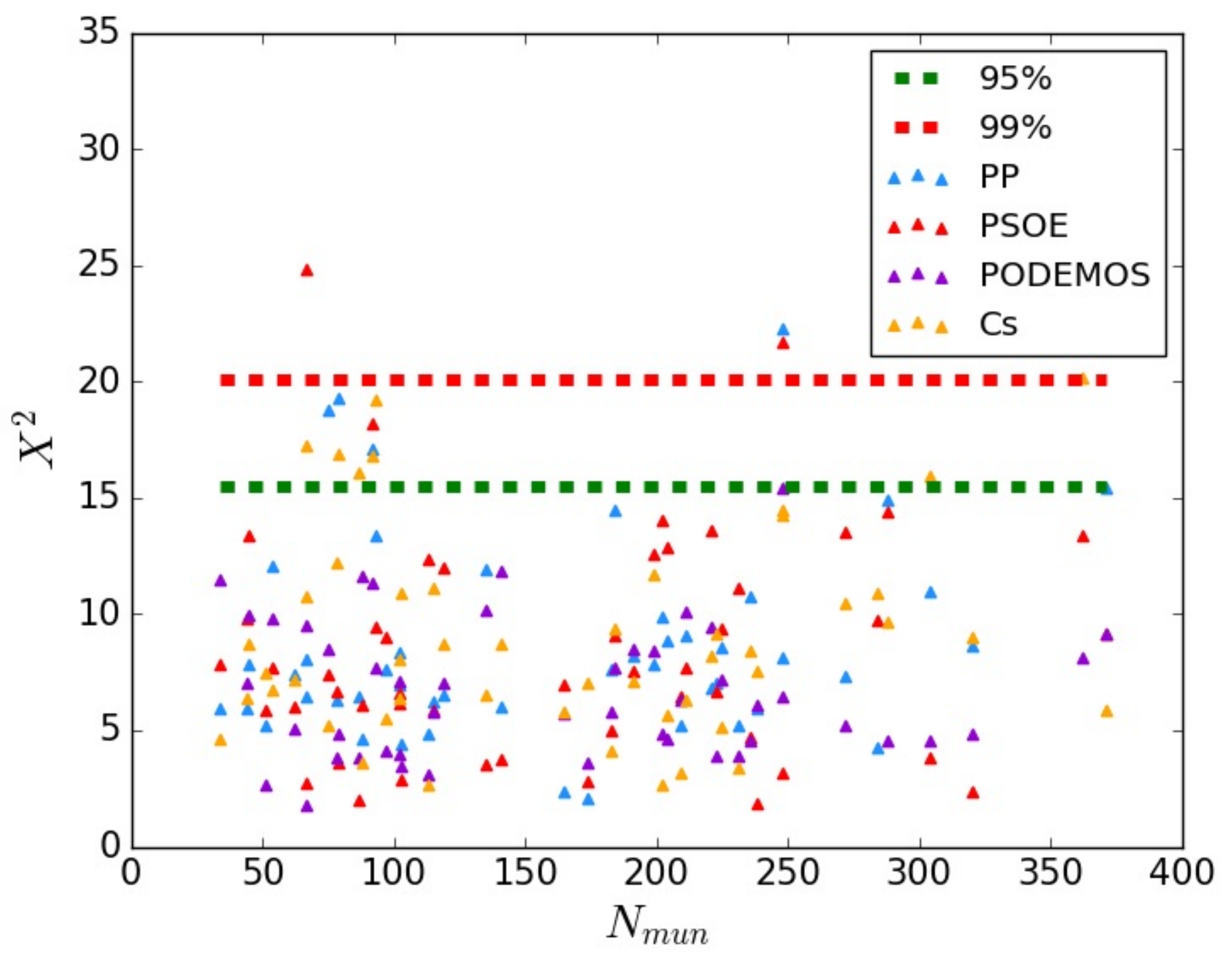}
\includegraphics[width=0.45\columnwidth]{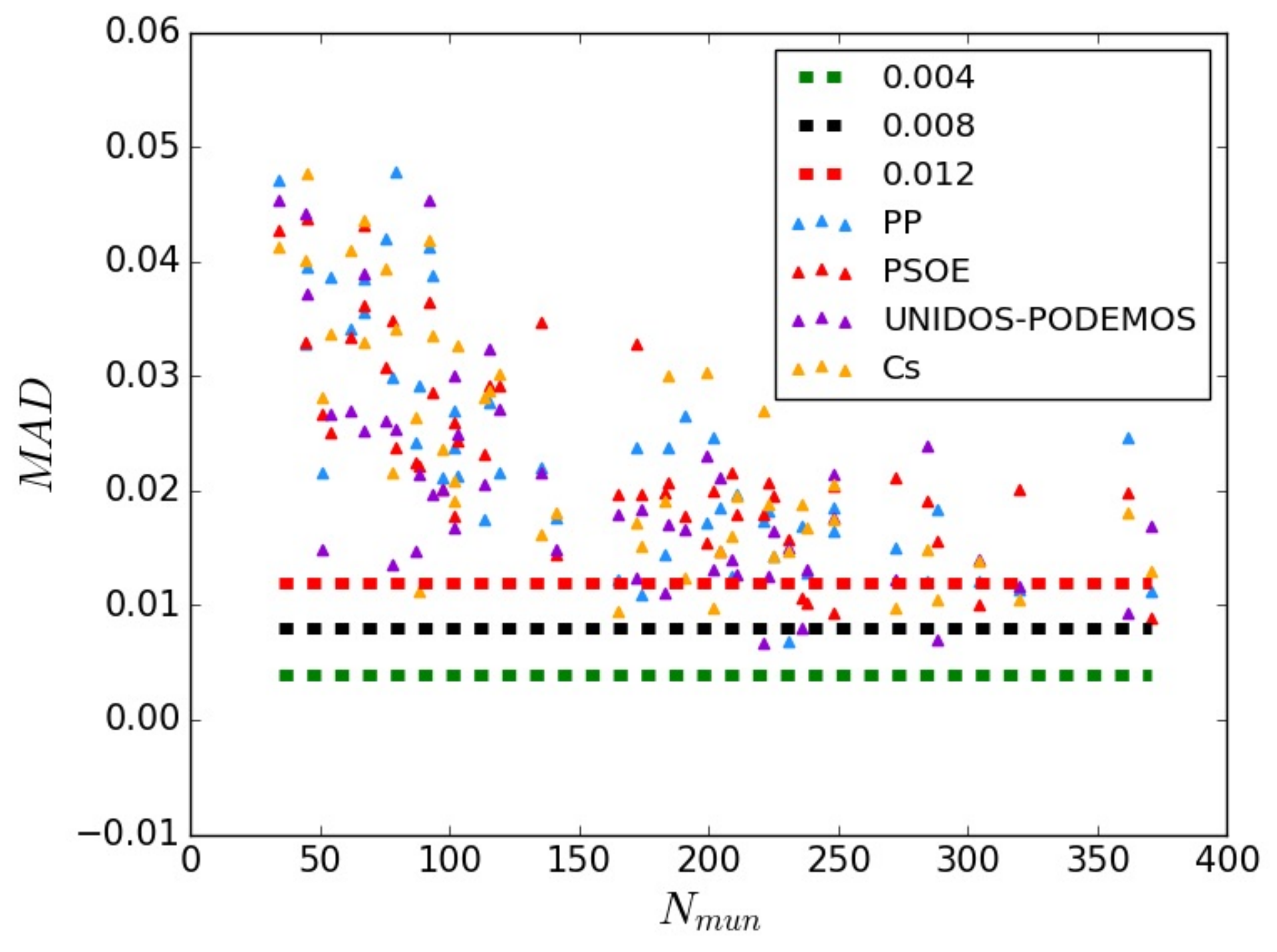}
\includegraphics[width=0.45\columnwidth]{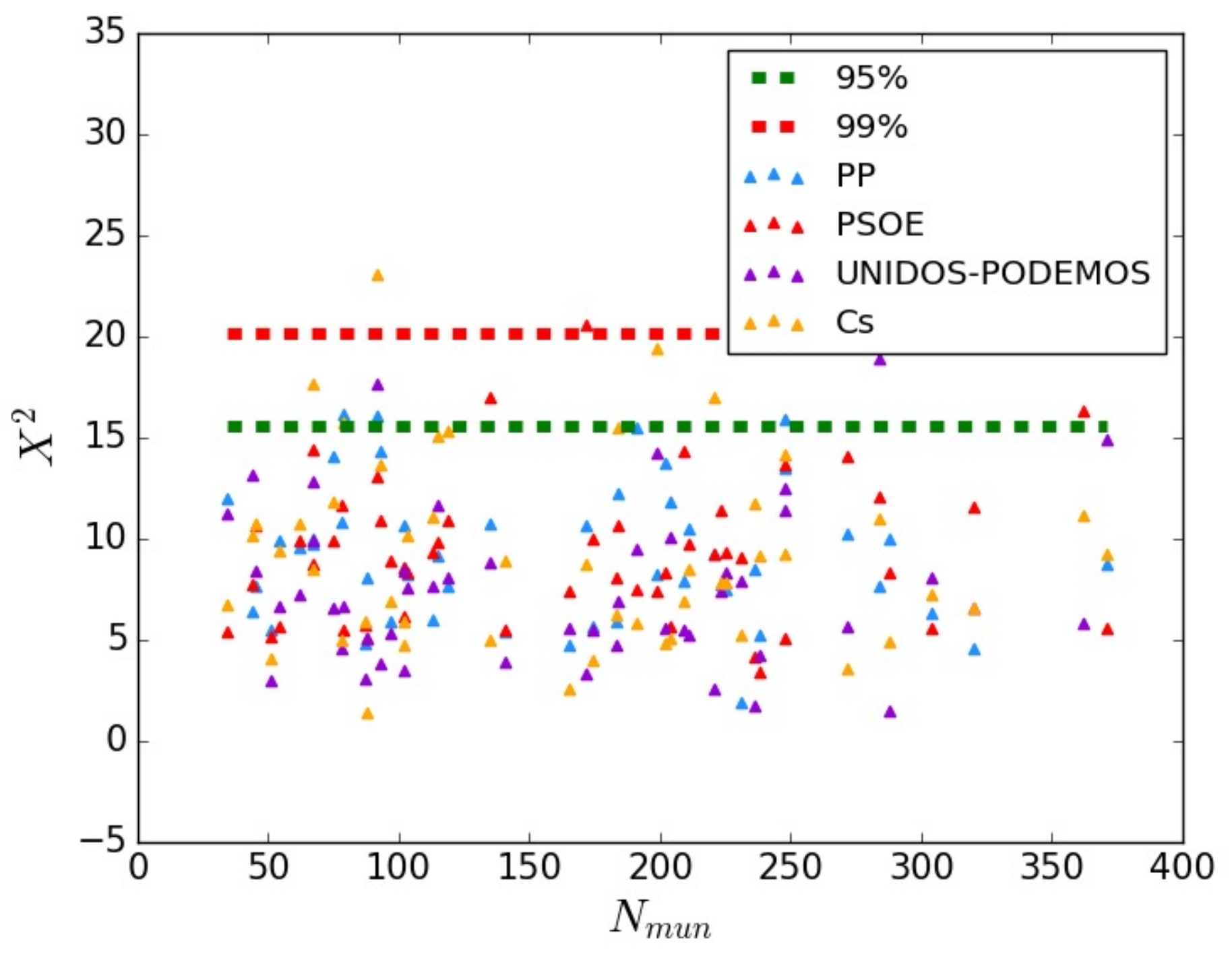}
\caption{Scatter plot of the MAD (left panels) and $\chi^2$ (right panels) statistics extracted from the 1BL test of each precinct as a function of the number of municipalities in each precinct for year 2015 (top panels) and  2016 (bottom panels). In the case of MAD, we find a negative correlation as expected, but this correlation is not enough to explain the systematic nonconformance to 1BL. In the case of $\chi^2$ there is no perceivable size effect.}
\label{fig:scatter_bl}
\end{figure*}

\begin{figure*}
\centering
\includegraphics[width=0.4\columnwidth]{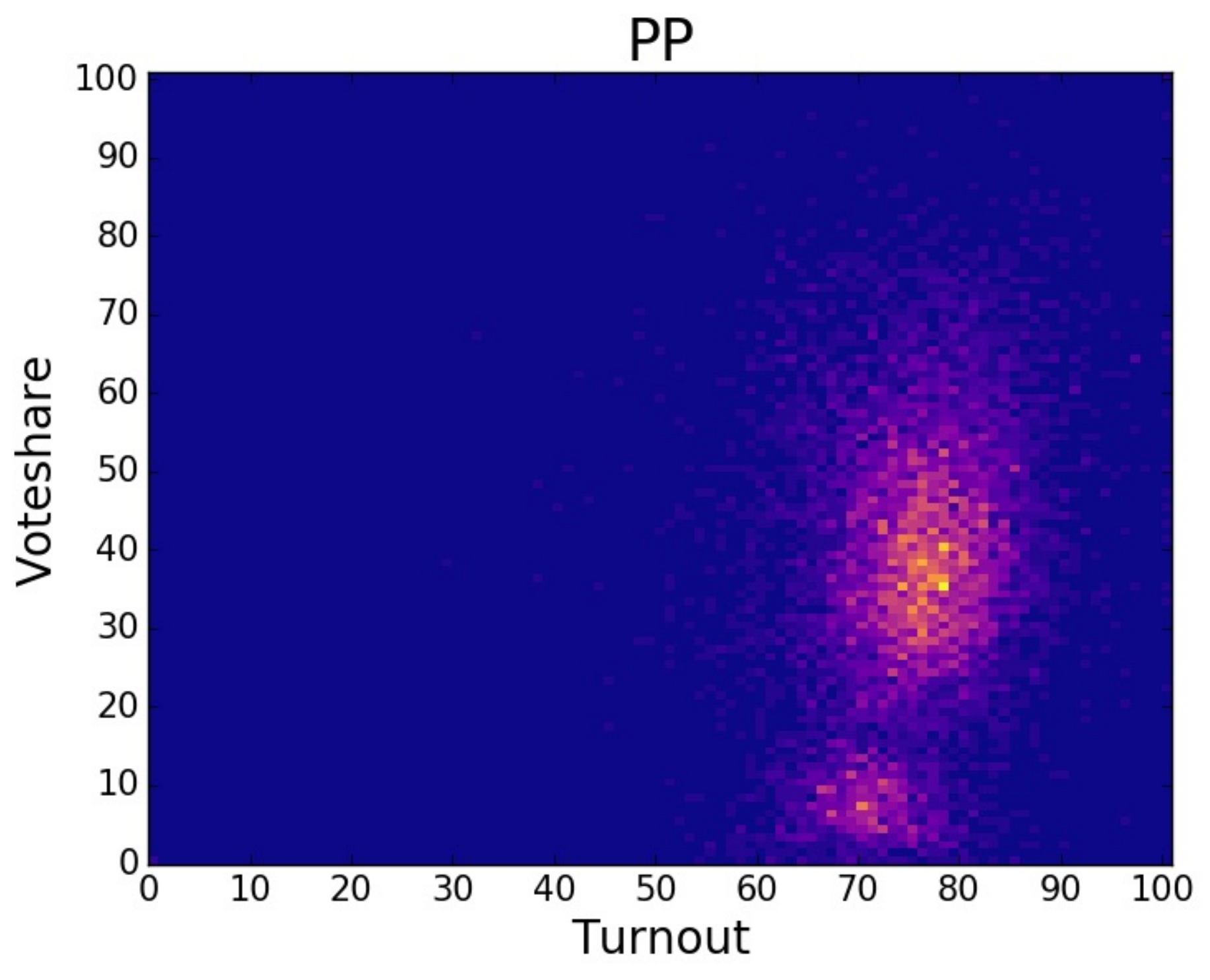}
\includegraphics[width=0.4\columnwidth]{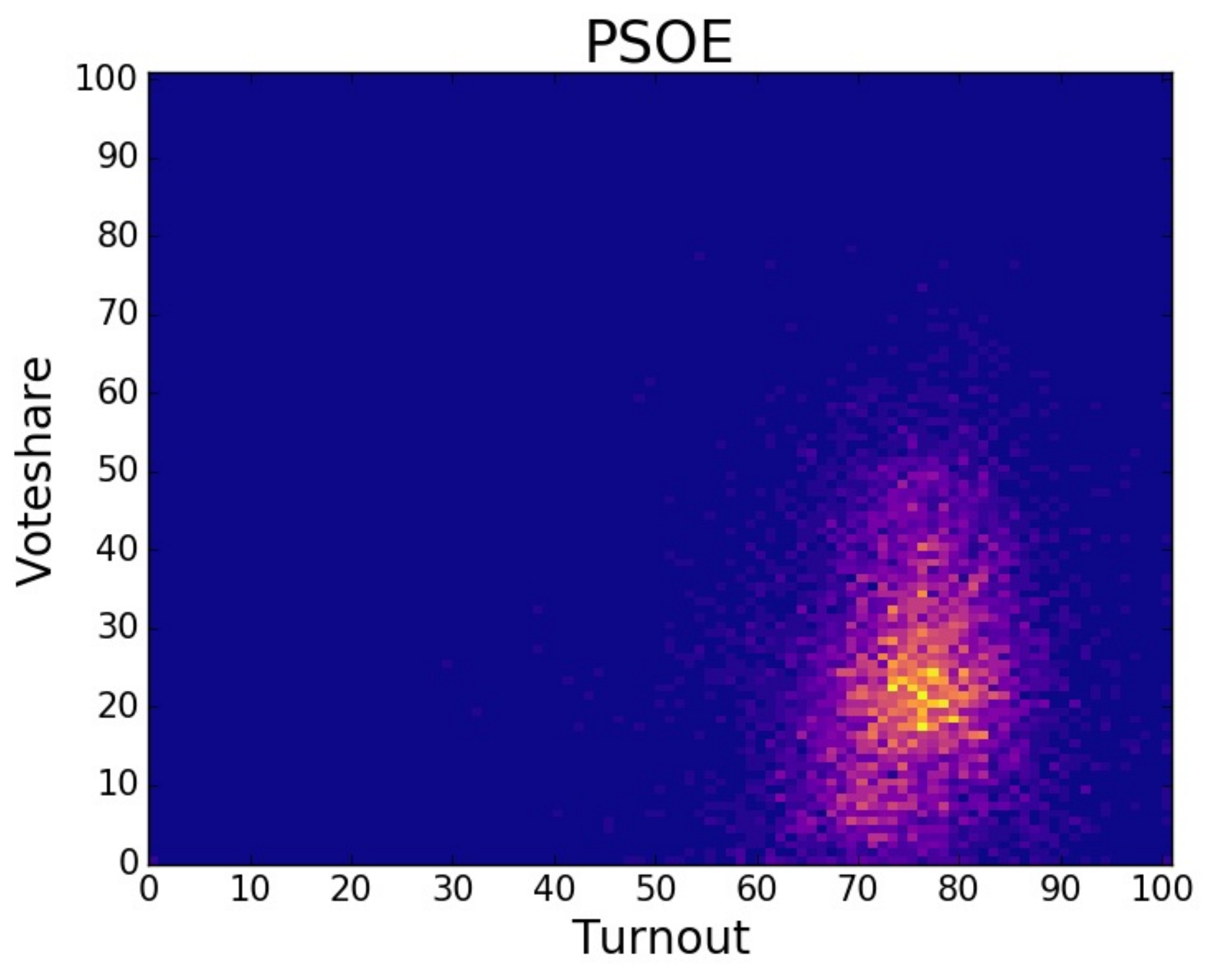}
\includegraphics[width=0.4\columnwidth]{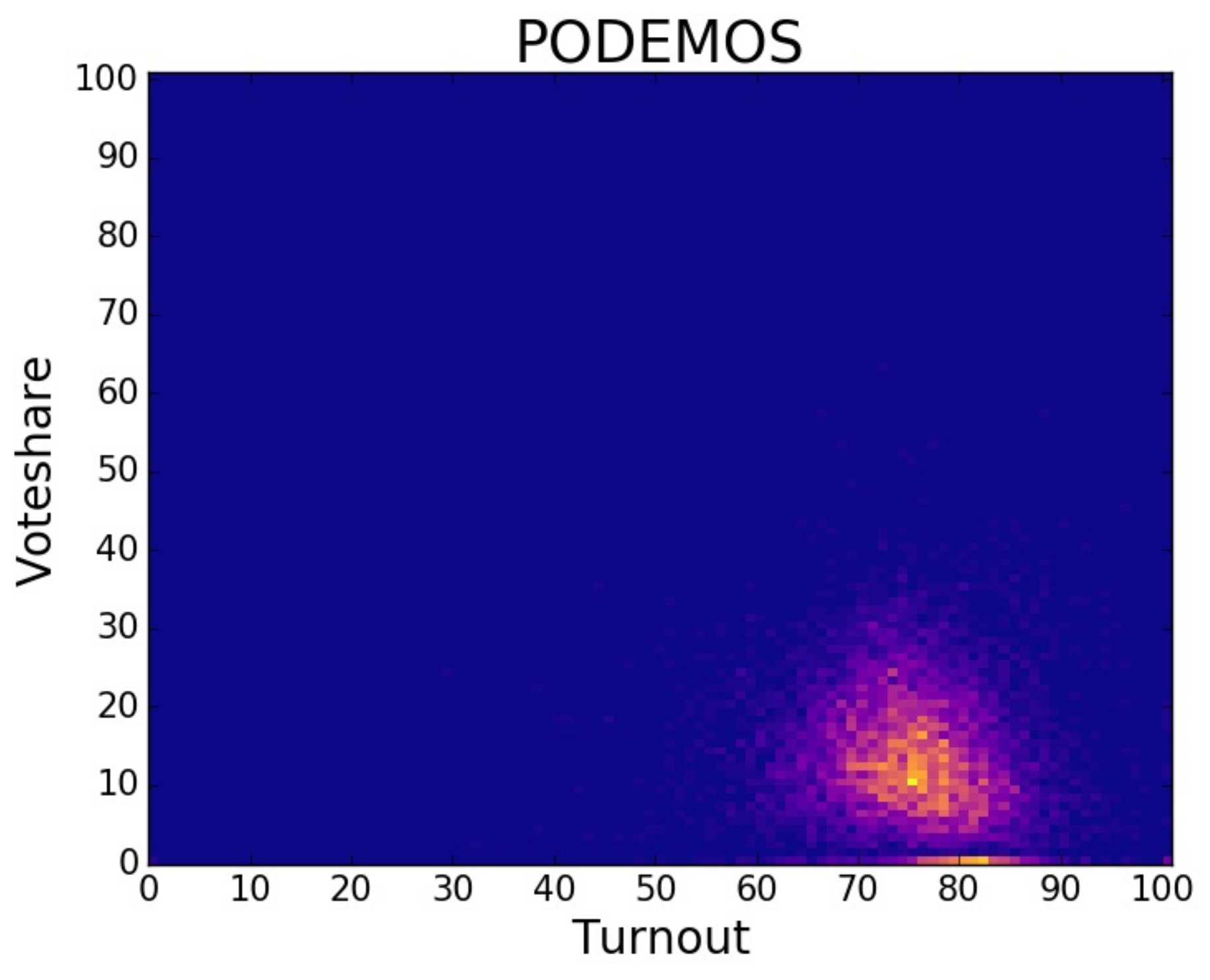}
\includegraphics[width=0.4\columnwidth]{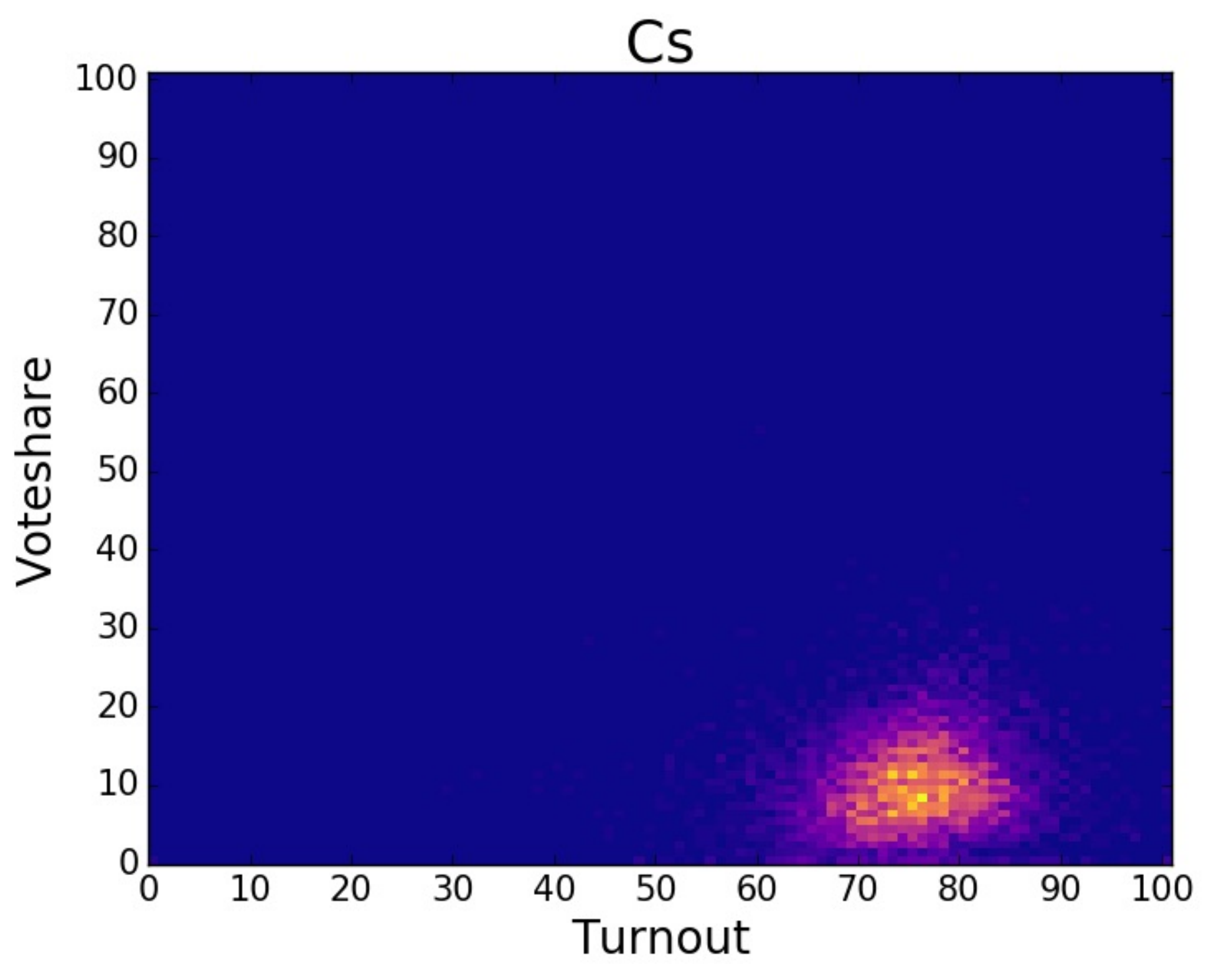}
\caption{Heatmaps plotting the percentage (in color scale) of municipalities where a given political party has received a certain percentage of votes, as a function of the relative participation. These are results associated to December 2015 elections. According to Klimek et al. \cite{klimek} a smear out of the cluster towards the top-right corner of the heat map is a sign of incremental fraud, whereas extreme fraud would occur for bimodal distributions where a cluster emerges at the top-right corner.}
\label{fig:heatmap_todos_2015}
\end{figure*}

\begin{table}[]
\centering
\begin{tabular}{l|l|l|l}
Precinct&$N_{\text{mun}}$&Range 2015&Range 2016\\
\hline
\hline
Cadiz&$44$&$383$--$162564$&$376$--$162111$\\
Tarragona&$184$&$36$--$89136$&$36$--$89031$\\
A-Coruna&$93$&$1020$--$196251$&$1007$--$196492$\\
Zaragoza&$304$&$12$--$98267$&$12$--$98399$\\
Valencia-Valencia&$284$&$37$--$43858$&$39$--$58015$\\
Leon&$211$&$42$--$101272$&$40$--$100862$\\
Avila&$248$&$13$--$43810$&$13$--$43759$\\
Gipuzkoa&$88$&$85$--$145679$&$85$--$145167$\\
Granada&$172$&$57$--$182735$&$119$--$182450$\\
La-Rioja&$174$&$5$--$108493$&$5$--$108755$\\
Lugo&$67$&$186$--$76951$&$183$--$77209$\\
Castellon-Castello&$135$&$16$--$116252$&$14$--$116049$\\
Jaen&$97$&$297$--$89693$&$285$--$89487$\\
Cordoba&$75$&$305$--$255629$&$301$--$255476$\\
Barcelona&$320$&$25$--$184321$&$23$--$182889$\\
Araba-Alava&$51$&$140$--$183368$&$136$--$183559$\\
Valladolid&$225$&$24$--$243129$&$20$--$242609$\\
Teruel&$236$&$10$--$25834$&$10$--$25959$\\
Ourense&$92$&$331$--$85240$&$314$--$85329$\\
Palencia&$191$&$13$--$63236$&$15$--$63072$\\
Navarra&$272$&$17$--$145462$&$16$--$145189$\\
Asturias&$78$&$146$--$223974$&$139$--$223268$\\
Huelva&$79$&$47$--$111520$&$45$--$111093$\\
Pontevedra&$62$&$535$--$232242$&$542$--$232465$\\
Soria&$183$&$8$--$28459$&$7$--$28540$\\
Madrid&$199$&$36$--$176867$&$38$--$176527$\\
Sevilla&$115$&$272$--$97848$&$266$--$98048$\\
Huesca&$202$&$18$--$38062$&$21$--$37988$\\
Illes-Balears&$67$&$192$--$275448$&$178$--$275883$\\
Lleida&$231$&$51$--$90807$&$48$--$90289$\\
Cantabria&$102$&$64$--$135418$&$66$--$135258$\\
Murcia&$45$&$492$--$308510$&$482$--$309387$\\
Malaga&$113$&$133$--$83852$&$142$--$84163$\\
Ciudad-Real&$102$&$81$--$57081$&$84$--$57248$\\
Cuenca&$238$&$10$--$40719$&$10$--$40722$\\
Caceres&$223$&$70$--$72783$&$75$--$74773$\\
Segovia&$209$&$17$--$38948$&$16$--$38867$\\
Guadalajara&$288$&$8$--$58795$&$7$--$59179$\\
Girona&$221$&$59$--$63288$&$62$--$63305$\\
Salamanca&$362$&$16$--$116942$&$17$--$117091$\\
Almeria&$103$&$70$--$139271$&$62$--$139412$\\
Bizkaia&$119$&$116$--$78875$&$111$--$78573$\\
Toledo&$204$&$6$--$61813$&$6$--$61731$\\
Santa-Cruz-de-Tenerife&$54$&$887$--$159534$&$919$--$159695$\\
Albacete&$87$&$54$--$130156$&$55$--$130572$\\
Alicante-Alacant&$141$&$44$--$234975$&$46$--$234691$\\
Las-Palmas&$34$&$508$--$292289$&$496$--$292504$\\
Badajoz&$165$&$63$--$111575$&$66$--$114755$\\
Zamora&$248$&$34$--$51358$&$34$--$51049$\\
Burgos&$371$&$5$--$134171$&$5$--$133923$\\
\hline      
\hline
\end{tabular}
\caption{List of precincts with their number of different municipalities and the voting population ranges (by voting population we mean the number of possible voters). The largest cities, such as Madrid, Barcelona, Bilbao, Sevilla, Valencia, Zaragoza and Malaga have been subsequently divided into electoral districts and we have treated these latter districts as municipalities.}
\label{table:precincts}
\end{table}

\noindent {\bf Availability of data and materials. }
The datasets supporting the conclusions of this article are available under request.\\

\noindent {\bf Acknowledgments. }
We thank M.Ant\`onia Tugores for her helpful assistance in the data gathering process, Luis F. Lafuerza for fruitful suggestions and D. Kobak for pointing out to recent works \cite{new, new2}. LL acknowledges funding from EPSRC Early Career Fellowship EP/P01660X/1.\\

\noindent {\bf Conflicts of interest. } The authors declare that there is no conflict of interest regarding the publication of this paper.\\
  
\noindent {\bf Author's contributions}
    JFG and LL designed the study, JFG performed the data gathering and analysis, JFG and LL interpreted the results, LL wrote the manuscript and JFG and LL revised the manuscript.

\bibliography{apssamp}


\end{document}